\documentclass[smallextended]{svjour3}	
\usepackage{url}
\usepackage{bm}		
\usepackage{graphicx}	
\usepackage{amsmath}	
\usepackage{amssymb}	
\usepackage{hyperref}

\journalname{General Relativity and Gravitation}

\begin{document}


\title{Hawking radiation inside a Schwarzschild black hole}

\author{Andrew J S Hamilton}
\institute{Andrew J S Hamilton \at
JILA \\
Box 440 \\
U. Colorado \\
Boulder \\
CO 80309 \\
USA \\
\email{Andrew.Hamilton@colorado.edu}
}	

\def\hksqrt{\mathpalette\DHLhksqrt}
\def\DHLhksqrt#1#2{\setbox0=\hbox{$#1\sqrt{#2\,}$}\dimen0=\ht0
\advance\dimen0-0.2\ht0
\setbox2=\hbox{\vrule height\ht0 depth -\dimen0}%
{\box0\lower0.4pt\box2}}

\newcommand{\simproptoinn}[2]{\mathrel{\vcenter{
  \offinterlineskip\halign{\hfil$##$\cr
    #1\propto\cr\noalign{\kern2pt}#1\sim\cr\noalign{\kern-2pt}}}}}
\newcommand{\simpropto}{\mathpalette\simproptoinn\relax}

\newcommand{\dd}{d}
\newcommand{\ddsq}{\dd^2\mkern-1.5mu}
\newcommand{\ddd}{\dd^3\mkern-1.5mu}
\newcommand{\dddd}{\dd^4\mkern-1.5mu}
\newcommand{\DD}{D}
\newcommand{\ee}{e}
\newcommand{\im}{i}
\newcommand{\Ei}{\textrm{Ei}}
\newcommand{\perpperp}{\perp\!\!\perp}
\newcommand{\ppartial}{\partial^2\mkern-1mu}
\newcommand{\nn}{\nonumber\\}
\renewcommand{\Re}{\textrm{Re}}
\renewcommand{\Im}{\textrm{Im}}

\newcommand{\aniso}{\textrm{aniso}}
\newcommand{\diag}{\textrm{diag}}
\newcommand{\edge}{\textrm{edge}}
\newcommand{\eff}{\textrm{eff}}
\newcommand{\emit}{\textrm{em}}
\newcommand{\expinf}{\xi}
\newcommand{\flux}{\textrm{flux}}
\newcommand{\HeunC}{H}
\newcommand{\hor}{\textrm{h}}
\newcommand{\iso}{\textrm{iso}}
\newcommand{\jel}{\text{\sl j}}
\newcommand{\Lz}{L}
\newcommand{\Msun}{\textrm{M}_\odot}
\newcommand{\uel}{\text{\sl u}}
\newcommand{\vel}{\text{\sl v}}
\newcommand{\inn}{\textrm{in}}
\newcommand{\obs}{\textrm{obs}}
\newcommand{\out}{\textrm{ou}}
\newcommand{\sep}{\textrm{sep}}
\newcommand{\stat}{\textrm{stat}}
\newcommand{\vac}{\textrm{vac}}
\newcommand{\unit}[1]{\, \textrm{#1}}

\newcommand{\be}{\bm{e}}
\newcommand{\bg}{\bm{g}}
\newcommand{\bJ}{\bm{J}}
\newcommand{\bk}{\bm{k}}
\newcommand{\bL}{\bm{L}}
\newcommand{\bp}{\bm{p}}
\newcommand{\bu}{\bm{u}}
\newcommand{\bv}{\bm{v}}
\newcommand{\bx}{\bm{x}}
\newcommand{\by}{\bm{y}}
\newcommand{\bgamma}{\bm{\gamma}}

\newcommand{\Apot}{{\cal A}}
\newcommand{\hatA}{\hat{A}}
\newcommand{\Br}{B}
\newcommand{\betar}{\lambda_r}
\newcommand{\Cx}{C_x}
\newcommand{\Cy}{C_y}
\newcommand{\Dx}{D_x}
\newcommand{\Dy}{D_y}
\newcommand{\Deltax}{\Delta_x}
\newcommand{\Deltaxinf}{\Delta_{x , \textrm{inf}}}
\newcommand{\Deltaxev}{\Delta_{x , \textrm{ev}}}
\newcommand{\Deltay}{\Delta_y}
\newcommand{\Er}{E}
\newcommand{\Fz}{{\tilde F}}
\newcommand{\starF}{\,{}^\ast\!F}
\newcommand{\Jx}{J_x}
\newcommand{\Jy}{J_y}
\newcommand{\Mass}{{\cal M}}
\newcommand{\Mbh}{M_\bullet}
\newcommand{\Mdot}{\dot{M}}
\newcommand{\KCarter}{{\cal K}}
\newcommand{\NUT}{{\cal N}}
\newcommand{\px}{p^x}
\newcommand{\Px}{P_x}
\newcommand{\Py}{P_y}
\newcommand{\QCarter}{{\cal Q}}
\newcommand{\Qelec}{Q}
\newcommand{\Qelecbh}{\Qelec_\bullet}
\newcommand{\Qmag}{{\cal Q}}
\newcommand{\DeltaQelec}{\Delta \Qelec}
\newcommand{\DeltaQmag}{\Delta \Qmag}
\newcommand{\rhosep}{\rho_\textrm{s}}
\newcommand{\rhox}{\rho_x}
\newcommand{\rhoy}{\rho_y}
\newcommand{\rem}{r_\textrm{em}}
\newcommand{\robs}{r_\textrm{obs}}
\newcommand{\Uinf}{U}
\newcommand{\Ur}{{\cal R}}
\newcommand{\Utheta}{\Theta}
\newcommand{\rc}{{\scriptstyle R}}
\newcommand{\tc}{{\scriptstyle T}}
\newcommand{\smallrc}{{\scriptscriptstyle R}}
\newcommand{\smalltc}{{\scriptscriptstyle T}}
\newcommand{\smallzero}{{\scriptscriptstyle 0}}
\newcommand{\Ux}{U_x}
\newcommand{\Uy}{U_y}
\newcommand{\xin}{x_\textrm{in}}
\newcommand{\Xx}{X_x}
\newcommand{\Xy}{X_y}
\newcommand{\Yx}{Y_x}
\newcommand{\Yy}{Y_y}
\newcommand{\Zx}{Z_x}
\newcommand{\Zy}{Z_y}
\newcommand{\omegax}{\omega_x}
\newcommand{\omegay}{\omega_y}

\newcommand{\kp}{k^\prime}
\newcommand{\rout}{r}
\newcommand{\rastout}{r^{\ast}}
\newcommand{\tout}{t}
\newcommand{\tauin}{\tau}
\newcommand{\tauout}{\tau}
\newcommand{\psiout}{\psi}
\newcommand{\omegain}{\nu}
\newcommand{\omegaout}{\omega}


\hyphenation{Schwarz-s-child}

\newcommand{\stpcollfig}{
    \begin{figure}[t!]
    \begin{center}
    \leavevmode
    \includegraphics[scale=.49]{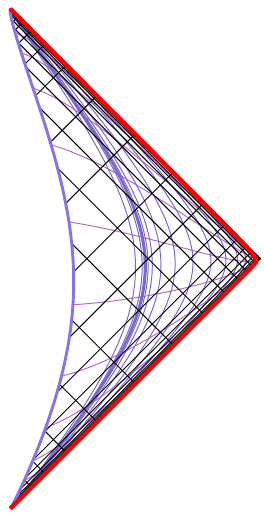}
    \hspace{0em}
    \includegraphics[scale=.49]{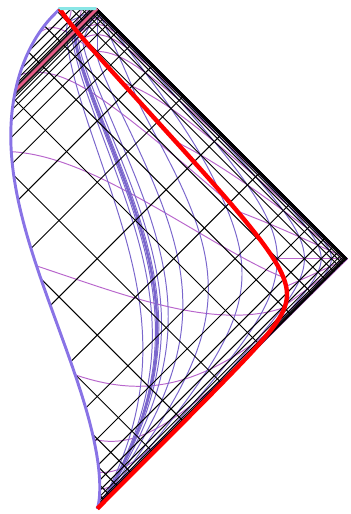}
    \hspace{-1em}
    \includegraphics[scale=.49]{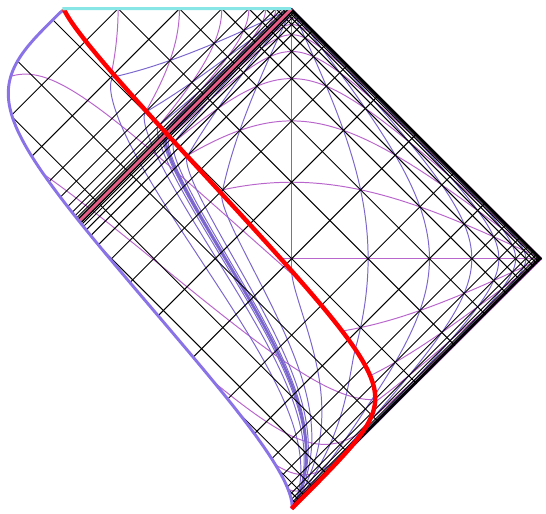}
    \hspace{-3em}
    \includegraphics[scale=.49]{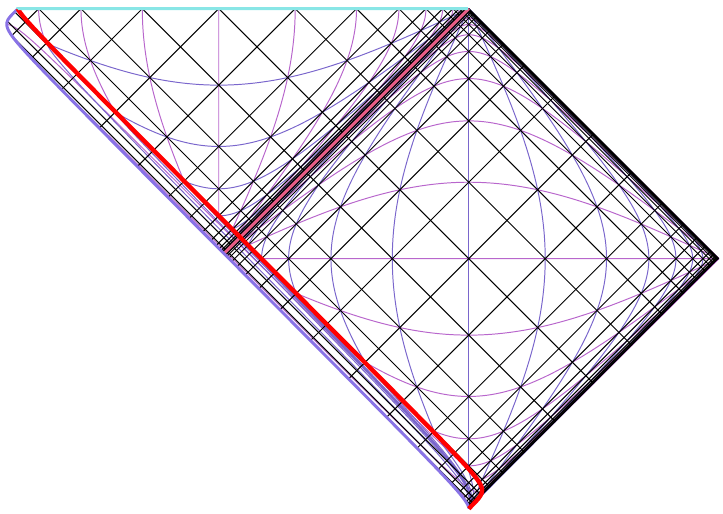}
    \hspace{-3.5em}
    \includegraphics[scale=.49]{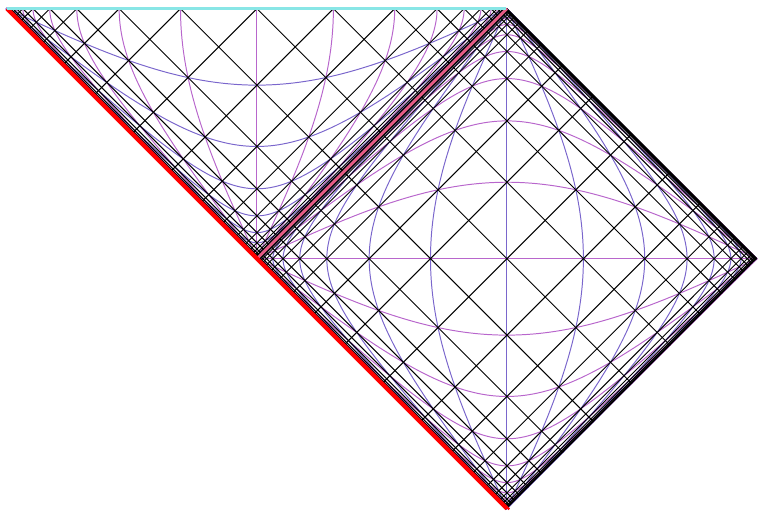}
    \caption[Penrose diagrams of Oppenheimer-Snyder collapse]{
    \label{stpcoll}
Sequence of Penrose diagrams illustrating
the Oppenheimer-Snyder collapse of a pressureless, spherical star
to a Schwarzschild black hole,
progressing in time of collapse from left to right.
On the left,
the collapse is to the future of
an observer at the centre of the diagram;
on the right, the collapse is to the past of
an observer at the centre of the diagram.
The diagrams are at times
$-32M$, $-8M$, $0M$, $8M$, and $32M$
relative to the middle diagram.
On the left
the Penrose diagram resembles that of Minkowski space,
while on the right the diagram resembles that of the Schwarzschild geometry.
The thick (red) line that goes from the bottom of the diagram (past infinity)
to top left (singularity formation)
is the surface of the collapsing star.
Thin (blue and purple) lines are lines of constant radius $r$ and time $t$,
the time $t$ being Schwarzschild time outside the surface of the star,
and cosmological (Friedmann) time inside the star.
An animated version of this diagram is at
\url{http://jila.colorado.edu/~ajsh/bh/collapse.html\#penrose}.
    }
    \end{center}
    \end{figure}
}

\newcommand{\penroseschwskyfig}{
    \begin{figure}[b!]
    \begin{center}
    \leavevmode
    \includegraphics[scale=.85]{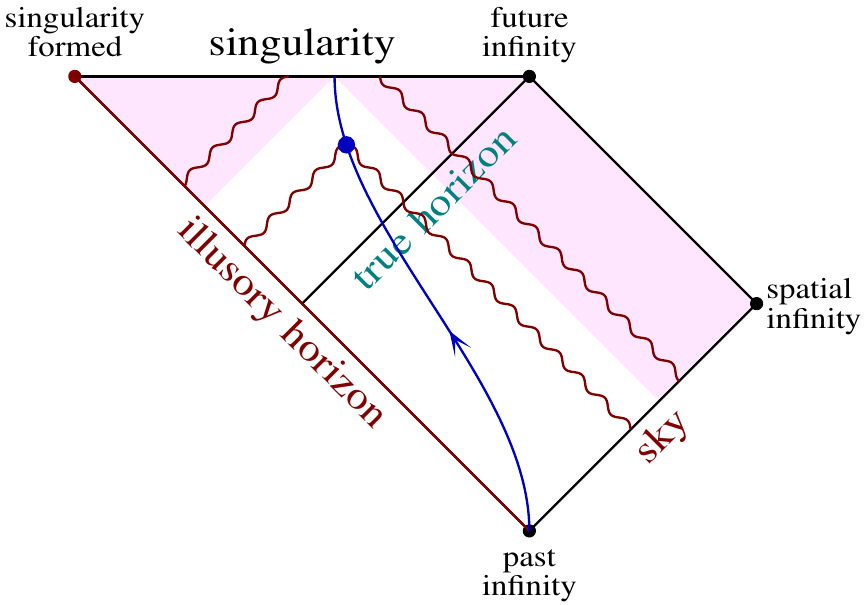}
    \caption[Penrose diagram]{
    \label{penroseschwsky}
Penrose diagram illustrating the trajectory of an observer
who falls to the singularity of a Schwarzschild black hole.
The observer sees Hawking radiation (wiggly lines) from the illusory horizon
below and from the sky above.
The lightly shaded region surrounds the causal diamond of the observer.
Hawking pair partners lie outside the causal diamond,
ensuring that there is no firewall contradiction.
    }
    \end{center}
    \end{figure}
}


\newcommand{\kappazerofig}{
    \begin{figure}[t!]
    \centering
    \includegraphics[scale=.5]{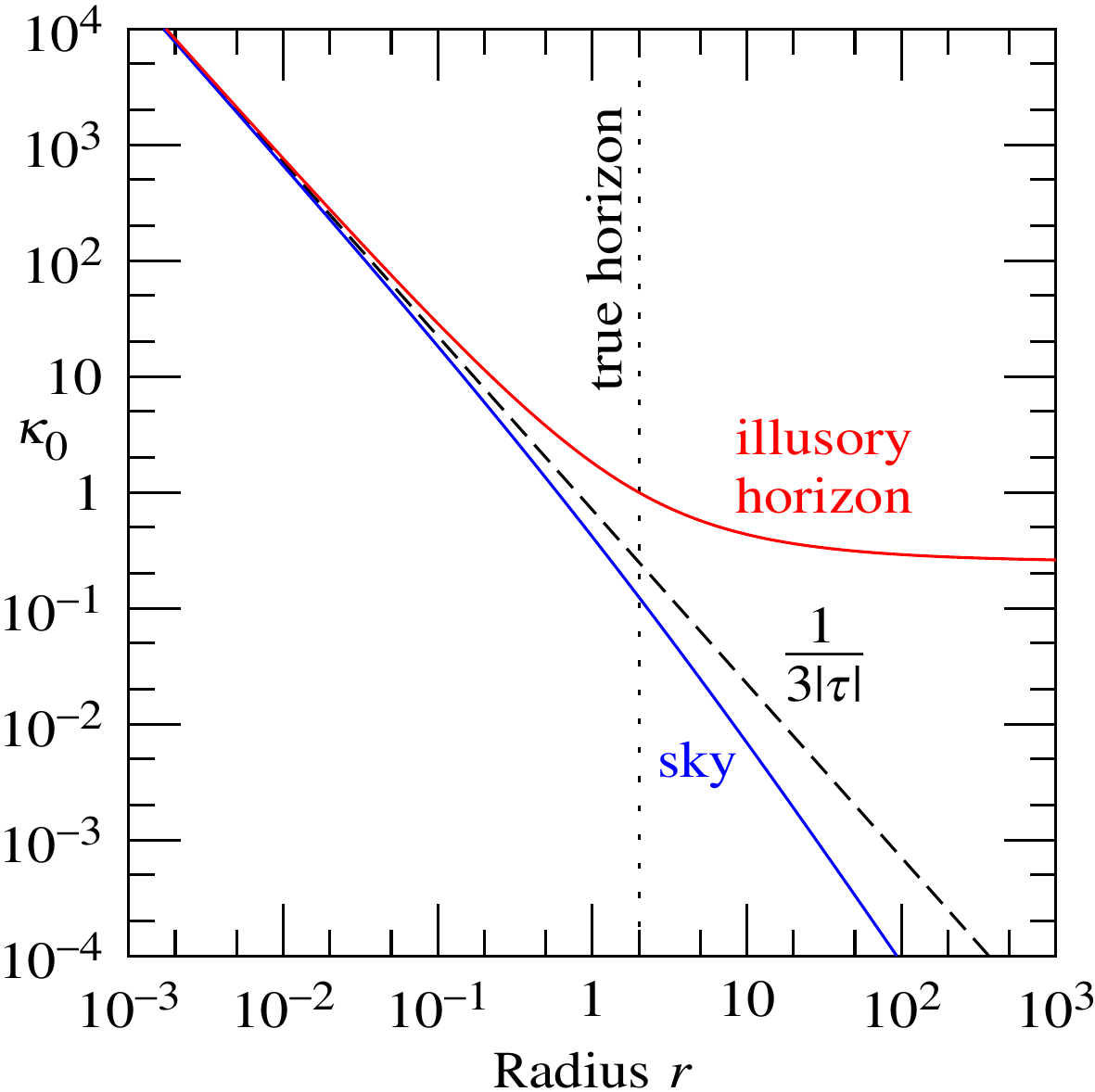}
    \caption[]{
    \label{kappa0}
Acceleration $\kappa_0$ on the illusory horizon directly below,
and in the sky directly above,
seen by a radially free-falling infaller at radius $r$.
The units are geometric ($c = G = M = 1$).
Both accelerations asymptote
to one third the reciprocal of the proper time $| \tau |$
left until the infaller hits the singularity,
indicated by the diagonal dashed line.
    }
    \end{figure}
}

\newcommand{\kappachifig}{
    \begin{figure*}[bt!]
    \centering
    \includegraphics[scale=.48]{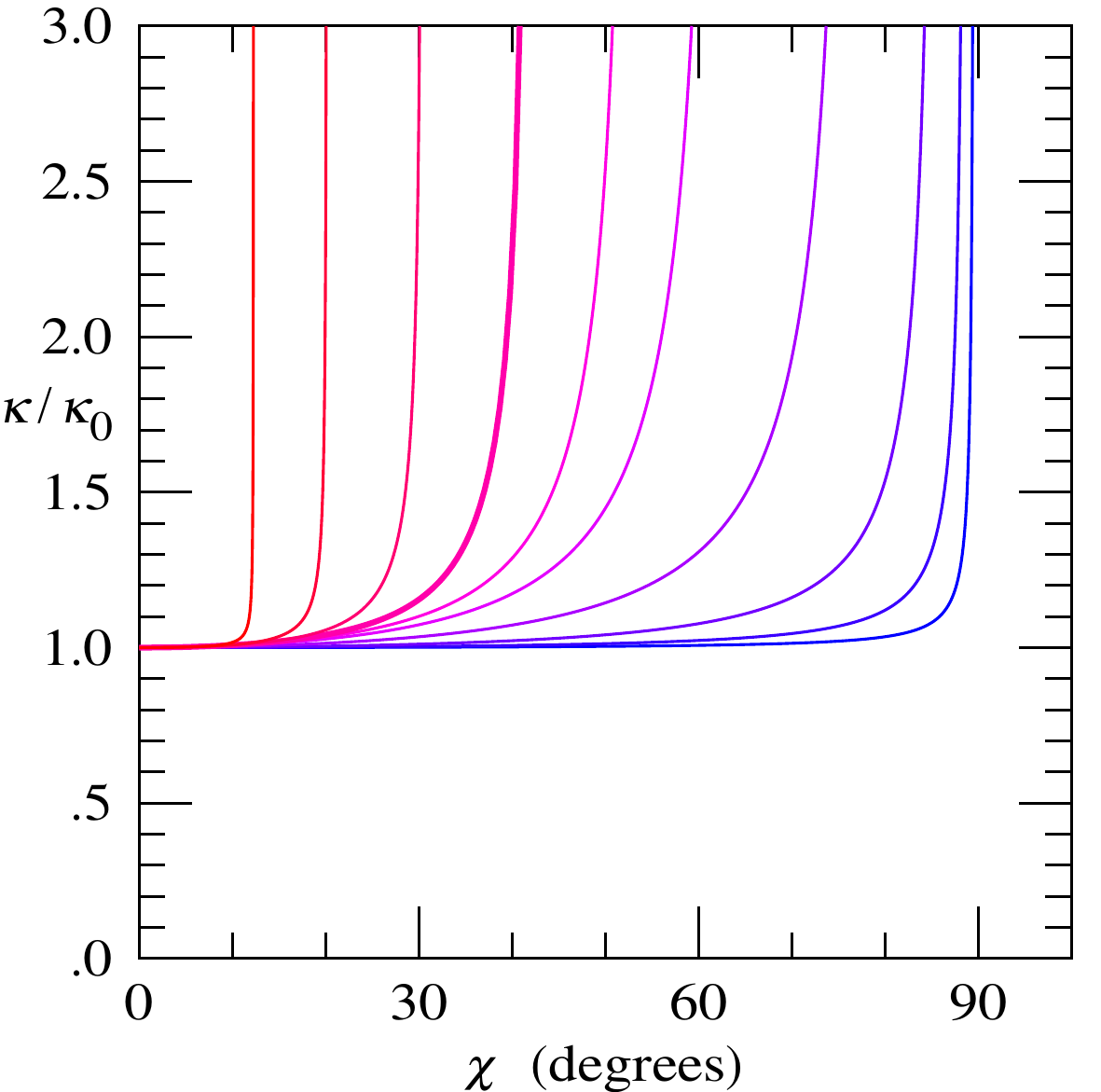}
    \includegraphics[scale=.48]{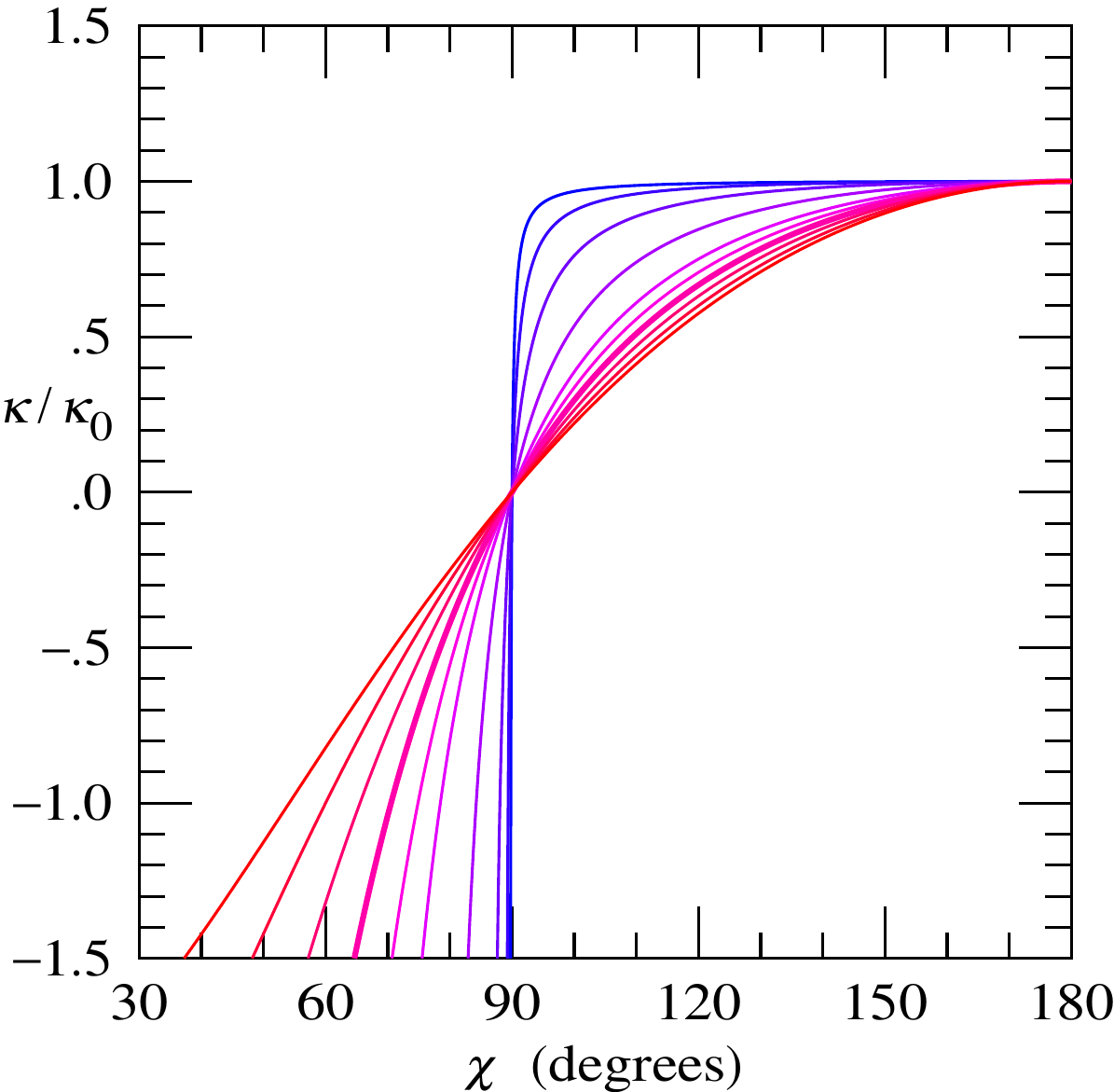}
    \caption[]{
    \label{kappachi}
Acceleration $\kappa$ on
(left) the illusory horizon
and (right) the sky above,
as seen by a radially free-falling non-rotating infaller,
relative to the acceleration $\kappa_0$ respectively
(left) directly downward and (right) directly upward,
as a function of the viewing angle $\chi$
($\chi = 0^\circ$ is directly downward,
$\chi = 180^\circ$ is directly upward).
The curves are as seen by the infaller at radius
(red to blue)
$16$, $8$, $4$, $2$ (true horizon, thick line),
$1$, $0.5$, $0.1$, $10^{-2}$, $10^{-3}$, and $10^{-4}$
geometric units
($c = G = M = 1$).
On the illusory horizon,
the acceleration is constant out to near
the perceived edge of the black hole, where the acceleration diverges.
On the sky,
the acceleration is positive (redshifting) in the upper hemisphere,
$\chi > 90^\circ$,
and negative (blueshifting) in the lower hemisphere,
$\chi < 90^\circ$.
Near the singularity (blue lines),
the illusory horizon and sky occupy the lower and upper hemispheres,
and the acceleration is almost constant over both hemispheres,
except for a thinning band near the equator.
    }
    \end{figure*}
}

\newcommand{\freqsfig}{
    \begin{figure*}[btp!]
    \centering
    \includegraphics[scale=.40]{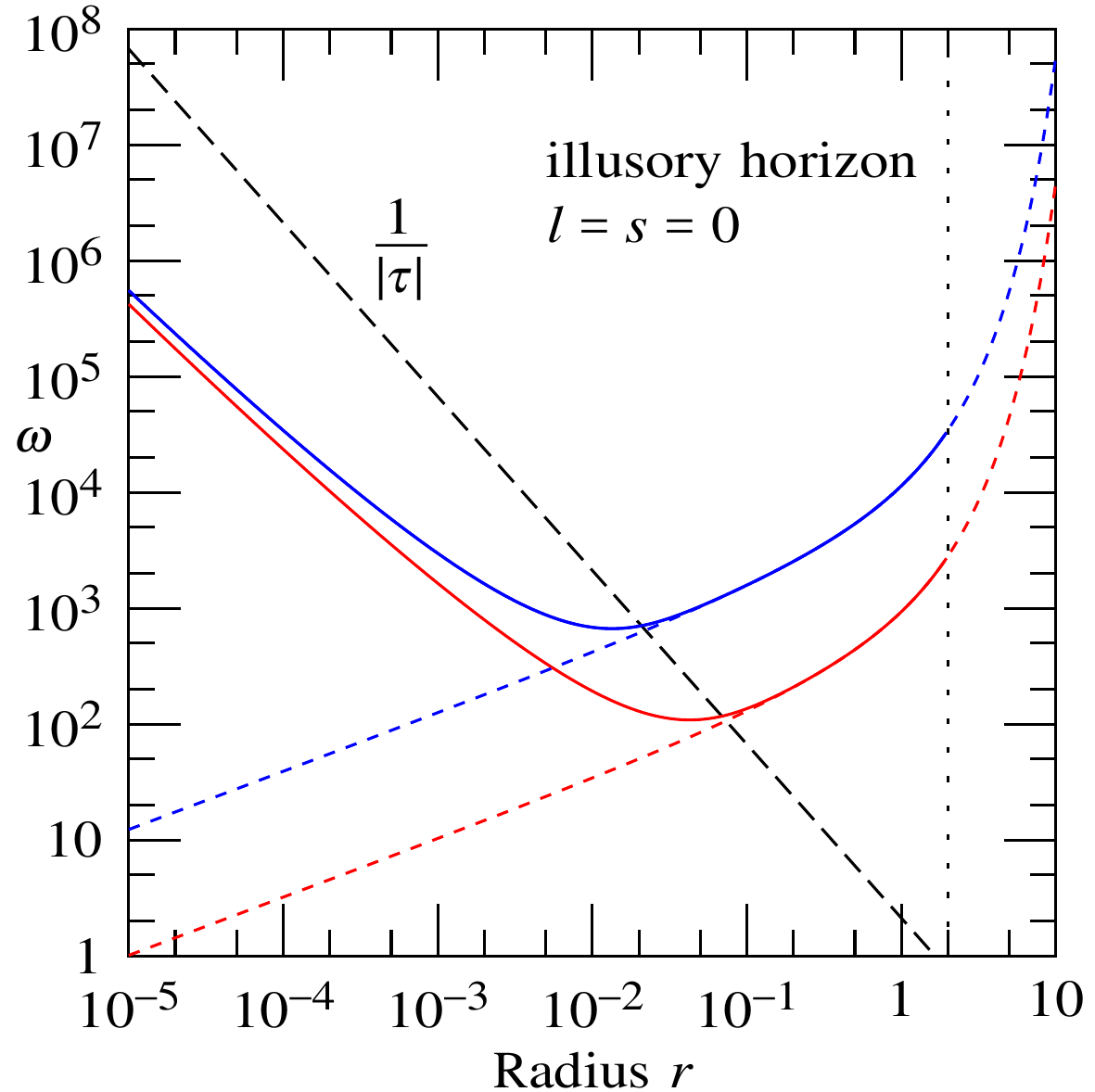}
    \includegraphics[scale=.40]{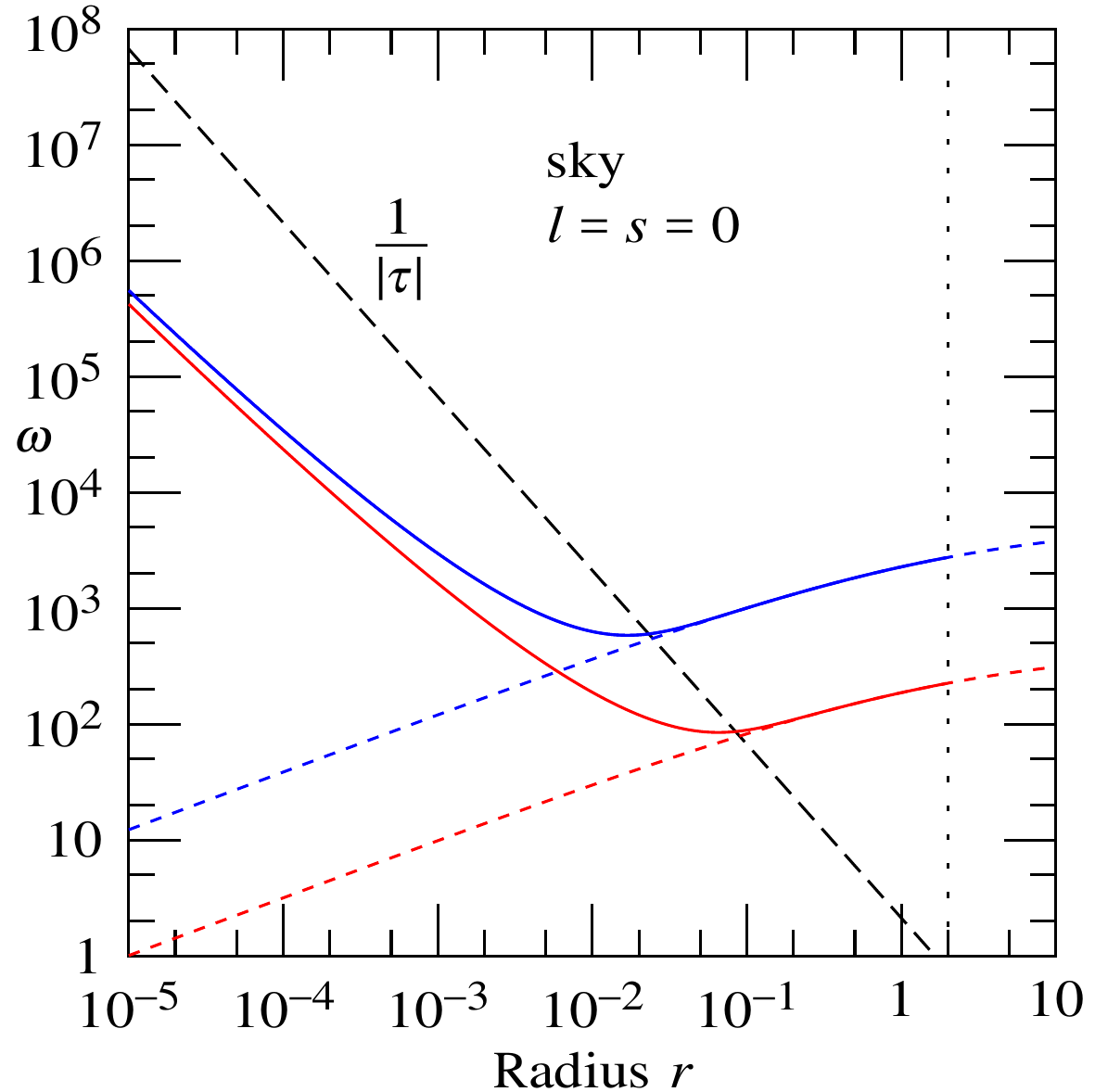}
    \includegraphics[scale=.40]{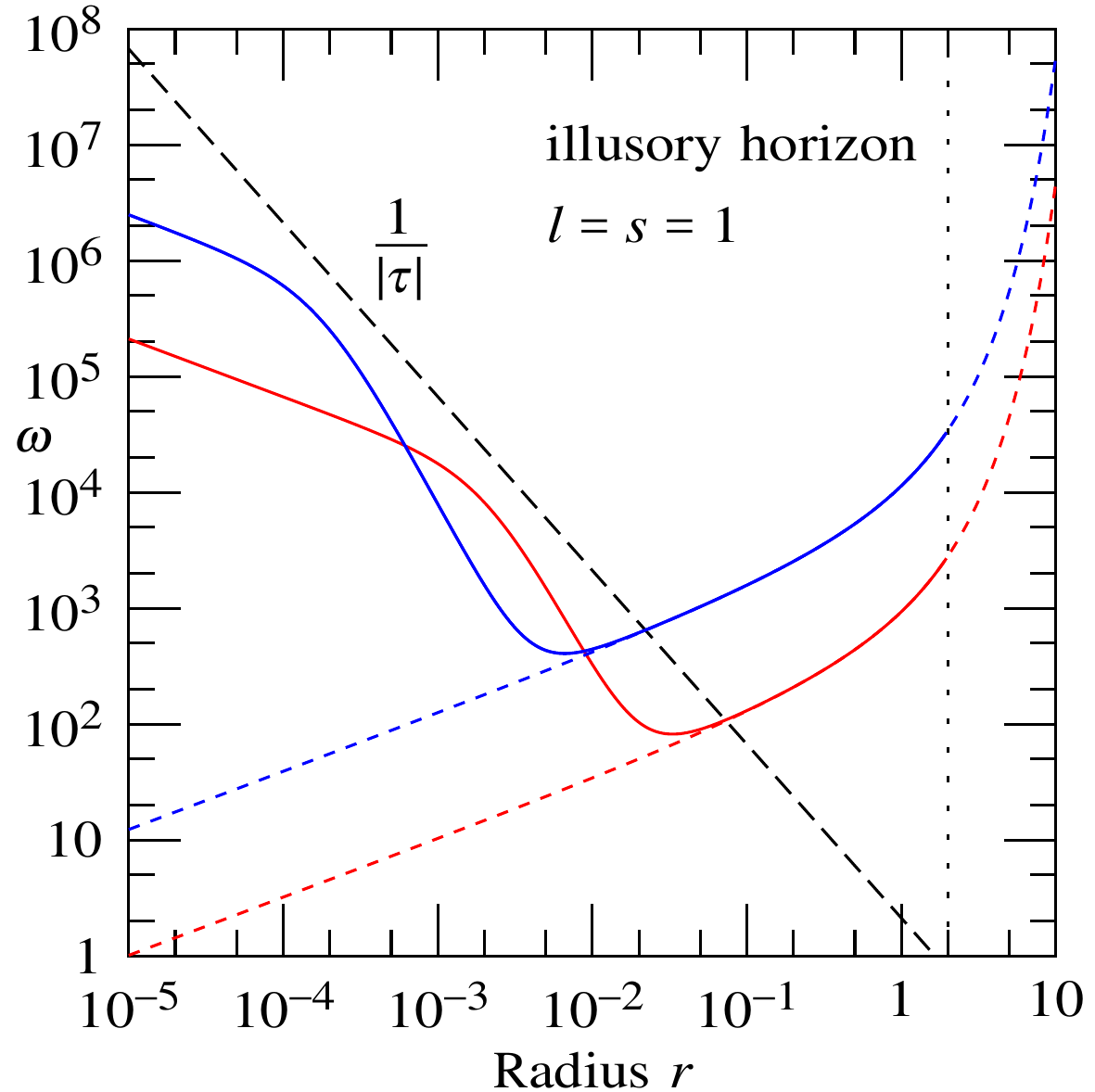}
    \includegraphics[scale=.40]{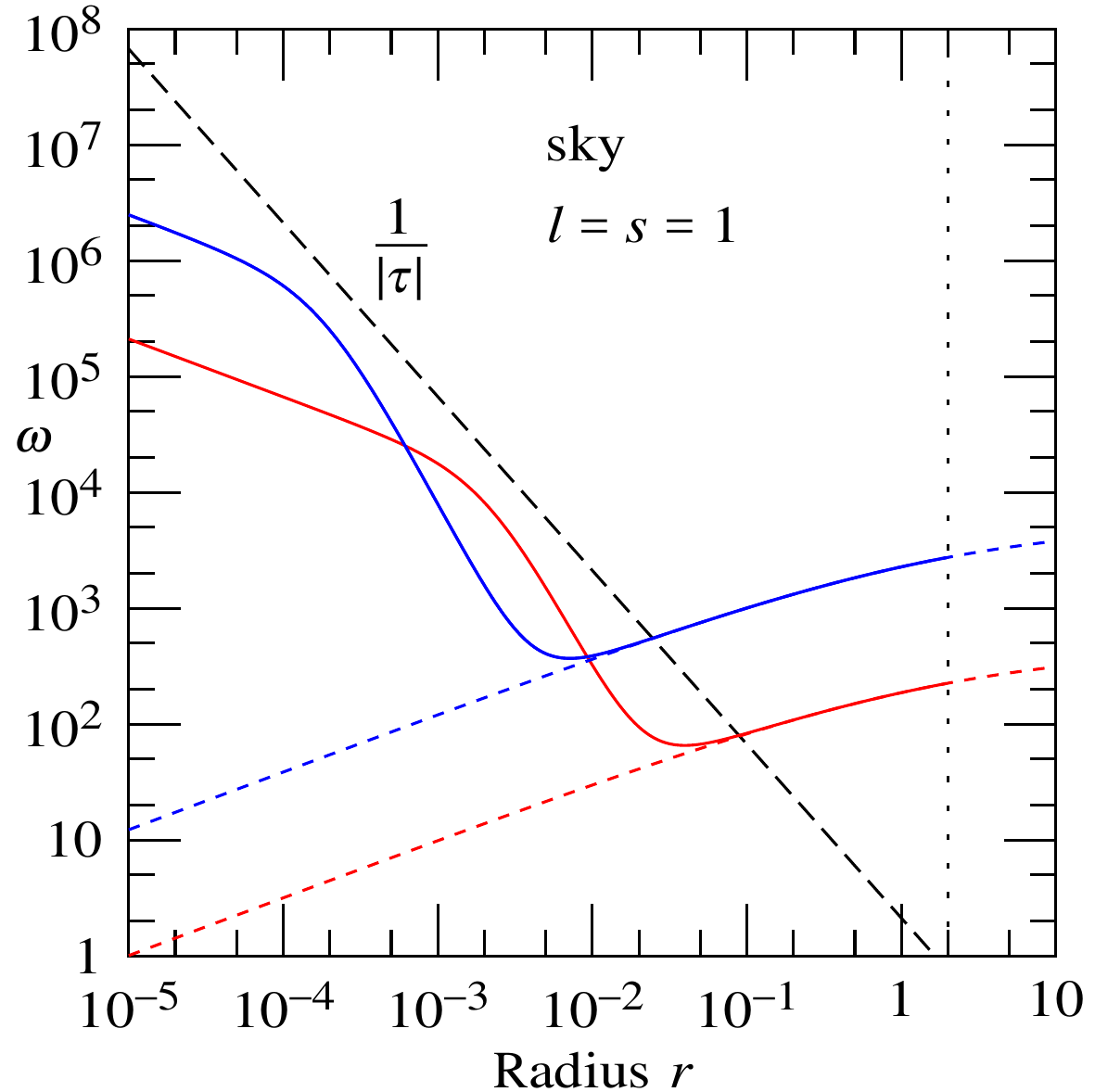}
    \includegraphics[scale=.40]{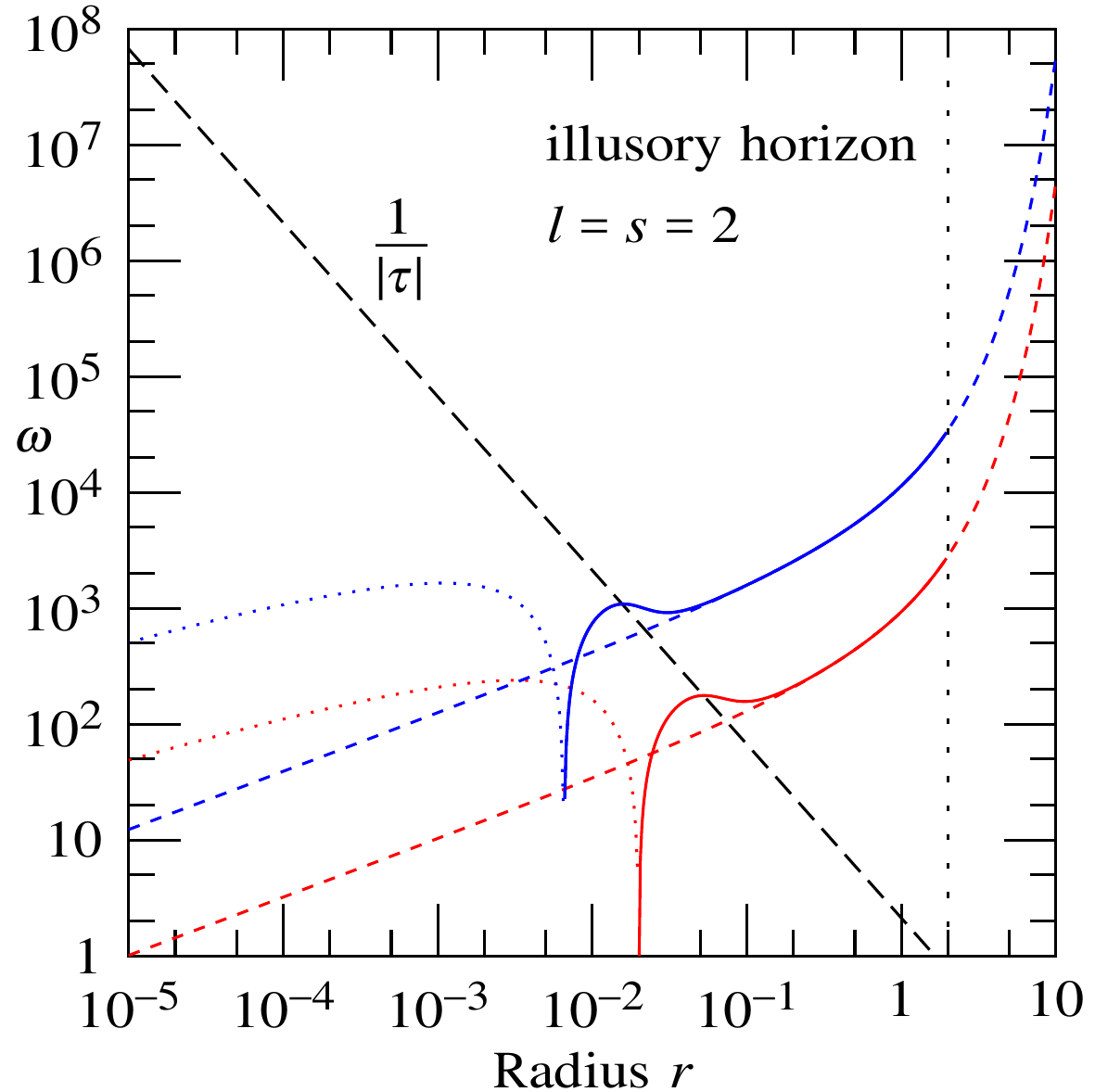}
    \includegraphics[scale=.40]{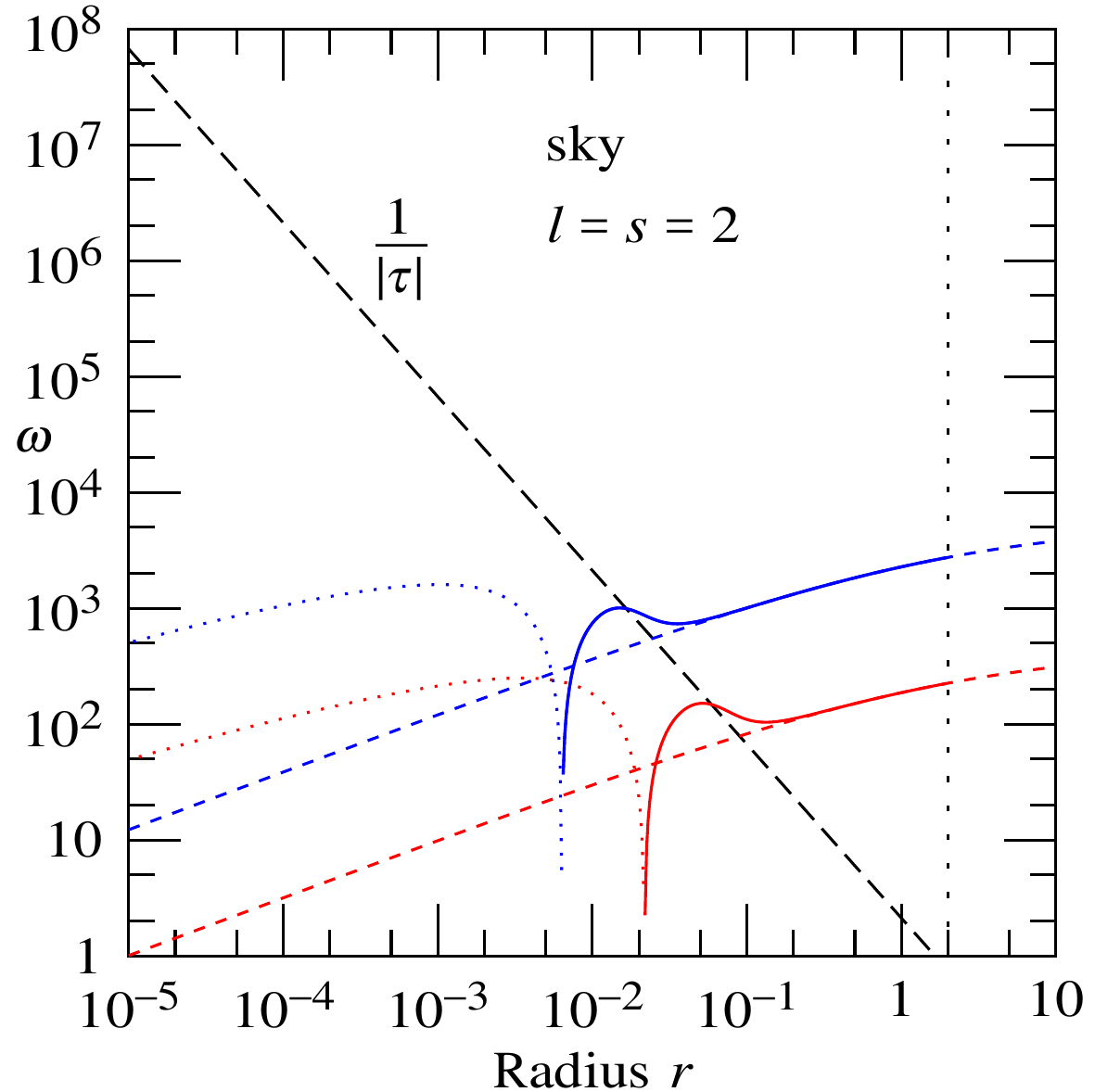}
    \caption[]{
    \label{freqs}
Frequency $\omegaout \equiv \dd \psiout / \dd \tauout$
of a wave $\varphi_\omegain \propto \ee^{- \im \psiout}$
of definite ``in'' frequency $\omegain$ from
(left)
the illusory horizon,
and
(right)
the sky above.
Results are shown for two representative ``in'' frequencies,
the second (red) frequency being a factor $\ee^{-5/2}$ lower than
the first (blue).
For each of the two ``in'' frequencies,
the frequency from the sky and illusory horizon
are mutually arranged to yield the same
observed frequency $\omegaout$ near the singularity.
The units are geometric ($c = G = M = 1$).
The wave has angular momentum $l$ and spin $s$ with
(top) $l = s = 0$ (scalar wave),
(middle) $l = s = 1$ (electromagnetic wave), and
(bottom) $l = s = 2$ (gravitational wave).
The solid (dotted where negative, which occurs for spin $s = 2$)
lines show the computed frequency $\omegaout$,
while the short dashed lines show
the frequency in the geometric-optics limit.
The diagonal (black) dashed line shows the inverse time
$1 / | \tau |$
left to hit the singularity.
The computed frequency agrees with the
geometric-optics limit as long as $| \omegaout \tauout | \gtrsim 1$
(above the diagonal).
The vertical dotted line marks the true horizon.
    }
    \end{figure*}
}

\newcommand{\kappasfig}{
    \begin{figure*}[bt!]
    \centering
    \includegraphics[scale=.40]{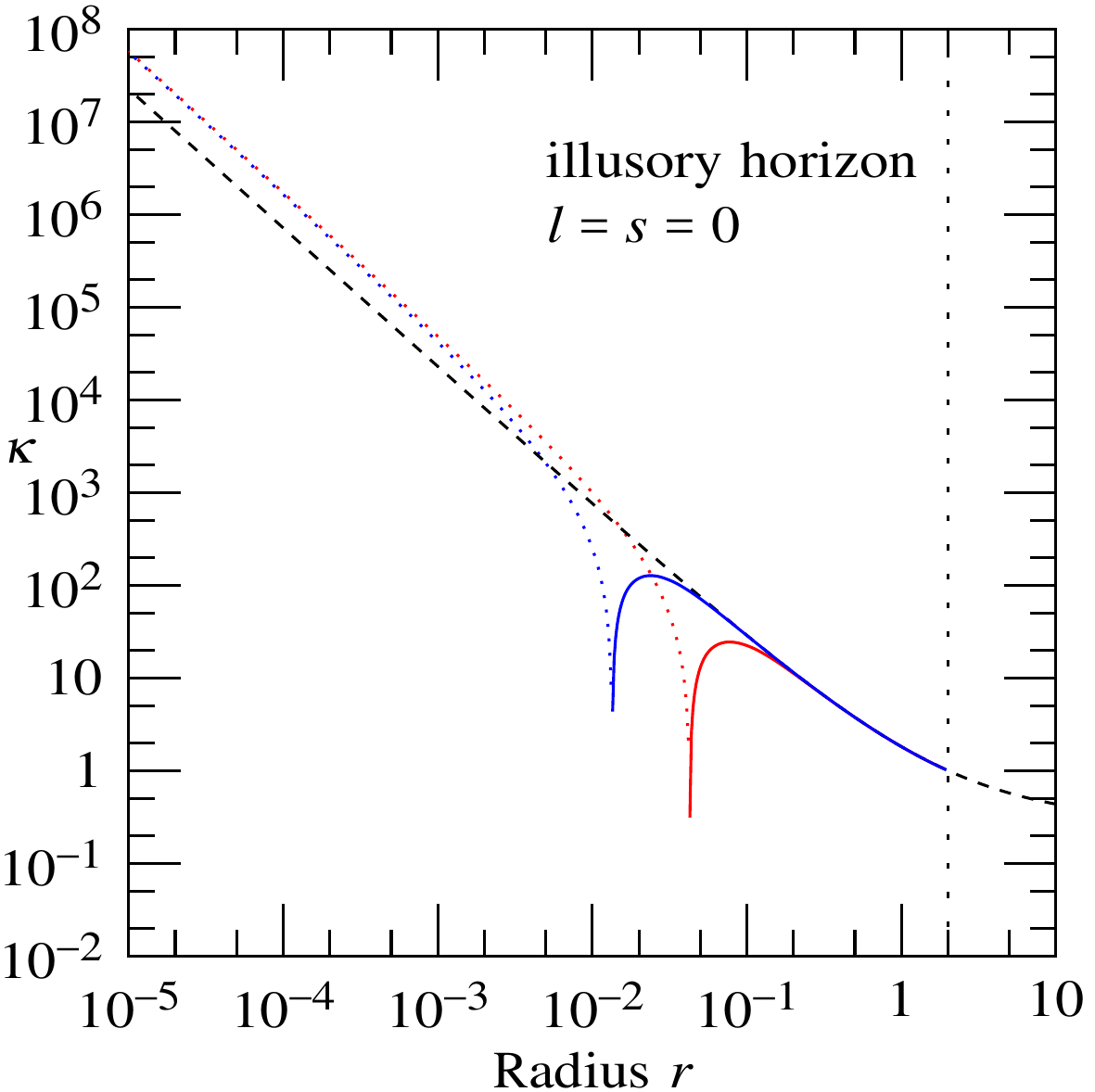}
    \includegraphics[scale=.40]{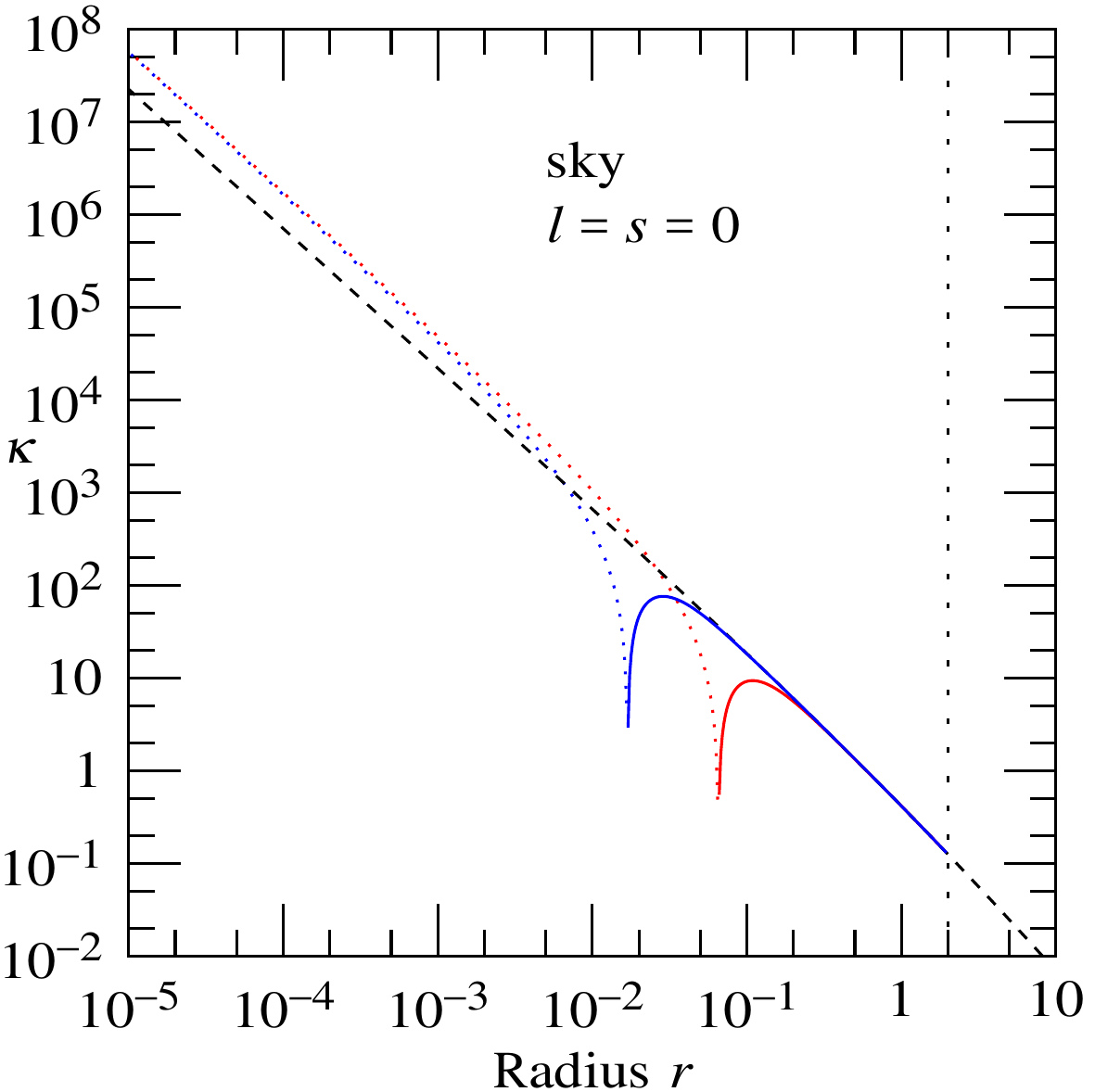}
    \includegraphics[scale=.40]{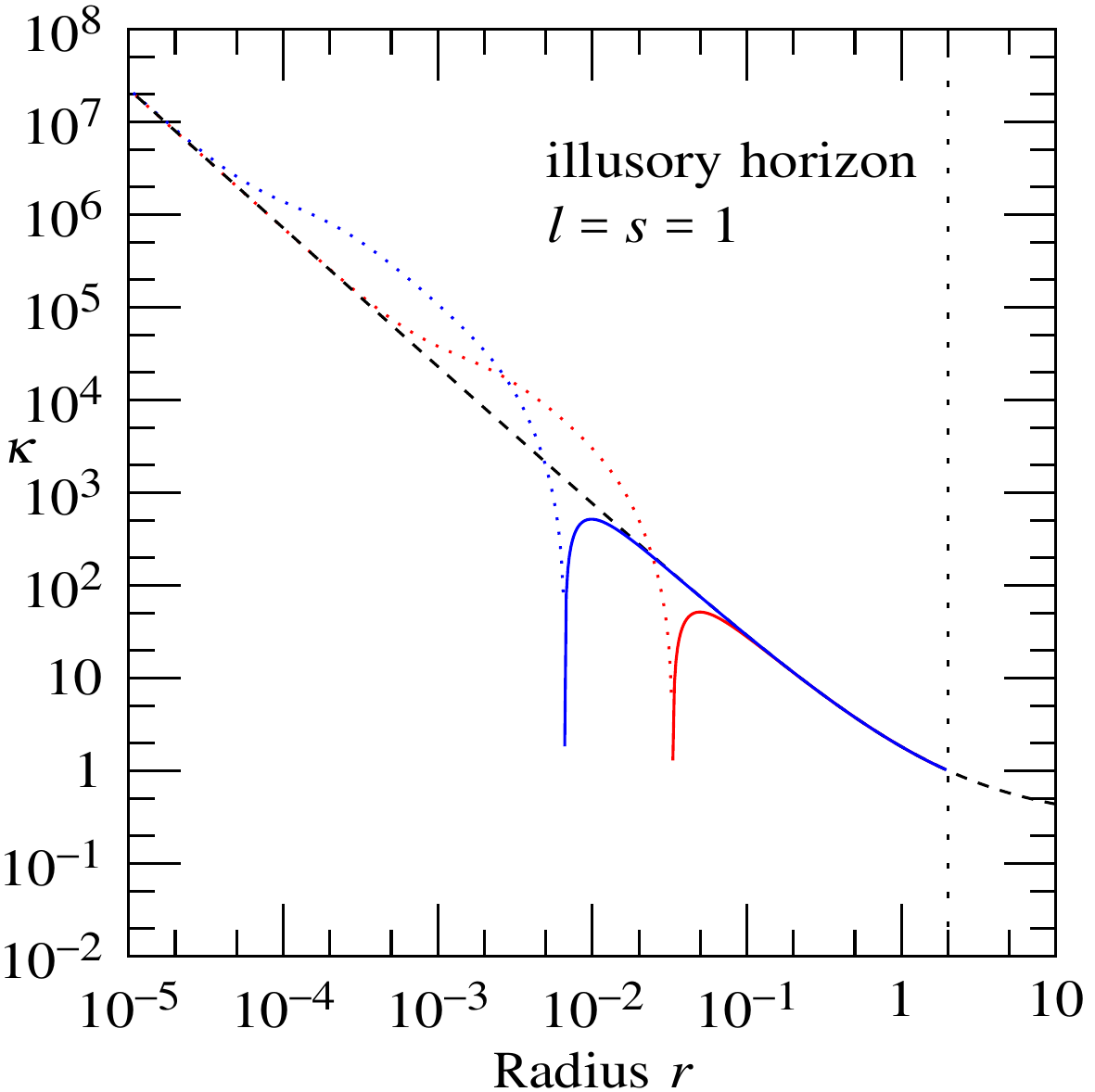}
    \includegraphics[scale=.40]{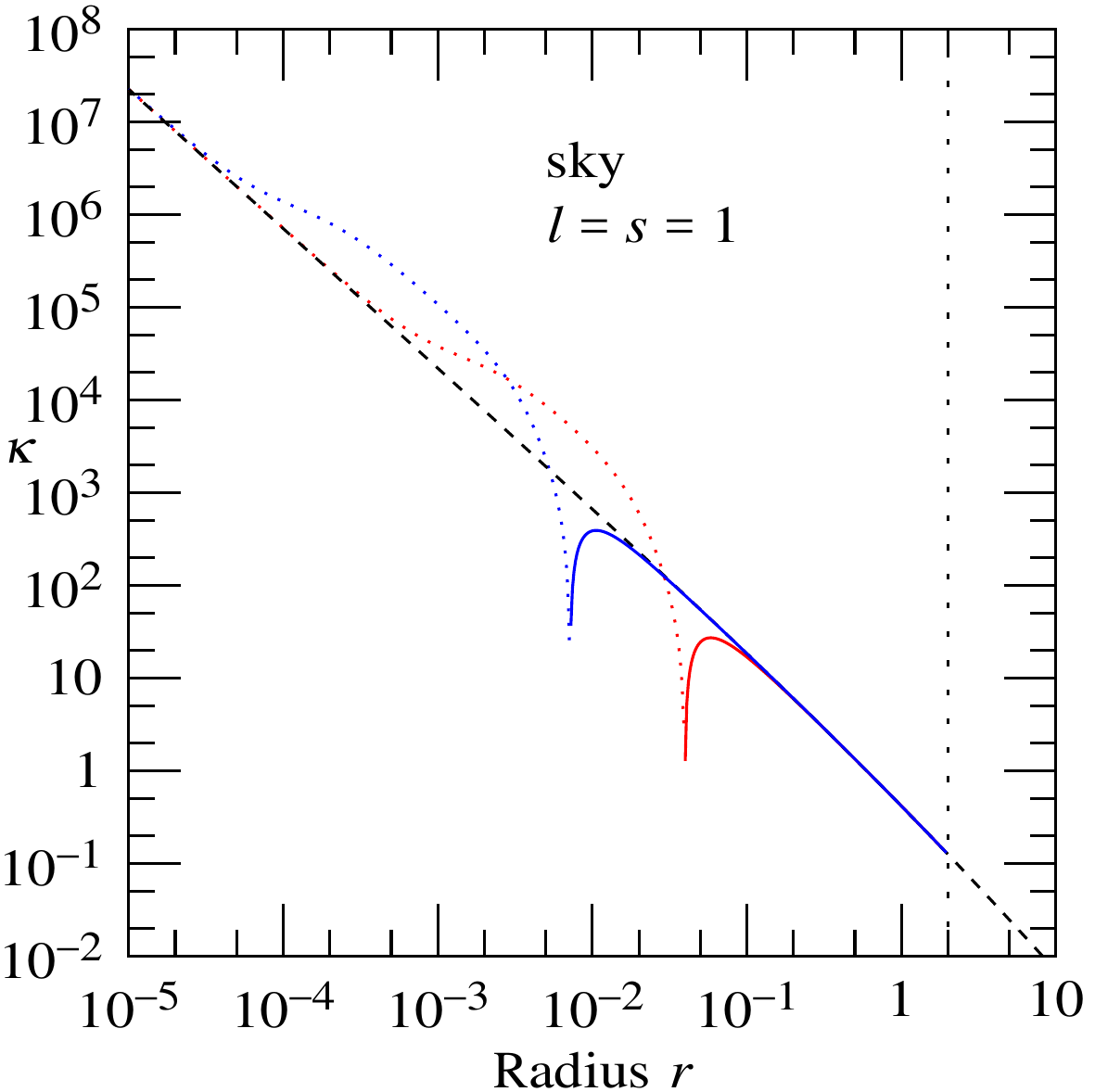}
    \includegraphics[scale=.40]{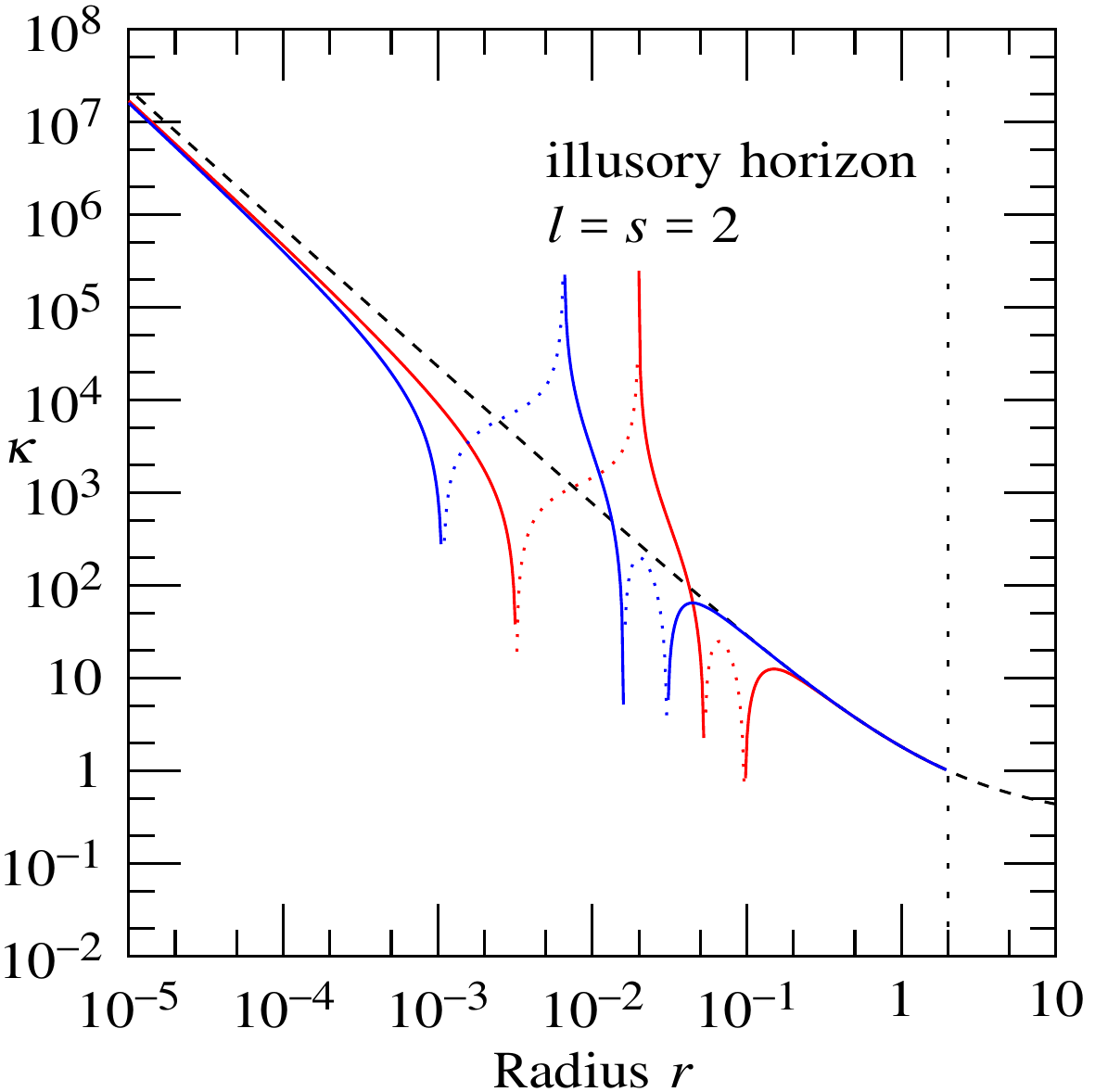}
    \includegraphics[scale=.40]{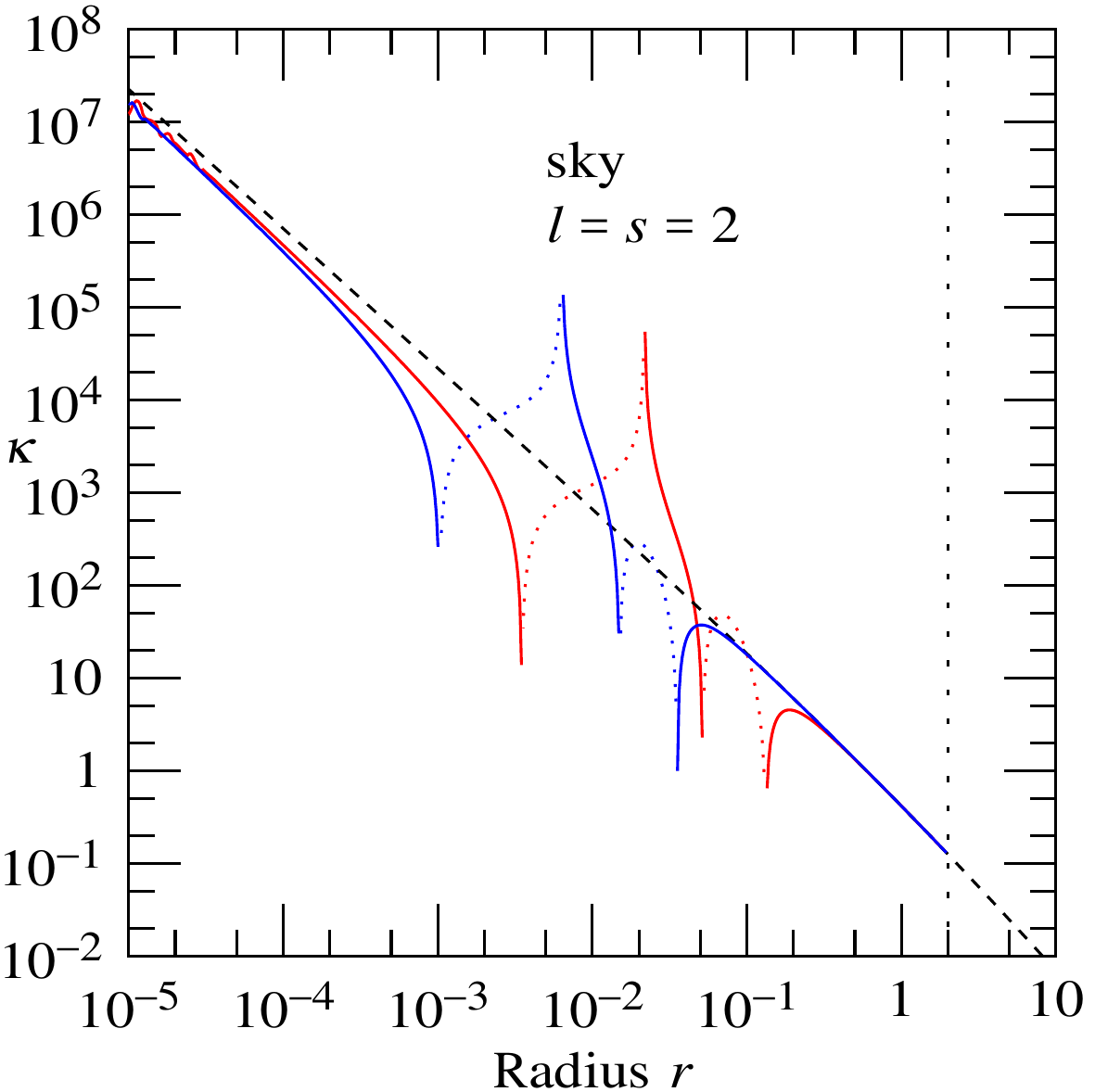}
    \caption[]{
    \label{kappas}
Observed acceleration $\kappa \equiv - \dd \ln \omegaout / \dd \tauout$
of the waves whose observed frequency $\omegaout$
are shown in Figure~\ref{freqs}.
The solid (dotted where negative)
lines show the numerically computed acceleration $\kappa$,
while the short dashed lines show the acceleration~(\ref{kappa0s})
in the geometric-optics limit.
    }
    \end{figure*}
}

\newcommand{\rhofig}{
    \begin{figure}[bt!]
    \centering
    \includegraphics[scale=.48]{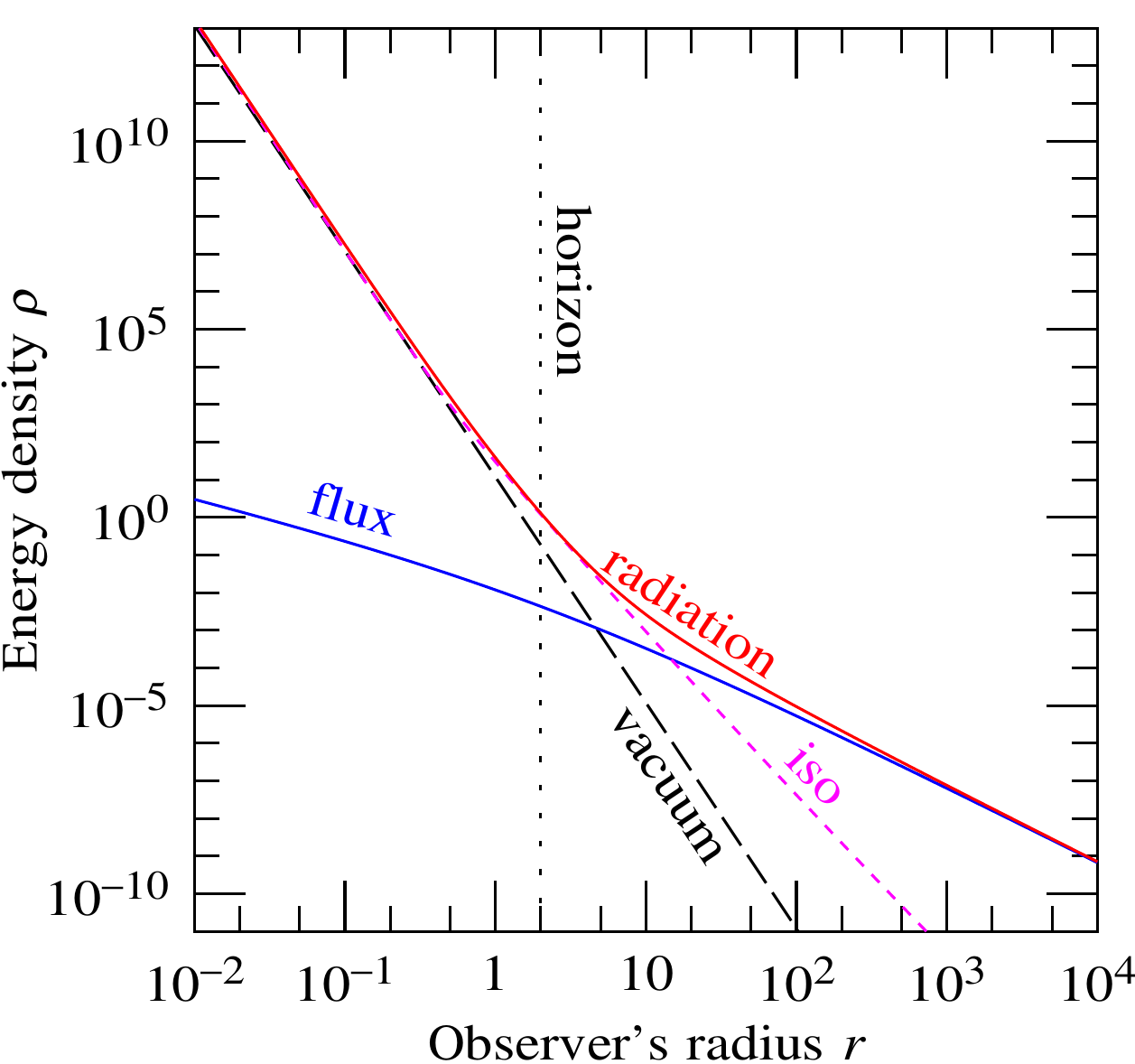}
    \caption[1]{
    \label{rho}
Energy densities of radiation $\rho_\flux + \rho_\aniso + \rho_\iso$
(solid red line)
and vacuum $\rho_\vac$
(long-dashed black line)
given by equations~(\ref{rhos}),
in units of $\rho_0$ defined by equation~(\ref{rho0}),
as a function of radius $r$ in units of $M$.
The radiation density is positive,
while the vacuum energy is negative.
Also shown are the
isotropic radiation density $\rho_\iso$ (dashed magenta line),
which is positive,
and the non-stationary energy flux $f_\flux$ (solid blue line),
which is positive (directed outward) and equal to minus $\rho_\flux$.
The adopted values of the constants $b$, $c$, and $e$
in equations~(\ref{rhos})
are $b = 10$, $c = \tfrac{5}{72}$, and $e = 1$.
The vertical dotted line marks the true horizon.
    }
    \end{figure}
}

\newcommand{\rhosingfig}{
    \begin{figure*}[bt!]
    \centering
    \includegraphics[scale=.48]{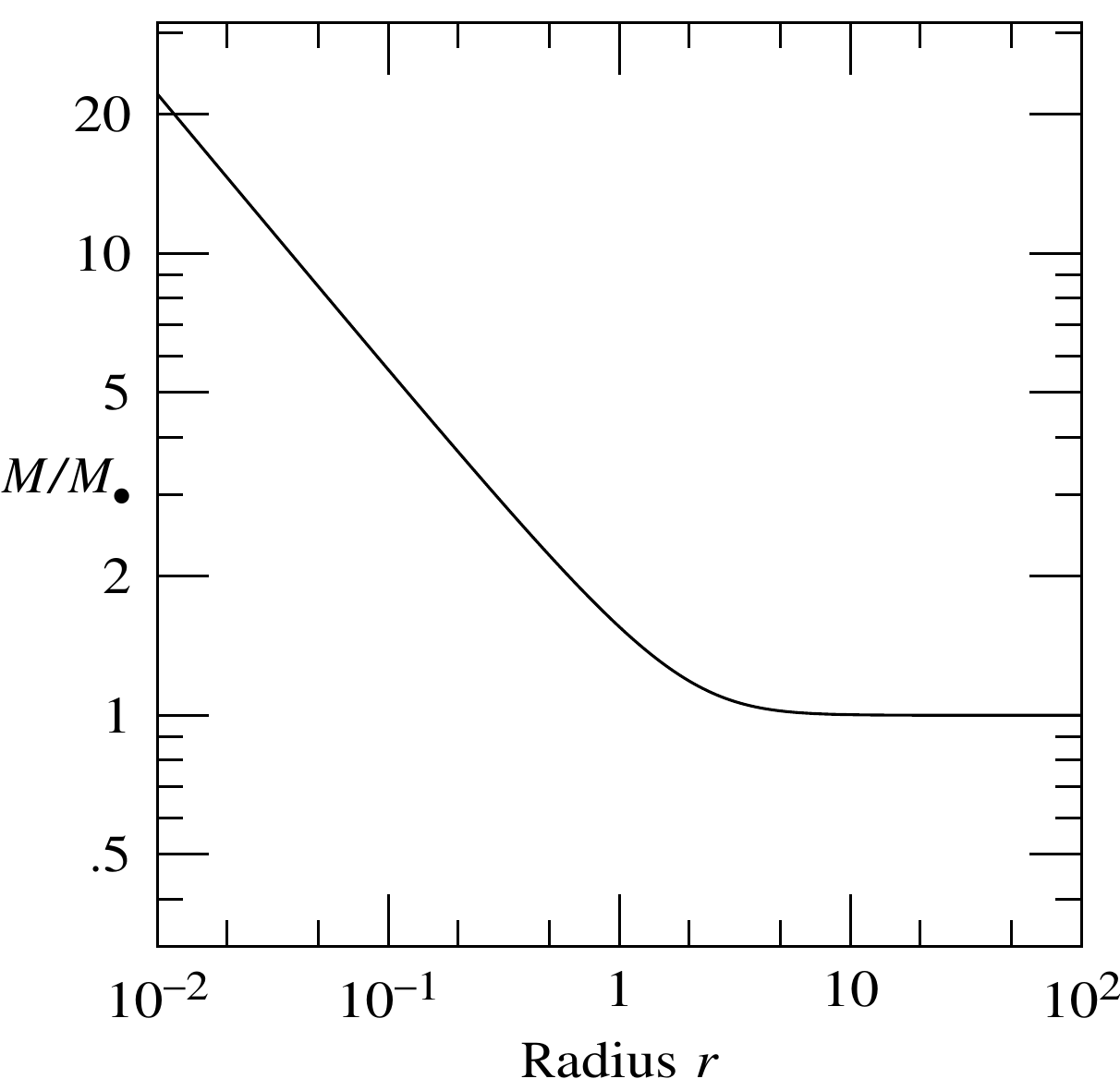}
    \includegraphics[scale=.48]{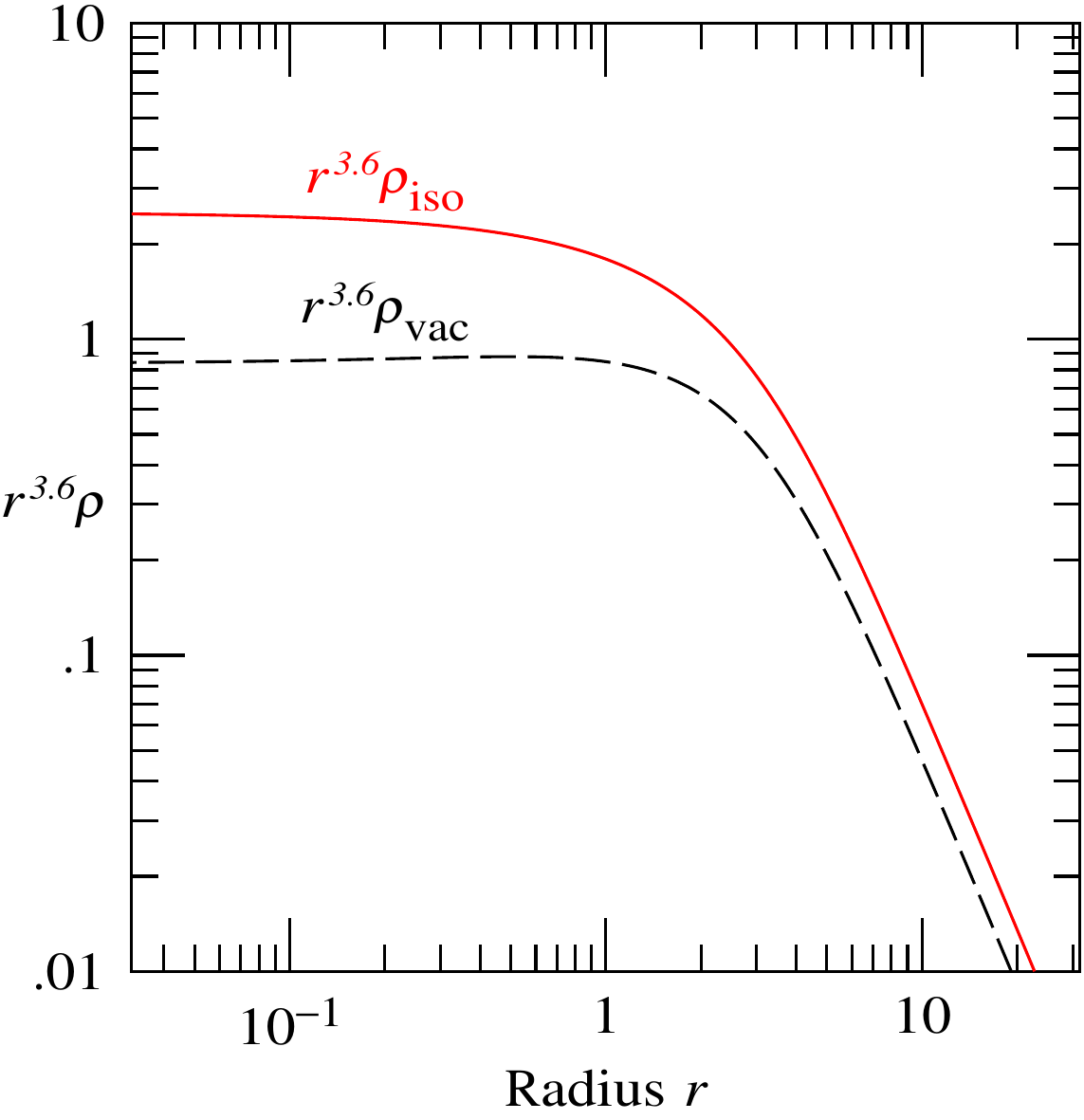}
    \caption[]{
    \label{rhosing}
(Left) Interior mass $M$,
and (right) energy densities $\rho_\iso$ of radiation
and $\rho_\vac$ of vacuum
when the back reaction of the quantum energy density on the geometry
is taken into account.
The radiation density $\rho_\iso$ is positive,
while the vacuum energy $\rho_\vac$ is negative.
The densities are multiplied by $r^{3.6}$
to bring out the fact that they diverge as
$\rho \propto M / r^3 \propto r^{-3.6}$
once back reaction sets in.
The anomaly coefficient $\alpha$
defined by equation~(\ref{TE})
has units of inverse density
(given in Planck units by equation~(\ref{qeff})).
The density $\rho$ plotted here is in units of the characteristic
density $1/\alpha$ at which back reaction sets in.
The plotted radius $r$ is in units of the characteristic
radius $( \alpha M )^{1/3}$
at which back reaction sets in.
    }
    \end{figure*}
}

\newcommand{\rhotwofig}{
    \begin{figure}[bt!]
    \centering
    \includegraphics[scale=.48]{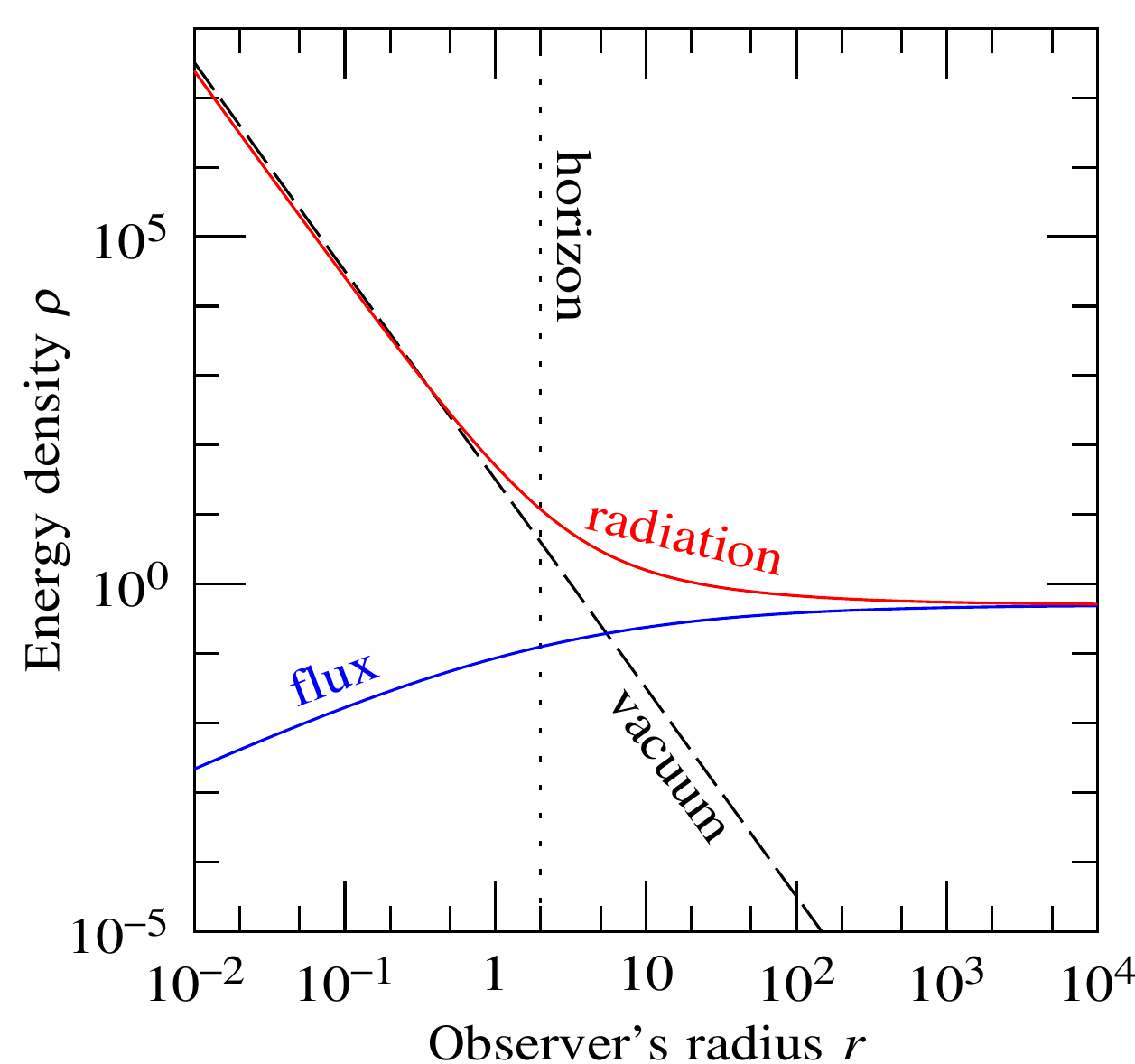}
    \caption[1]{
    \label{rho2}
Similar to Figure~\ref{rho}
but in 1+1 rather than 3+1 spacetime dimensions.
Energy densities of radiation $\rho_\flux + \rho_\stat$
(solid red line)
and vacuum $\rho_\vac$
(long-dashed black line)
given by equations~(\ref{rhos2}),
in units of $\rho_0$ defined by equation~(\ref{rho0}),
as a function of radius $r$ in units of $M$.
The radiation density is positive,
while the vacuum energy is negative.
Also shown is
the non-stationary energy flux $f_\flux$ (solid blue line),
which is positive (directed outward) and equal to minus $\rho_\flux$.
The vertical dotted line marks the true horizon.
    }
    \end{figure}
}

\newcommand{\kappathetafig}{
    \begin{figure}[bt!]
    \centering
    \includegraphics[scale=.48]{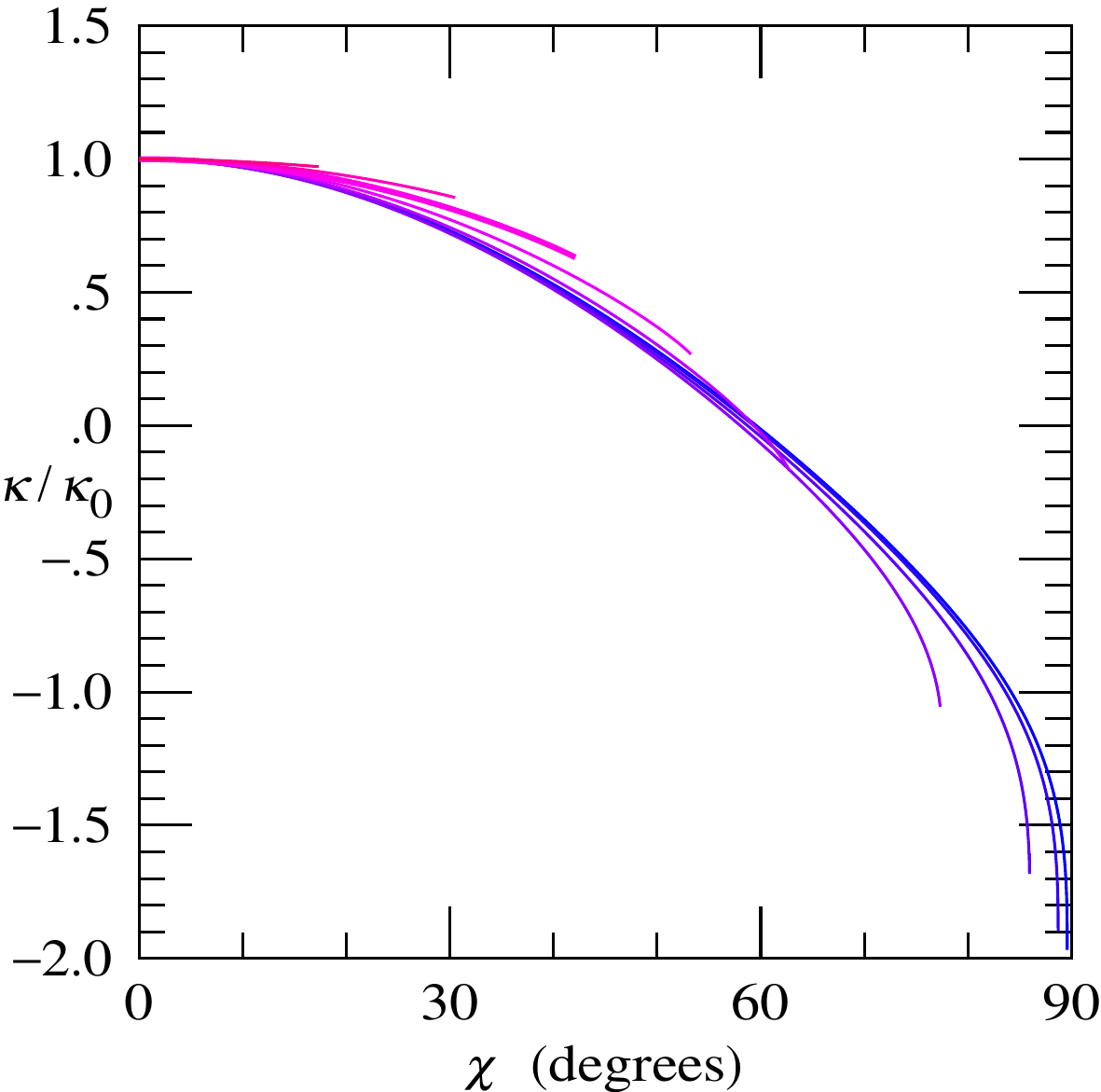}
    \includegraphics[scale=.48]{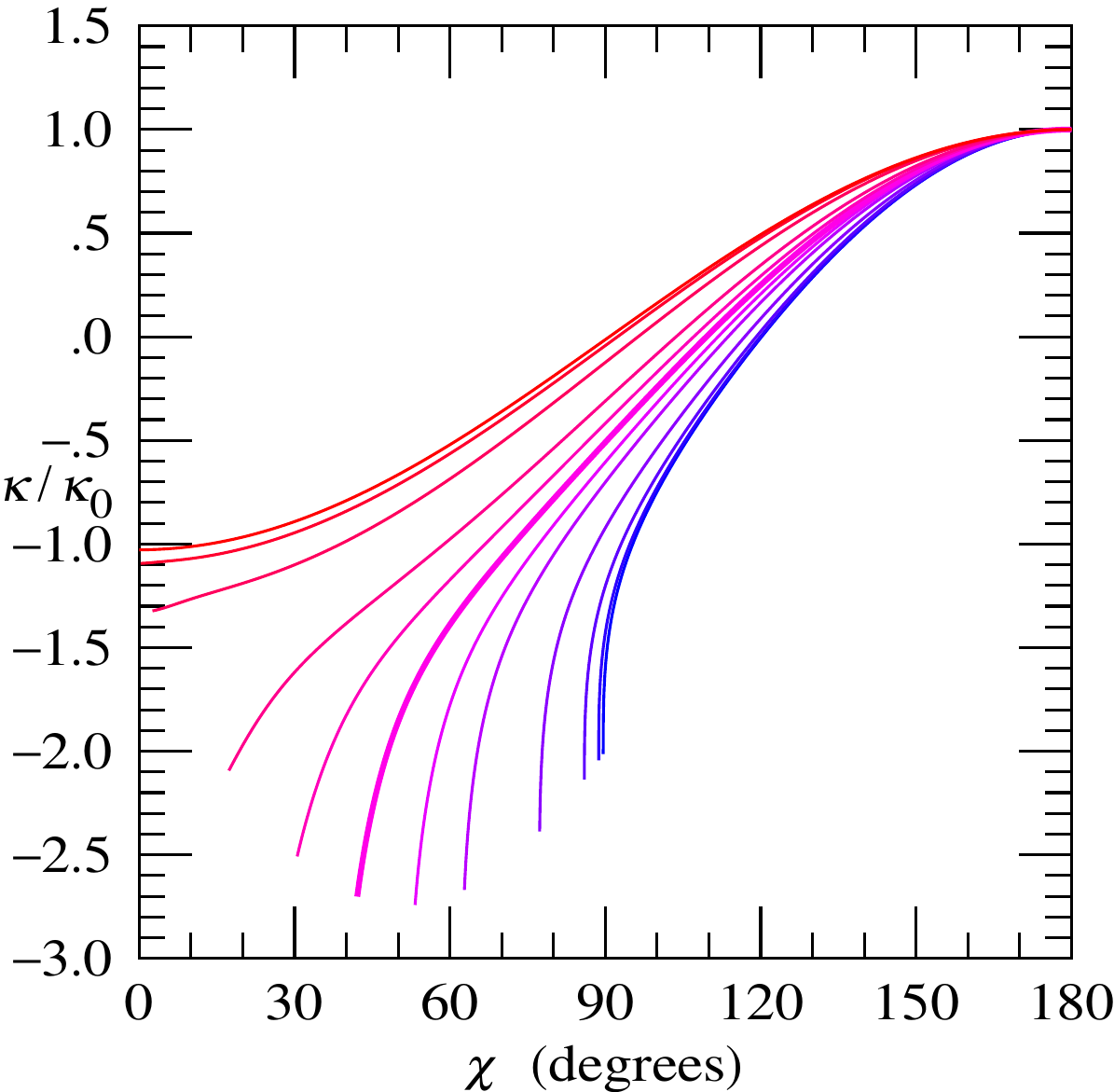}
    \caption[]{
    \label{kappathetas}
Acceleration $\kappa$ on
(left) the illusory horizon
and (right) the sky above,
as seen by a radially free-falling infaller,
relative to the acceleration $\kappa_0$ respectively
(left) directly downward and (right) directly upward.
This is similar to Figure~\protect\ref{kappachi},
but whereas in that Figure the observer was staring
fixedly in the same direction $\chi$,
here the observer is staring at a fixed location
(latitude) on the illusory horizon or sky above.
The observer here is thus rotating their view,
and they see a different acceleration from the
non-rotating observer of Figure~\protect\ref{kappachi}.
The curves are as seen by the infaller from
(top to bottom, red to blue)
radius
$10^4$, $10^3$, $10^2$, $10$, $4$, $2$ (true horizon, thick line),
$1$, $0.5$, $0.1$, $10^{-2}$, $10^{-3}$, and $10^{-4}$
geometric units.
    }
    \end{figure}
}


\date{\today}

\maketitle

\begin{abstract}
The boundary of any observer's spacetime is the boundary
that divides what the observer can see from what they cannot see.
The boundary of an observer's spacetime
in the presence of a black hole
is not the true (future event) horizon of the black hole,
but rather the illusory horizon,
the dimming, redshifting surface of the star that collapsed to the
black hole long ago.
The illusory horizon is the source of Hawking radiation
seen by observers both outside and inside the true horizon.
The perceived acceleration (gravity) on the illusory horizon
sets the characteristic frequency scale of Hawking radiation,
even if that acceleration varies dynamically,
as it must do from the perspective of an infalling observer.
The acceleration seen by a non-rotating free-faller
both on the illusory horizon below and in the sky above is calculated
for a Schwarzschild black hole.
Remarkably,
as an infaller approaches the singularity,
the acceleration becomes isotropic,
and diverging as a power law.
The isotropic, power-law character of the Hawking radiation,
coupled with conservation of energy-momentum,
the trace anomaly,
and the familiar behavior of Hawking radiation far from the black hole,
leads to a complete description of the quantum energy-momentum
inside a Schwarzschild black hole.
The quantum energy-momentum near the singularity diverges as $r^{-6}$,
and consists of relativistic Hawking radiation
and negative energy vacuum in the ratio $3 : -2$.
The classical back reaction of the quantum energy-momentum on the geometry,
calculated using the Einstein equations,
serves merely to exacerbate the singularity.
All the results are consistent with traditional calculations of the
quantum energy-momentum in 1+1 spacetime dimensions.
\end{abstract}


\section{Introduction}


The generalized laws of black hole thermodynamics
introduced by
\cite{Bekenstein:1973ur}
and
\cite{Hawking:1974}
are generally considered to be a
robust feature of quantum gravity,
that any successful final theory should predict.
According to these laws,
an observer outside the horizon of a black hole
will observe the black hole to emit radiation with a temperature
proportional to the gravity at the horizon,
and the entropy of the black hole equals one quarter
of its horizon area in Planck units.
The notion that black holes are thermodynamic objects,
with a definite temperature and a definite number of states,
has led to some of the most fertile ideas in contemporary physics,
including the information paradox
\cite{Hawking:1976ra},
holography
\cite{Susskind:1993if},
and AdS-CFT
\cite{Maldacena:1997re}.

There is a large literature on Hawking radiation and
black hole thermodynamics
from the perspective of observers outside the horizon,
but surprisingly little from the perspective of observers
who fall inside the horizon.
Part of the problem is that it is technically challenging.
The full apparatus of quantum field theory in curved spacetime is cumbersome
(e.g.\ \cite{Candelas:1980,Birrell:1982,Hofmann:2002va,Brout:1995rd},
and hampered by
the limited analytical understanding of the radial eigenfunctions
of the wave equation in static black hole spacetimes
(including the Schwarzschild geometry),
which are confluent Heun functions
\cite{Fiziev:2006tx,Fiziev:2009kh,Fiziev:2014bda}
(see Appendix~\ref{heun-sec}).

A complete solution is known in 1+1 spacetime dimensions,
where the trace anomaly and conservation of energy-momentum
serve to determine the complete renormalized quantum energy-momentum tensor
in any prescribed geometry
\cite{Davies:1976,Davies:1977,Birrell:1982,Chakraborty:2015nwa}.
At the end of this paper, \S\ref{dim2-sec},
the results derived in this paper for 3+1 dimensions
are shown to be consistent with the known results in 1+1 dimensions.

\cite{Hodgkinson:2012mr}
make some headway by considering
a 2+1 dimensional Ba\"nados-Teitel\-boim-Zanelli (BTZ) black hole,
for which the propagator of a scalar field
can be written down as an explicit infinite sum
\cite{Carlip:1995qv}.
\cite{Hodgkinson:2012mr} find that
the first few or several terms of the propagator
yield a satisfactory approximation to the response of an Unruh-deWitt detector
outside and near the horizon of the black hole.
Unfortunately, the number of terms needed increases near the
(conical) singularity, thwarting a successful computation there.

\cite{Hodgkinson:2014iua}
report a similar approach to the Schwarzschild geometry,
computing the propagator (Wightman function) entirely numerically.
The paper's conclusion states:
``the Wightman function is divergent at short distances, and while it is known
how the divergent parts come to be subtracted in the expressions
for the transition probability and transition rate, the challenge in numerical
work is to implement these subtractions term by term in a mode sum.
For a radially infalling geodesic in Schwarzschild, a subtraction procedure
in the Hartle-Hawking state is presented in \cite{Hodgkinson:2013tsa},
and a numerical evaluation of the transition rate is in progress.
We hope to report on the results of this evaluation in a future paper.''

\cite{Juarez-Aubry:2017kqm}
considers the response of an Unruh-deWitt detector
freely falling inside a 1+1 Schwarzschild black hole,
finding a switching rate proportional to $r^{-3/2}$
near the singularity (his equation~(4.53)),
consistent with the present paper.

\cite{Saini:2016rrt}
consider a massless scalar field
in the background geometry of a collapsing spherical shell in 3+1 dimensions.
By restricting to spherically symmetric (zero angular momentum) eigenmodes,
\cite{Saini:2016rrt}
are able to find exact solutions for the behavior of scalar modes
as seen by radial free-fallers at various times after collapse of the shell.
\cite{Saini:2016rrt}
conclude that the evolution is unitary and non-thermal.
To the extent that the emission can be characterized by a temperature,
``As the shell is collapsing
to a point and approaching 
singularity, the temperature
grows without limits,''
consistent with the results of the present paper.

\cite{Hiscock:1997jt}
collect from the literature
approximate expressions for the quantum energy-momentum
$\langle T_{kl} \rangle$
of scalar, spinor, and vector fields,
massless and massive,
in the Schwarzschild geometry in the Hartle-Hawking state.
The Hartle-Hawking state
represents a black hole in a reflecting cavity;
or equivalently, a black hole in unstable thermal equilibrium
with a thermal bath whose temperature at infinity is the Hawking temperature.
\cite{Hiscock:1997jt}
apply their results to the Schwarzschild interior,
but fall short of reaching definitive conclusions.


One fruitful approach to the calculation of the expectation value
$\langle T_{kl} \rangle$
of the quantum energy-momentum tensor was pioneered by
\cite{Christensen:1977}
(see also \cite{Visser:1997sd}),
who pointed out that in a stationary, spherical spacetime
some of the technical difficulties can be side-stepped
by imposing covariant conservation of energy-momentum
(which
$\langle T_{kl} \rangle$
should satisfy),
which reduces the number of components of
$\langle T_{kl} \rangle$
to be calculated from 4 to 2,
and by invoking the trace anomaly (or conformal anomaly),
which relates the trace of the quantum energy-momentum
to the Riemann curvature tensor in a general curved spacetime.
\cite{Christensen:1977}
did not apply their results to the Schwarzschild interior.

\cite{Bardeen:2014uaa} 
has discussed the interior of a Schwarzschild black hole
in the context of recent ideas in quantum gravity.
Reviewing
\cite{Christensen:1977}'s approach,
he writes:
``Can quantum back-reaction drastically alter the geometry deep
inside the horizon and prevent formation of a singularity?
The correct effective energy-momentum tensor deep inside the horizon
is not known, except for the conformal anomaly piece.
However, when gravitons are included, the conformal anomaly strongly
dominates the rest of the energy-momentum tensor in the vicinity of the
horizon, and quite possibly would do so even in the deep interior.''




The author suspects that there is another barrier to
understanding Hawking radiation inside a black hole,
a barrier that is conceptual rather then technical.
Most physicists would agree that Hawking radiation is
emitted from the black hole's horizon,
and that the states of the black hole are encoded on the horizon.
But what happens when an infaller free-falls through the horizon?
Do they find Hawking radiation there?
Do they find the states of the black hole there?

What happens becomes apparent from
general relativistic ray-traced visualizations
of what it looks like falling into a black hole
\cite{Hamilton:2010my}.
When an observer watches a black hole formed from gravitational collapse,
they are watching not the future event horizon of the black hole,
but rather the redshifting, dimming surface of the star
or whatever else collapsed into the black hole long ago.
\cite{Hamilton:2010my} dubbed this redshifting, dimming surface
the ``illusory horizon,''
because it gives the appearance of a horizon,
but it is not a true event horizon
(neither future nor past;
a future event horizon is
defined by
\cite{Hawking:1973}
to be the boundary of the past lightcone of the extension of
an observer's worldline into the indefinite future;
a past horion is the same with past $\leftrightarrow$ future).
When an observer falls through the ``true'' future event horizon,
the surface of no return,
the observer does not catch up with the illusory horizon,
the redshifted image of the collapsed star.
Rather,
as the visualizations of \cite{Hamilton:2010my} show,
the illusory horizon remains ahead of the infaller,
still redshifting and dimming away.
An observer inside the true horizon sees the true horizon above,
and the illusory horizon below.
They are not the same thing, despite having the same radial coordinate.

The illusory horizon is the boundary of an observer's spacetime.
It divides what an observer can see from what they cannot see,
which remains true inside as well as outside the true horizon.
The illusory horizon, not the true horizon,
is the source of Hawking radiation seen by an observer.
If it is accepted that the illusory horizon is the source of Hawking radiation,
then it becomes possible to calculate the Hawking radiation seen
by an observer inside the true horizon.
This is the purpose of the present paper,
to calculate the Hawking radiation,
and the expectation value of the quantum energy-momentum,
inside the true horizon of a Schwarzschild black hole.


The inadequacy of the future event horizon
as a source of Hawking radiation
was recognized by
\cite{Susskind:1993if},
who introduced the ``stretched horizon,''
a timelike surface located one Planck area above the event horizon.
Unfortunately, the concept of the stretched horizon fails
for observers who fall through the horizon.
Moreover the idea that the stretched horizon lives literally
just above the true horizon is misleading:
it suggests that an infaller might go down and touch the stretched horizon.
The name illusory horizon is better.
Like a mirage or a rainbow, it is real, yet always out of reach.


Unless otherwise stated,
the units in this paper are Planck,
$c = \hbar = G = 1$.

\section{The illusory horizon}

This section discusses the illusory horizon
as a prelude to the main text starting with \S\ref{accel}.

\stpcollfig

\penroseschwskyfig

Figure~\ref{stpcoll}
illustrates a sequence of Penrose diagrams
of the Oppenheimer-Snyder collapse of a uniform, pressureless, spherical star
to a Schwarzschild black hole.
Penrose diagrams are commonly sketched,
but one should remember that they are genuine spacetime diagrams,
and can be drawn accurately
with a specific choice of time and space coordinates.
It can be a useful exercise to draw Penrose diagrams accurately,
because sometimes sketched diagrams can be misleading.
The Penrose diagrams in
Figure~\ref{stpcoll}
are drawn in Penrose coordinates defined in Appendix~\ref{penrosecoords},
equation~(\ref{rtPenrose}) and~(\ref{fPenrose}).
The Penrose diagrams in
Figure~\ref{stpcoll}
are drawn from the perspective of observers at several different times
during the collapse of the black hole.
The diagram evolves
from something that resembles Minkowski space well before collapse,
to something that resembles the Schwarzschild geometry well after collapse.

An important point to notice
is that the surface of the collapsed star
in the Penrose diagram
asymptotes to the place where the antihorizon, or past horizon,
would be in the analytically extended Schwarzschild solution.
This is the illusory horizon.

Figure~\ref{penroseschwsky}
illustrates a Penrose diagram of Hawking radiation seen by
an observer who falls to the singularity of a Schwarzschild black hole

\section{On the calculation of Hawking radiation}
\label{accel}

Quantum field theory (QFT) in curved spacetime predicts that
an accelerating observer,
or an inertial observer watching an accelerating emitter,
perceives spontaneous quanta of radiation that would classically be absent.
The relative acceleration between emitter and observer
causes positive frequency modes in the emitter's frame
to transform into a mix of positive and negative frequency modes
in the observer's frame.
In QFT, negative frequency modes signal particle creation.

A star that collapses to a black hole
asymptotically approaches a stationary state.
If the collapsed star has zero angular momentum,
then the stationary state is the Schwarzschild geometry.
But the black hole never quite achieves the stationary state:
an outside observer at rest watching the collapsing star
sees it freeze at the illusory horizon,
redshifting and dimming exponentially into the indefinite future.
Long after collapse,
the ratio
of observed
$\omega_\obs$
to emitted
$\omega_\emit$
frequencies
of radiation is exponentially tiny,
and the rate of exponential redshift
at the illusory horizon
with respect to the observer's proper time $\tau$
approaches a constant,
the acceleration $\kappa$,
\begin{equation}
  {\dd \ln ( \omega_\obs / \omega_\emit ) \over \dd \tau}
  =
  - \kappa
  \ .
\end{equation}
The acceleration $\kappa$ equals the gravity at the black hole's horizon,
and is independent of the original state of motion (orbital parameters)
of whatever fell into the black hole long ago.
For a Schwarzschild black hole of mass $M$,
the acceleration at the illusory horizon seen by a distant observer at rest is
\begin{equation}
  \kappa
  =
  \frac{1}{4 M}
  \ .
\end{equation}
QFT predicts that the observer will see the black hole emit
\cite{Hawking:1975b}
radiation with a thermal spectrum at temperature $T$
proportional to the acceleration,
\begin{equation}
\label{Tkappa}
  T =
  {\kappa \over 2\pi}
  \ .
\end{equation}
Equation~(\ref{Tkappa}) is quite general:
an observer who watches a bifurcation horizon
at which the acceleration $\kappa$ appears constant over
at least several $\ee$-folds of redshift
will see approximately thermal radiation
at the temperature given by equation~(\ref{Tkappa})
\cite{Visser:2001kq}.

When an observer falls through the true horizon of a black hole,
they do not catch up with the star that collapsed long ago.
Rather, they continue to see the illusory horizon ---
the dimming, redshifting surface of the collapsed star ---
ahead of them, still dimming and redshifting away.
Consequently an infalling observer will continue to see
Hawking radiation from the illusory horizon
even after they have fallen through the true horizon.

An observer who free-falls into a black hole
also sees the sky above redshifting
as they accelerate away from the sky into the black hole.
Consequently the infalling observer will also see
Hawking radiation from the sky above.

As discussed in the Introduction,
a rigorous computation of Hawking radiation seen by an infaller
is mathematically and numerically challenging,
e.g.\ \cite{Hodgkinson:2014iua}.
However,
progress can be made by going to the geometric-optics limit,
as shown in the next section \S\ref{geometricoptics-sec}.
The reliability of the geometric-optics limit is tested numerically in
\S\ref{waves-sec}.


\section{Geometric-optics limit}
\label{geometricoptics-sec}

In the geometric-optics limit,
light rays move along null geodesics.
Geometric optics is valid for light at wavelengths much shorter than
any other lengthscale in the problem,
namely
the lengthscale over which the wave properties vary,
and the curvature lengthscale
\cite{MTW:1973}.
In the geometric-optics limit,
it is straightforward to calculate the acceleration $\kappa$
on either the illusory horizon below or in the sky above
seen by an infaller.

To avoid ambiguity with radiation that a non-inertial observer would see
\cite{Sriramkumar:1999nw},
I impose that the infaller be freely-falling and non-rotating
(a non-rotating observer stares fixedly in the same inertial direction,
whereas a rotating observer swivels their eyeballs).
For simplicity, I take the observer to free-fall radially
from zero velocity at infinity.
The infaller's viewing direction is defined by the angle $\chi$ relative
to the direction directly towards the black hole.

\kappazerofig

In the Schwarzschild geometry,
the acceleration $\kappa_0$
(the subscript $0$ denotes light rays of zero angular momentum)
seen by a radial free-infaller
from the illusory horizon directly below at $\chi = 0^\circ$,
and also from the sky directly above at $\chi = 180^\circ$,
takes a simple analytic form as a function of
the radial position $r$ of the infaller
(the technical details of the calculation are relegated
to Appendix~\ref{tech}, eq.~\ref{kappa0omega}):
\begin{equation}
\label{kappa0s}
  \kappa_0
  =
  \left\{
  \begin{array}{ll}
  \displaystyle
  {\bigl( 1 + \sqrt{r / ( 2 M )} \bigr)
  \bigl( 1 + r / ( 2 M ) \bigr)
  \over 4 M \, \bigl( r / ( 2 M ) \bigr)^{3/2}}
  &
  \mbox{below}
  \ ,
  \\[3ex]
  \displaystyle
  {1 \over 4 M \bigl( r / ( 2 M ) \bigr)^{3/2} \bigl( 1 + \sqrt{r / ( 2 M )} \bigr)}
  &
  \mbox{above}
  \ .
  \end{array}
  \right.
\end{equation}
The accelerations $\kappa_0$ below and above tend to the same diverging limit
near the singularity,
\begin{equation}
\label{kappa0sing}
  \kappa_0
  \rightarrow
  {1 \over 4 M} \left( {2 M \over r} \right)^{3/2}
  =
  {1 \over 3 | \tau |}
  \quad
  \mbox{as $r \rightarrow 0$}
  \ ,
\end{equation}
where $|\tau|$ is the proper time
left before the infaller hits the central singularity
(the minus sign is so that $\tau$ increases as the infaller's time goes by),
\begin{equation}
\label{tau}
  \tau
  =
  -
  {4 M \over 3}
  \left( {r \over 2 M} \right)^{3/2}
  \ .
\end{equation}
The two accelerations~(\ref{kappa0s}),
along with the asymptotic limit~(\ref{kappa0sing}),
are plotted in Figure~\ref{kappa0}.

Figure~\ref{kappa0}
shows that both the illusory horizon below
and the sky above appear increasingly redshifted
as the infaller falls.
The infaller will therefore see Hawking radiation
not only from the illusory horizon below,
but also from the ``cosmological horizon'' in the sky above.
When the observer is far from the black hole,
the perceived acceleration on the illusory horizon is $1/(4M)$,
while the acceleration in the sky goes to zero.

In reality the distant universe will not be
the asymptotically Minkowski space of the Schwarzschild geometry,
but rather will be whatever the true cosmology is.
Regardless of the cosmology,
near the black hole the perceived acceleration on the sky
will be dominated by the observer's infall into the black hole,
so the precise details of the cosmology are unimportant.

Equation~(\ref{kappa0sing}) says that the acceleration $\kappa_0$
tends to a third the inverse proper time $|\tau|$
left before the infaller hits the central singularity.
The fact that the redshift in both directions diverges near the singularity
can be attributed to the diverging tidal force from the black hole.
The condition for the observer to see thermal radiation
--- that the acceleration be approximately constant over
at least several $\ee$-folds ---
is not achieved when the infaller is near or inside the black hole.
Thus the Hawking radiation that the infaller sees is not thermal.

\kappachifig

It is impossible to arrange the motion of the infaller
to eliminate acceleration from both below and above.
By accelerating inward or outward,
the infaller can reduce the perceived acceleration below or above,
but that merely increases the acceleration in the opposite direction.
Consequently it is impossible to eliminate all Hawking radiation
by adjusting the motion of the infaller.
The infaller will necessarily see Hawking radiation.
It is sometimes asserted that an observer who free-falls through
the horizon sees no Hawking emission.
It is true that the free-faller will see no Hawking radiation
from the true horizon.
But the infaller will see Hawking radiation
from the illusory horizon below, and from the sky above.

The acceleration seen by the infaller in a general viewing direction
does not have a simple expression such as equation~(\ref{kappa0s}),
but can be calculated straightforwardly if laboriously
(see Appendix~\ref{tech}).
Figure~\ref{kappachi}
shows the acceleration $\kappa$ on the illusory horizon
and on the sky
seen by the non-rotating infaller as a function of the viewing angle $\chi$,
normalized to the acceleration $\kappa_0$ respectively directly below
and directly above.
The left panel of the Figure shows that the perceived acceleration
on the illusory horizon is approximately constant
over the disk of the black hole out to near its apparent edge.
Near the apparent edge,
the acceleration increases rapidly,
diverging to infinity at the edge
(the infinity is an artifact of the exact Schwarzschild geometry;
in a real black hole formed a finite time in the distant past,
the acceleration at the edge would be large but finite).
The right panel of
Figure~\ref{kappachi}
shows that the acceleration on the sky
is positive (redshifting) in the upper hemisphere, $\chi > 90^\circ$,
negative (blueshifting) in the lower hemisphere, $\chi < 90^\circ$,
reflecting the fact that the freely-falling infaller is accelerating
away from the sky above and towards the black hole below.

A striking feature of
Figure~\ref{kappachi}
is that as the infaller approaches the singularity,
the acceleration appears to be almost constant over almost all
the illusory horizon and almost all the sky;
the acceleration varies with direction
only over a thinning band near the equator.
The near constancy of the acceleration can be attributed
to the diverging tidal force near the singularity,
which aberrates null rays away from vertical directions
towards horizontal directions.
Moreover,
as seen in Figure~\ref{kappa0},
the acceleration near the singularity
asymptotes to the same value, $\kappa \rightarrow 1 / ( 3 |\tau| )$,
on both illusory horizon and sky.

The fact that the perceived acceleration
goes over to a (time-varying) constant
over almost all the illusory horizon and sky
as the infaller approaches the singularity
suggests that Hawking radiation seen by the infaller
will become isotropic near the singularity.
This idea is a key ingredient underlying the calculation
of the quantum energy-momentum
in \S\ref{traceanomaly-sec}.

The acceleration shown in Figure~\ref{kappachi}
is as perceived by a non-rotating infaller,
who stares fixedly in the same direction $\chi$ as they fall inward.
This is not the same as an observer
who stares at the same angular position (latitude) $\theta$
on the illusory horizon or sky above.
The acceleration perceived by the latter observer,
who is necessarily rotating
(swiveling their eyeballs),
is different,
as shown in Appendix~\ref{tech}
(Figure~\ref{kappathetas}).
This emphasizes that Hawking radiation is observer-dependent.
As remarked above,
the least contaminated view
is that of an inertial observer, who neither accelerates nor rotates.
Near the singularity,
a real observer can hardly help being inertial:
the non-gravitational forces that they can command to accelerate or rotate
cannot compete with the diverging gravitational forces.

An infaller who observes fixedly in the same viewing direction $\chi$
pans over a range of angular positions $\theta$
on the illusory horizon or sky
as they fall inward.
Correct application of QFT requires that each in-mode
that the observer is viewing is the same over the illusory horizon or sky.
This can be achieved by considering each in-mode to be emitted
spherically symmetrically
in the frame of a spherically symmetric shell.
In the case of the illusory horizon,
from the perspective of an emitter close to the horizon,
only photons emitted within a tiny interval of directly outward
can make it to an outside observer,
or survive long enough to be seen by an infaller who falls in
much later.
Although the photons are emitted almost directly outward,
a later observer sees them at various angles.
The condition that the observer is viewing the same mode is that
the observed phase of the wave emitted from different points be the same.
This requires that the emitted affine distance
$\lambda_\emit$
along the path of the light ray between emitter and observer be the same.
The emitted affine distance $\lambda_\emit$
is defined to be the affine parameter $\lambda$
normalized so that it measures proper distance in the locally inertial
frame of the emitter.
Affine distances in different frames differ only by a normalization factor
(they are all proportional to an affine parameter),
proportional to the frequency of the wave.
The emitted affine distance $\lambda_\emit$
is related to the observed affine distance $\lambda_\obs$ by
\begin{equation}
\label{lambdaem}
  \lambda_\emit
  =
  {\omega_\emit \over \omega_\obs} \lambda_\obs
  \ .
\end{equation}
The observed affine distance $\lambda_\obs$
along a ray from any point on the illusory horizon to the observer
is straightforward to calculate.
The condition that $\lambda_\emit$ be the same over the illusory horizon
at fixed observer position
thus translates into a condition on the blueshift factor
$\omega_\obs / \omega_\emit$ as the angular position of the emitter
is varied over the illusory horizon.
Appendix~\ref{tech} contains further details.

\subsection{Anisotropy near the singularity}

It has been argued above that Hawking radiation becomes
isotropic near the singularity,
a fact that will be central to the determination
in \S\ref{traceanomaly-sec}
of the quantum energy-momentum tensor near the singularity.
However,
as Figure~\ref{kappachi} shows,
the Hawking radiation becomes anisotropic away from the singularity,
and the question arises of how fast this anisotropy develops.

An approximate answer
comes from noticing that the edge of the black hole subtends
a perceived angle $\chi_\edge$
given by solving equation~(\ref{chi}) with
photon angular momentum equal to that at the photon sphere,
$J = 3 \sqrt{3} M$.
The angular size of the black hole
perceived by the radially infalling observer is
\begin{equation}
\label{chiedge}
  \chi_\edge
  \approx
  {\pi \over 2}
  -
  \left( {r \over 2 M} \right)^{1/2}
  \ .
\end{equation}
Equation~(\ref{chiedge}) will be used in \S\ref{conserveT-sec}
to infer the sub-leading behavior of the quantum energy-momentum
near the singularity, equation~(\ref{prhoedge}).

\section{Waves}
\label{waves-sec}

The calculation in the previous section
invoked the classical, geometric-optics limit
where light moves along null geodesics,
which is valid in the limit of high frequencies.
A correct calculation of Hawking radiation requires
following wave modes of finite frequency.
The purpose of this section is to compute a small sample of waves
so as to test the geometric-optics approximation.
For simplicity,
I consider only modes of low angular momentum
(with various spins $s$).
I compute the waves numerically from the
wave equation~(\ref{phieq}) in the Schwarzschild geometry.
Details of the wave equation and its numerical solution
are deferred to Appendix~\ref{waveeqsoln-sec}.

\freqsfig

Hawking radiation from the illusory horizon originates as
outgoing waves of definite ``in'' frequency $\omegain$
in the inertial frame of an infaller at the true horizon.
The strategy to compute these waves is to consider the wave
in the inertial frames of two infallers,
an ``in'' faller who falls in first,
and an ``out'' faller who falls in some (long) time later.
The inertial frame of the first infaller defines in-modes
of definite frequency $\omegain$
that the second infaller observes some time later.

Hawking radiation from the sky above originates as
waves of definite ``in'' frequency $\omegain$
in the rest frame at infinity.
Such waves are necessarily ingoing when they fall through the true horizon.
The waves are eigenfunctions of the wave equation,
which are confluent Heun functions
\cite{Fiziev:2006tx,Fiziev:2009kh}
(see Appendix~\ref{heun-sec}).
As with waves from the illusory horizon,
I compute waves from sky numerically from the wave equation~(\ref{phieq}),
which in this case reduces to an ordinary differential equation.
In principle it would be possible to compute the waves from
the series expansion~(\ref{heunseries}) of the Heun function,
but because an observer well inside the black hole sees
waves highly redshifted compared to when they were emitted,
the rest-frame ``in'' frequencies of relevance are high,
requiring a large number of terms of the series expansion.
Near cancelation of many terms leads to loss of numerical precision.
I checked that, for lower frequencies than those of relevance,
the numerical solution of the wave equation
agrees with that computed from the series expansion.

By definition,
the phase $\psi$ of
a wave $\varphi_\omegain$ of pure frequency $\omegain$ in the ``in'' frame
increases uniformly with proper time in that frame.
By constrast
an infaller observing the wave sees the phase changing at a non-uniform rate.
The phase $\psi$ of a wave seen by the infalling observer is defined by
\begin{equation}
  \varphi_\omegain
  =
  | \varphi_\omegain |
  \ee^{- \im \psi}
  \ ,
\end{equation}
that is, the phase is minus the imaginary part of the logarithm
of the wavefunction.
The frequency $\omegaout$ observed by the infaller is defined to be
\begin{equation}
  \omegaout \equiv
  {\dd \psiout \over \dd \tauout}
  \ .
\end{equation}

Figure~\ref{freqs} shows the frequency observed by an infaller
for each of two representative ``in'' frequencies $\omegain$.
The second (red) frequency is a factor $\ee^{-5/2} \approx 1/12$
lower than the first (blue).
The actual ``in'' frequency is indeterminate,
since a wave emitted time $t$ earlier
with a frequency $\ee^{\kappa_\hor t}$ higher
yields the same observed frequency $\omegaout$
(as long as the wave was emitted sufficiently long before it was observed).
The top, middle, and bottom panels
of Figure~\ref{freqs}
show waves of spin $s = 0$, $1$, and $2$,
each with the smallest possible angular momentum, $l = s$.

The Figure shows that the frequency $\omegaout$ agrees with that calculated
in the high-frequency, geometric-optics limit provided that
$| \omegaout \tauout | \gg 1$.
At lower frequencies,
$| \omegaout \tauout | \lesssim 1$,
the wave ceases to redshift in accordance with the geometric-optics limit,
and instead starts blueshifting, or redshifting more slowly.
This means that the geometric-optics limit is reliable only as long as
the observed frequency $\omegaout$
is higher than the inverse proper time left $1/|\tauout|$
before hitting the singularity.
This makes physical sense.

\kappasfig

Figure~\ref{kappas}
shows the acceleration
\begin{equation}
  \kappa \equiv
  - {\dd \ln \omegaout \over \dd \tauout}
\end{equation}
of the waves whose frequencies $\omegaout$ are shown in Figure~\ref{freqs}.
For infallers inside the true horizon,
the observed acceleration $\kappa$
deviates from the geometric-optics limit~(\ref{kappa0s}).
Notwithstanding the excursions from the geometric-optics limit at
intermediate radii,
the geometric optics limit sets the approximate characteristic value of
the acceleration at all radii.
(The acceleration passes through zero when the observed frequency
passes through an extremum;
and for $s = 2$, the acceleration also passes through infinity,
which happens when the observed frequency passes through zero,
that is, the phase goes through an extremum.)
To the extent that the acceleration sets the characteristic frequency
of Hawking radiation observed by an infaller,
the waves expected to dominate Hawking radiation
are those whose observed frequency satisfies the resonance condition
$\omegaout \sim \kappa$.
In summary,
the waves expected to dominate Hawking radiation are those satisfying
\begin{equation}
\label{omegadom}
  \omegaout \sim \kappa \sim {1 \over 3|\tauout|}
  \ .
\end{equation}
The conclusion is that the situation is more complicated
than in the geometric-optics limit,
but the geometric-optics limit still sets
the characteristic frequency scale of Hawking radiation.

Figure~\ref{kappas}
shows that near the singularity the acceleration asymptotes
to a value proportional to $1/|\tau|$,
but with a spin-dependent coefficient,
\begin{equation}
\label{kappa0spin}
  \kappa
  \rightarrow
  {2 s - 3 \over 3 |\tau|}
  \ .
\end{equation}
Equation~(\ref{kappa0spin})
can be derived analytically from the generic behavior of waves
near the singularity,
as shown in Appendix~\ref{asymptotesing-sec},
equation~(\ref{kappa0H}).
It is worth noting that the acceleration is negative (blueshifting)
or positive (redshifting)
as the spin is less than or greater than $\tfrac{3}{2}$.
Thus only gravitational waves, $s = 2$,
continue to redshift near the singularity.
For gravitational waves,
the acceleration near the singularity is the same as that in
the geometric optics limit, equation~(\ref{kappa0sing}).

It should be remarked that the behavior~(\ref{kappa0spin})
is of subdominant relevance to Hawking radiation,
because, as is apparent from Figures~\ref{freqs} and~(\ref{kappas}),
the observed frequency where~(\ref{kappa0spin}) holds
is much smaller than the acceleration,
$|\omegaout| \ll |\kappa|$,
so the resonance condition $|\omegaout| \sim |\kappa|$ is not satisfied.
The waves expected to dominate Hawking radiation near the singularity
originate from ``in'' frequencies much higher than those shown in
Figure~\ref{freqs},
such that condition~(\ref{omegadom}) is satisfied.

\section{Energy-momentum tensor}

\subsection{Trace anomaly}
\label{traceanomaly-sec}

The arguments of \S\ref{geometricoptics-sec} have indicated
that a freely-falling, non-rotating observer
who falls inside the horizon of a Schwarzschild black hole
necessarily sees Hawking radiation from both the illusory horizon below
and the sky above.
The Hawking radiation should contribute
a non-zero expectation value
$\langle T_{kl} \rangle$
of the energy-momentum,
which in turn should back react on the geometry of the black hole,
\S\ref{backreaction-sec}.

Calculating the quantum expectation value
$\langle T_{kl} \rangle$
of the energy-momentum in a curved spacetime is not easy
\cite{Candelas:1980,Howard:1984}.

However,
as reviewed by
\cite{Duff:1993wm},
the trace of the expectation value of the energy-momentum,
the so-called trace anomaly, or conformal anomaly, is calculable.
On general grounds, the trace anomaly
$T \equiv g^{\mu\nu} \langle T_{\mu\nu} \rangle$
must take the form
\cite[eq.~(21)]{Duff:1993wm}
\begin{equation}
\label{anomaly}
  T
  =
  \alpha_F F
  +
  \alpha_E E
  +
  \alpha_R \square R
  \ ,
\end{equation}
where
$F$ is the squared Weyl tensor
$C_{klmn}$,
and $E$ is the Euler density
(whose integral yields the Euler characteristic),
which in terms of the Riemann tensor $R_{klmn}$,
Ricci tensor $R_{kl}$, and Ricci scalar $R$ are
\begin{subequations}
\begin{align}
  F
  &
  \equiv
  R_{klmn} R^{klmn}
  -
  2
  R_{kl} R^{kl}
  +
  \tfrac{1}{3}
  R^2
  =
  C_{klmn} C^{klmn}
  \ ,
\\
  E
  &\equiv
  R_{klmn} R^{klmn}
  -
  4
  R_{kl} R^{kl}
  +
  R^2
  \ .
\end{align}
\end{subequations}
The coefficients $\alpha_F$, $\alpha_E$, $\alpha_R$
in equation~(\ref{anomaly})
depend on the number of particle species.
\cite{Duff:1993wm}
gives
$\alpha_R = \tfrac{2}{3} \alpha_F$,
but different methods of renormalization yield different results
for this coefficient
\cite{Hawking:2000bb,Asorey:2003uf}.
In any case,
$\alpha_R$ can be adjusted arbitrarily by adding an $R^2$ term
to the Hilbert Lagrangian.
For $\alpha_F$ and $\alpha_E$,
\cite[eqs.~(30) and~(31)]{Duff:1993wm}
states that for massless species
\begin{subequations}
\begin{align}
  \alpha_F
  &=
  {1 \over 1920 \pi^2}
  \left(
  n_0 + 3 n_{1/2} + 12 n_1
  \right)
  \ ,
\\
  \alpha_E
  &=
  -
  {1 \over 5760 \pi^2}
  \left(
  n_0 + \tfrac{11}{2} n_{1/2} + 62 n_1
  \right)
  \ ,
\end{align}
\end{subequations}
where $n_0$ is the number of real scalars
(1 degree of freedom per scalar),
$n_{1/2}$ is the number of Majorana spinors
(2 helicities per spinor;
note
\cite{Duff:1993wm}'s $n_{1/2}$ is the number of Dirac spinors,
with 4 helicities per spinor),
and
$n_1$ is the number of vector particles
(2 helicities per vector).

\cite{Christensen:1978}
state that,
in theories that are one-loop finite such as ordinary gravity and supergravity,
when matter terms in the Lagrangian are included,
the trace anomaly becomes proportional to the Euler term $E$ alone,
with overall coefficient $\alpha$
equal to the sum of $\alpha_F$ and $\alpha_E$
(see \cite[eq.~(35)]{Duff:1993wm}),
\begin{equation}
\label{TE}
  T = ( \alpha_F + \alpha_E ) E = \alpha E
  \ .
\end{equation}
The coefficient $\alpha$ is
\begin{equation}
\label{qeff}
  \alpha
  =
  {q_\eff \over 2880 \pi^2}
  \ , \quad
  q_\eff
  \equiv
  \sum_s q_s n_s
  \ ,
\end{equation}
where $n_s$ is the number of species of spin $s$,
and $q_s$ are spin-dependent coefficients.
For massless species,
the coefficients $q_s$ are
\cite[Table~2]{Christensen:1978}
(counting 1 degree of freedom for each scalar,
2 helicities per species otherwise)
\begin{equation}
\label{qsmassless}
  q_s = \{ 1 , \tfrac{7}{4} , -13 , -\tfrac{233}{4} , 212 \}
  \ 
  \mbox{for}
  \ 
  s = \{ 0 , \tfrac{1}{2} , 1 , \tfrac{3}{2} , 2 \}
  \ .
\end{equation}
For massive species
\cite[Table~3]{Christensen:1978}
(counting $n_s = 2 s + 1$ degrees of freedom per species)
\begin{equation}
\label{qsmassive}
  q_s = \{ 1 , \tfrac{7}{4} , -12 , -\tfrac{226}{4} , 200 \}
  \ 
  \mbox{for}
  \ 
  s = \{ 0 , \tfrac{1}{2} , 1 , \tfrac{3}{2} , 2 \}
  \ .
\end{equation}
In supergravity there are further contributions to $q_\eff$
from higher-derivative scalars, given in Table~1 of
\cite{Duff:1993wm}
(see also \cite{Asorey:2003uf}).

In a Schwarzschild black hole,
the classical Ricci tensor $R_{kl}$ and Ricci scalar $R$ vanish,
so the only contribution to the trace anomaly is from the squared Weyl tensor,
$C_{klmn} C^{klmn} = 48 M^2 / r^6$,
yielding, if~equation~(\ref{TE}) is correct,
\begin{equation}
\label{anomalyschw}
  T
  =
  \alpha
  C_{klmn} C^{klmn}
  =
  {q_\eff \over 60 \pi^2}
  {M^2 \over r^6}
  \ .
\end{equation}

\cite{Christensen:1977}
were the first to apply the trace anomaly~(\ref{anomalyschw}),
coupled with energy-momentum conservation
(which 
$\langle T_{kl} \rangle$
should satisfy),
and the behavior of Hawking flux at infinity
to derive
$\langle T_{kl} \rangle$
outside a Schwarzschild black hole.
However,
the trace anomaly, spherical symmetry,
conservation of energy-mom\-ent\-um,
and asymptotic behavior at infinity
are insufficient to determine the inside energy-momentum tensor uniquely.

The arguments of \S\ref{geometricoptics-sec}
supply two additional ingredients that resolve the ambiguity:
first,
the Hawking energy-momentum becomes isotropic near the singularity;
and second,
the Hawking energy-momentum has a power-law behavior with
proper time $\tau$ near the singularity,
equation~(\ref{kappa0sing}).
The arguments indicate that the Hawking energy-momentum tensor
near the singularity should approximate an isotropic, relativistic fluid,
with energy density going as
\begin{equation}
\label{rhoHsingprop}
  \rho_\iso \sim \kappa^4
  \propto
  | \tau |^{-4}
  \propto
  {M^2 \over r^6}
  \ ,
\end{equation}
and isotropic pressure $p_\iso = \tfrac{1}{3} \rho_\iso$.
It is encouraging that the predicted Hawking energy density~(\ref{rhoHsingprop})
depends on mass $M$ and radius $r$ in the same way as the trace anomaly,
equation~(\ref{anomalyschw}),
suggesting a relation between the two.

\subsection{Conservation of energy-momentum}
\label{conserveT-sec}

The expectation value $\langle T_{kl} \rangle$ of the quantum
contribution to the energy-mom\-en\-t\-um tensor in a curved spacetime
must satisfy conservation of energy-mom\-en\-t\-um
\cite{Christensen:1977}.
To allow calculation of the back reaction of the energy-momentum
on the spacetime, \S\ref{backreaction-sec},
it is convenient to use the following general form of
a spherically symmetric line-element,
a generalization of the Gullstrand-Painlev\'e version~(\ref{gp})
of the Schwarzschild line-element,
\begin{equation}
\label{ff}
  \dd s^2
  =
  - \, \dd t^2
  +
  {1 \over \beta_1^2}
  \left(
  \dd r - \beta_0 \, \dd t
  \right)^2
  +
  r^2
  \dd o^2
  \ .
\end{equation}
Through
$g_{\mu\nu} = \eta_{mn} e^m{}_\mu e^n{}_\nu$,
the line element~(\ref{ff}) defines not only a metric $g_{\mu\nu}$
but also a vierbein $e^m{}_\mu$
and associated locally inertial tetrad
$\bgamma_m \equiv e_m{}^\mu \be_\mu$
in terms of the basis of tangent vectors $\be_\mu$.
The tetrad defined by the line element~(\ref{ff}) is
locally inertial
($\bgamma_m \cdot \bgamma_n = \eta_{mn}$),
and freely falling.
The line element may be constructed from a general spherically symmetric
metric in ADM form with lapse $\alpha$
(that is, $\dd t \rightarrow \alpha \, \dd t$ in the line element~(\ref{ff})),
and noticing that the acceleration of the tetrad frame is
$\beta_1 \partial \ln \alpha / \partial r$,
which must vanish if the tetrad is everywhere in free-fall.
Imposing the boundary condition $\alpha = 1$ at spatial infinity
(as permitted by the gauge freedom in the choice of lapse $\alpha$)
fixes $\alpha = 1$ everywhere.
The coefficients $\beta_0$ and $\beta_1$
in the line element~(\ref{ff}) comprise the components
of a tetrad-frame 4-vector $\beta_m = \partial_m r$,
where $\partial_m$ is the tetrad-frame directed derivative.
In the special case of Gullstrand-Painlev\'e (i.e.\ Schwarzschild),
$\beta_1$ equals one,
and $\beta_0$ is minus the Newtonian escape velocity,
equation~(\ref{betagp}) with $\beta_0 = \beta$.
In general, the coefficients $\beta_m$
could be functions of time $t$ and radius $r$.
In the present case, the black hole
is almost stationary, and it suffices to consider $\beta_m$
to be functions only of radius $r$.
The interior (Misner-Sharp) mass $M$, a scalar, is defined by
\begin{equation}
\label{M}
  1 - {2 M \over r}
  \equiv
  \beta_m \beta^m
  =
  - \, \beta_0^2 + \beta_1^2
  \ .
\end{equation}
The advantage of the free-fall line element~(\ref{ff})
is first that it remains well-behaved inside as well as outside the horizon,
and second that, according to the arguments of \S\ref{geometricoptics-sec},
it is with respect to a free-fall frame that
the energy-momentum tensor of Hawking radiation near the singularity
is isotropic.

As originally demonstrated by
\cite{Christensen:1977},
and further explicated by
\cite{Visser:1997sd},
the most general spherically symmetric energy-momentum tensor
that is quasi-stationary and covariantly conserved
depends on two arbitrary functions
$R(r)$ and $P(r)$
and two constants of integration
$C^+$ and $C^-$.
The calculation of covariant conservation of energy-momentum is most elegant
when carried out in a Newman-Penrose double-null tetrad,
related to the locally inertial, free-fall tetrad
$\{ \bgamma_0 , \bgamma_1 , \bgamma_2 , \bgamma_3 \}$ by
\begin{equation}
  \bgamma_{\overset{\scriptstyle{v}}{\scriptstyle{u}}}
  =
  \tfrac{1}{\sqrt{2}} ( \bgamma_0 \pm \bgamma_1 )
  \ , \quad
  \bgamma_\pm
  =
  \tfrac{1}{\sqrt{2}} ( \bgamma_2 \pm \im \bgamma_3 )
  \ .
\end{equation}
The Newman-Penrose tetrad-frame components of
a spherical, quasi-stationary, conserved energy-momentum tensor
in the free-fall frame satisfy
(do not confuse the function $R(r)$ with the Ricci scalar $R$)
\begin{subequations}
\label{conservedT}
\begin{align}
\label{conservedTvv}
  T^{vv}
  &\equiv
  \tfrac{1}{2} ( \rho + p ) + f
\nonumber
\\
  &=
  {( \beta_1 - \beta_0 )^2 \over 2 ( \beta_1^2 + \beta_0^2 )}
  ( R + P )
  +
  {C^+ \over r^2}
  \left( {\beta_1 \over \beta_1 + \beta_0} \right)^2
  \ ,
\\
\label{conservedTuu}
  T^{uu}
  &\equiv
  \tfrac{1}{2} ( \rho + p ) - f
\nonumber
\\
  &=
  {( \beta_1 + \beta_0 )^2 \over 2 ( \beta_1^2 + \beta_0^2 )}
  ( R + P )
  +
  {C^- \over r^2}
  \left( {\beta_1 \over \beta_1 - \beta_0} \right)^2
  \ ,
\\
\label{conservedTvu}
  T^{vu}
  &\equiv
  \tfrac{1}{2} ( \rho - p )
\nonumber
\\
  &=
  \tfrac{1}{2} ( R - P )
  \ ,
\\
\label{conservedTpm}
  T^{+-}
  &\equiv
  p_\perp
\nonumber
\\
  &=
  {1 \over 4 r}
  \Biggl\{
  {\beta_1^2 \over \beta_1^2 - \beta_0^2}
  {\dd \over \dd r}
  \left[
  r^2
  {( \beta_1^2 - \beta_0^2 )^2 \over \beta_1^2 ( \beta_1^2 + \beta_0^2 )}
  ( R + P )
  \right]
  -
  {\dd \over \dd r}
  \left[
  r^2
  ( R - P )
  \right]
  \Biggr\}
  \ .
\end{align}
\end{subequations}
The quantities $\rho$, $p$, $p_\perp$, and $f$
are the proper energy density, radial pressure, transverse pressure,
and energy flux in the free-fall frame.
If the constants $C^+$ and $C^-$ both vanish,
then $R = \rho$ is the energy density,
and $P = p$ is the radial pressure.
Finiteness of the outgoing energy flux $T^{vv}$ at the horizon,
where $\beta_0 + \beta_1 = 0$,
forces the constant $C^+$ to vanish,
\begin{equation}
\label{Cp}
  C^+
  =
  0
  \ .
\end{equation}
The Einstein tensor corresponding to the line element~(\ref{ff}) satisfies
$G^{vv} / G^{uu} = ( \beta_0 - \beta_1 )^2 / ( \beta_0 + \beta_1 )^2$,
which would then seem to imply that
the constants $C^+$ and $C^-$ must be the same
(hence zero, since $C^+ \,{=}\, 0$).
But this conclusion is an artifact of taking
the line element~(\ref{ff}) to be exactly stationary,
which would prohibit a net radial flux of energy.
In the situation under consideration, the spacetime is almost but
not exactly stationary,
and a net radial flux of energy is permitted,
that is, $C^-$ can be non-zero.

In \S\ref{backreaction-sec},
$\beta_0$ and $\beta_1$ will be solved using the Einstein equations
to determine the back reaction of $\langle T_{kl} \rangle$ on the geometry.
Until then, let the back reaction be ignored,
so that the geometry is Schwarzschild,
where $\beta_1 = 1$ and $\beta_0 = - \sqrt{2 M / r}$.

In the Schwarzschild spacetime,
the trace anomaly is proportional to $r^{-6}$,
equation~(\ref{anomalyschw}).
Imposing isotropy,
$p_\perp = p$,
and keeping only terms behaving as $r^{-6}$ as $r \rightarrow 0$,
equations~(\ref{conservedT}) imply
that the conserved energy density $\rho$ and pressure $p$ in the
free-fall frame are, in terms of the trace anomaly $T$,
\begin{equation}
\label{rhopschwsing}
  \{ \rho , p \}
  =
  \{ \tfrac{1}{8} T , \tfrac{3}{8} T \}
  \ .
\end{equation}
It has been argued above,
equation~(\ref{rhoHsingprop}),
that the energy-momentum must include
a Hawking component,
an isotropic relativistic fluid,
necessarily with zero trace.
Equation~(\ref{rhopschwsing}) shows that such a relativistic fluid
by itself
does not satisfy energy-momentum conservation in the Schwarzschild geometry,
so some other component of energy-momentum must also be present.
A plausible additional component
--- perhaps the only plausible additional component ---
is vacuum energy,
which has an isotropic equation of state with
pressure equal to minus energy density,
$p_\vac = - \rho_\vac$.
Only the vacuum component contributes to the trace anomaly.
Equation~(\ref{rhopschwsing}) can be accomplished
by a sum of isotropic Hawking and vacuum components with
\begin{equation}
\label{rhoHvsing}
  \rho_\iso
  =
  3 \rho
  =
  \tfrac{3}{8} T
  \ , \quad
  \rho_\vac
  =
  - 2 \rho
  =
  - \tfrac{1}{4} T
  \ .
\end{equation}
If the trace anomaly $T$ is positive,
then the Hawking component has positive energy $\rho_\iso$,
while the vacuum component has negative energy $\rho_\vac$.
The physical interpretation is that Hawking particles acquire
their energy from the vacuum,
which decays to lower energy.
The picture is reminiscent of reheating following inflation in cosmology,
where vacuum energy decays to particle energy,
releasing entropy.

If this interpretation is correct,
then equation~(\ref{rhoHvsing}) with the anomaly~(\ref{anomalyschw})
implies that the energy density of Hawking radiation near the singularity is
\begin{equation}
\label{rhoHsing}
  \rho_\iso
  =
  {q_\eff \over 160 \pi^2}
  {M^2 \over r^6}
  \ .
\end{equation}
Now the anomaly coefficient $q_\eff$
depends on the particle content of the theory,
including not only massless but also massive particles,
equations~(\ref{qsmassless}) and~(\ref{qsmassive})
(M.\ J.\ Duff, private communication 2016, confirms this statement).
\cite[Table~1]{Duff:1993wm}
notes that $q_\eff$ vanishes in maximal ($N = 8$) supergravity
in 4 spacetime dimensions with all particles taken massless.
Unfortunately,
absent a knowledge of the correct theory of everything,
determining what $q_\eff$ actually is in reality is elusive.

Instead I follow the lead of
\cite{Christensen:1977},
asking what value of $q_\eff$ is required
in order to reproduce the correct flux of Hawking radiation at infinity.
The Hawking flux at infinity
should look like an outgoing flux of relativistic particles,
decaying as $f \rightarrow r^{-2}$ as $r \rightarrow \infty$.
The only such solution to equations~(\ref{conservedT})
(given that $C^+ = 0$ to ensure finiteness at the horizon)
is one with $C^-$ non-vanishing,
which gives
\begin{equation}
\label{fC}
  f
  =
  -
  {C^- \over 2 \bigl( r + \sqrt{2 M r} \bigr)^2}
  \quad
  \mbox{as $r \rightarrow \infty$}
  \ .
\end{equation}
The coefficient $C^-$ must be negative to ensure that the
flux of energy is pointed away from the black hole.
The density and pressure associated with the solution~(\ref{fC})
(with $R = P = 0$)
satisfy
$\rho = p = - f$
(and $p_\perp = 0$),
whereas an outgoing flux of relativistic particles should have
\begin{equation}
\label{rhopC}
  \rho = p = f
  \quad
  \mbox{as $r \rightarrow \infty$}
  .
\end{equation}
Therefore it is necessary to adjoin an additional traceless solution
of equations~(\ref{conservedT}) satisfying
$\rho = p \rightarrow - C^- / r^2$ as $r \rightarrow \infty$.
Such a solution exists, equation~(\ref{rhopower}).

As
\cite{Christensen:1977}
point out,
because the emitting disk of the black hole is finite,
the transverse pressure $p_\perp$ should fall off as $r^{-4}$ at infinity.
The black hole can be thought of as a fuzzy emitting disk,
with effective area
$\langle r_\bullet^2 \rangle$
and $r^2$-weighted area
$\langle r_\bullet^4 \rangle$
as seen from afar.
The dimensionless constants $a$ and $b$ may be defined by
\begin{equation}
  \langle r_\bullet^2 \rangle
  \equiv
  a M^2
  \ , \quad
  {\langle r_\bullet^4 \rangle \over
  \langle r_\bullet^2 \rangle}
  \equiv
  b M^2
  \ .
\end{equation}
In the high-frequency, geometric-optic limit,
the black hole is a uniformly emitting disk of radius $3 \sqrt{3} M$,
so the effective area and $r^2$-weighted areas would be
$\langle r_\bullet^2 \rangle = 27\pi M^2$
and
$\langle r_\bullet^4 \rangle = \langle r_\bullet^2 \rangle^2$,
implying $a = b = 27 \pi$.
In general,
the effective area should lie somewhere between zero and
the geometric-optics limit, and
$\langle r_\bullet^4 \rangle \leq \langle r_\bullet^2 \rangle^2$,
implying
\begin{equation}
  0 < a \leq 27 \pi
  \ , \quad
  0 < b \leq a
  \ .
\end{equation}
The radial and transverse pressures should fall off with radius as
\begin{equation}
\label{pperpp}
  p \propto
  {\langle r_\bullet^2 \rangle \over r^2}
  \ , \quad
  p_\perp \propto
  {\langle r_\bullet^4 \rangle \over r^4}
  \ , \quad
  {p_\perp \over p}
  \rightarrow
  {b M^2 \over r^2}
  \quad
  \mbox{as $r \rightarrow \infty$}
  \ .
\end{equation}

Finally,
there is a question of how rapidly the Hawking radiation
deviates from isotropy near the singularity.
Anisotropy in the pressure arises from a concentration
of Hawking radiation above and below relative to transverse directions.
Figure~\ref{kappachi}
suggests that the concentration of Hawking radiation from the sky above
is comparable to that from the illusory horizon below.
If the black hole and sky are modeled as fuzzy disks
of effective angular radius $\chi_\eff$
of the order of, perhaps slightly smaller than,
the angular size $\chi_\edge$ of the black hole,
equation~(\ref{chiedge}),
then
\begin{equation}
\label{chieff}
  \chi_\eff
  \approx
  {\pi \over 2} - e \sqrt{r \over 2 M}
  \ ,
\end{equation}
where $e$ is some constant of order, perhaps slightly greater than, unity.
If so, then the ratio $p / \rho$ of radial pressure to energy density is
\begin{equation}
\label{prhoedge}
  {p \over \rho}
  \approx
  \frac{1}{3}
  \left( {1 - \cos^3\!\chi_\eff \over 1 - \cos \chi_\eff} \right)
  \approx
  \frac{1}{3}
  \left(
  1 + e \sqrt{r \over 2 M}
  \right)
  \ .
\end{equation}

Solutions exist for the conserved energy-momentum
that satisfy equations~(\ref{conservedT})
and all the conditions imposed so far,
namely the trace anomaly condition~(\ref{anomalyschw}),
the asymptotic flux conditions~(\ref{fC}) and~(\ref{rhopC}),
the asymptotic transverse-to-radial pressure condition~(\ref{pperpp}),
isotropy near the singularity,
anisotropy near the singularity growing as~(\ref{prhoedge}),
and the condition that the energy-momentum be that
of a relativistic gas of Hawking radiation plus vacuum energy.
The last condition implies that only the vacuum contributes
to the trace of the energy-momentum.
The solutions are,
aside from the flux term~(\ref{fC}),
sums of integral and half-integral powers of inverse radius $r$.
The solution for the energy-momentum equals a sum of four components,
the Hawking flux component determined by equation~(\ref{fC}),
an anisotropic relativistic component,
an isotropic relativistic component,
and a vacuum component,
\begin{align}
\label{rhocomp}
  \{ \rho , p , p_\perp \}
  =
  \rho_\flux \{ 1 , 1 , 0 \}
  +
  \rho_\aniso \{ 1 , 1 , 0 \}
  +
  \rho_\iso \{ 1 , \tfrac{1}{3} , \tfrac{1}{3} \}
  +
  \rho_\vac \{ 1 , -1 , -1 \}
  \ .
\end{align}
The energy densities of the 4 components are
\begin{subequations}
\label{rhos}
\begin{align}
  \rho_\flux
  &=
  -
  f_\flux
  =
  - \rho_0
  {c M^2 \over \bigl( r + \sqrt{2 M r} \bigr)^2}
  \ ,
\\
  \rho_\aniso
  &=
  \rho_0
  \left(
  {2 c M^2 \over r^2} + {12 c M^3 \over r^3}
  +
  {2 c ( 20 - b ) M^4 \over r^4}
  \right.
\\
\nonumber
  &\quad\quad\quad
  \left.
  -
  \,
  {243 e M^{9/2} \over 14 \sqrt{2} \, r^{9/2}}
  +
  {6 ( 72 c - 5 ) M^5 \over r^5}
  +
  {27 e M^{11/2} \over \sqrt{2} \, r^{11/2}}
  \right)
  \ ,
\\
  \rho_\iso
  &=
  \rho_0
  \left(
  {3 b c M^4 \over r^4}
  +
  {405 e M^{9/2} \over 14 \sqrt{2} \, r^{9/2}}
  +
  {18 ( 3 - 32 c) M^5 \over r^5}
  -
  {243 e M^{11/2} \over 7 \sqrt{2} \, r^{11/2}}
  +
  {18 M^6 \over r^6}
  \right)
  \ ,
\\
  \rho_\vac
  &=
  -
  \rho_0
  {12 M^6 \over r^6}
  \ ,
\end{align}
\end{subequations}
where $\rho_0$ is the constant
\begin{equation}
\label{rho0}
  \rho_0
  \equiv
  {\alpha \over M^4}
  \ ,
\end{equation}
with $\alpha$ the anomaly coefficient given by equation~(\ref{qeff}).
The constant $b$ is determined by equation~(\ref{pperpp}),
the constant $e$ by equation~(\ref{chieff}),
and the negative constant $C^-$ in the flux component~(\ref{fC})
has been replaced by $-2 \rho_0 c$ with $c$ a positive constant.
The fact that the vacuum component is proportional to the trace anomaly
(that is, it has only an $r^{-6}$ part)
is a consequence of the assumption that all energy-momenta
other than the vacuum are relativistic, therefore traceless.
The energy flux contributed by
the anisotropic and isotropic relativistic components is
(the flux contributed by the vacuum is zero)
\begin{equation}
\label{fH}
  f_{\aniso + \iso}
  =
  {\sqrt{2 M r} \over r + 2 M}
  ( \rho + p )_{\aniso + \iso}
  \ ,
\end{equation}
which falls off as $f \propto r^{-5/2}$ as $r \rightarrow \infty$,
so does not contribute to the asymptotic flux at infinity.
The only contributor to the flux at infinity is $\rho_\flux$,
for which the asymptotic flux is
\begin{equation}
\label{fflux}
  f
  \rightarrow
  f_\flux
  \rightarrow
  {\rho_0 c M^2 \over r^{2}}
  \quad
  \mbox{as $r \rightarrow \infty$}
  \ .
\end{equation}
The different behavior of the energy fluxes $f_\flux$ and $f_\aniso$
is the reason for distinguishing the flux and anisotropic components
$\rho_\flux$ and $\rho_\aniso$
despite their having the same equation of state
$\{ \rho , p , p_\perp \} \propto \{ 1 , 1 , 0 \}$.

\cite{Visser:1997sd}
has emphasized that the flux $f_\flux$ of energy over the horizon
is an ingoing flux of negative energy
($C^- = -2 \rho_0 c$ is negative).
There appear to be other contributions to the flux from
equations~(\ref{conservedT}),
namely equation~(\ref{fH}),
but the latter contributions are associated with the fact
that the tetrad frame is infalling,
and an infaller will see an apparent outward flux
even when the energy-momentum tensor is stationary.
That the contributions~(\ref{fH}) do not correspond
to a real outward flux follows from the fact that for a strictly
stationary spacetime the ratio
of outgoing to ingoing energy densities
would be
$T^{vv} / T^{uu} = ( \beta_0 - \beta_1 )^2 / ( \beta_0 + \beta_1 )^2$,
which the $R + P$ terms
in equations~(\ref{conservedTvv}) and~(\ref{conservedTuu})
satisfy,
but the $C^-$ term does not (given that $C^+ = 0$).
This is another reason
to distinguish the flux and anisotropic components
$\rho_\flux$ and $\rho_\aniso$:
only the former yields a genuine non-stationary flux of radiation.

\rhofig

Figure~\ref{rho}
shows the radiation and vacuum densities from equations~(\ref{rhos}).
The radiation density goes smoothly from being highly
anisotropic at large radii to isotropic at small radii.
This is consistent with the arguments of \S\ref{geometricoptics-sec},
which indicate that the Hawking radiation
is isotropic near the singularity,
but lop-sided further out,
the Hawking radiation from the sky being less than the Hawking
radiation from the illusory horizon.

Strictly, the solution~(\ref{rhos}) for the energy-momentum
holds only as long as the anomaly coefficient $q_\eff$ defined by
equation~(\ref{qeff}) is constant.
However, the solution should be a good approximation provided that
$q_\eff$ varies slowly with $r$.

It should be commented that there are other solutions
for conserved energy-momentum satisfying
equations~(\ref{conservedT})
that are not integral or half-integral powers of radius,
equations~(\ref{rhopower}),
but these solutions are independent of the integral or half-integral
power-law solutions~(\ref{rhos}),
and are not needed here.
These other power-law solutions would be needed
if the anomaly coefficient $q_\eff$ were not constant,
but rather some other power of radius.
More generally, one would expect that energy-momentum-conserving
solutions would exist for any arbitrarily varying
anomaly coefficient $q_\eff$.

The asymptotic Hawking flux~(\ref{fflux})
is proportional to the product of the anomaly coefficient $q_\eff$
and the constant $c$.
To relate $q_\eff$ to the asymptotic Hawking flux,
a value of $c$ is needed.
A plausible constraint is that the $r^{-5}$ contribution
to the Hawking radiation should range between being maximally isotropic
($r^{-5}$ contribution to $\rho_\aniso$ vanishes)
to maximally anisotropic
($r^{-5}$ contribution to $\rho_\iso$ vanishes),
which imposes
\begin{equation}
\label{crange}
  0.069
  =
  \frac{5}{72}
  \leq
  c
  \leq
  \frac{3}{32}
  =
  0.094
  \ .
\end{equation}

The asymptotic Hawking flux~(\ref{fflux})
should be equal to a thermal flux of radiation
at the Hawking temperature $1 / ( 8\pi M )$,
\begin{equation}
\label{fth}
  f
  \rightarrow
  {n_\eff \pi^2 \over 30}
  {a M^2 \over 4\pi r^2} \left( {1 \over 8\pi M} \right)^4
  \ ,
\end{equation}
where $n_\textrm{eff}$ is the effective number of relativistic particle species,
\begin{equation}
  n_\eff
  =
  \sum_s g_s n_s
  \ ,
\end{equation}
with $g_s$ the number of helicity states per species,
which is $g_s = 1$ for real scalars,
and $g_s = 2$ for $s \geq \tfrac{1}{2}$.
The calculation of the effective number $n_\eff$
of relativistic species is familiar from cosmology,
at least up to energies $\sim 1 \unit{TeV}$ accessible to experiment
\cite{Kolb:1990vq}:
$n_\eff$ varies from $4$ (2 helicities for each of photons and gravitons)
at the lowest energies (if there are no massless neutrinos),
up to
$n_\eff = 2 ( 14 + \frac{7}{8} 45 ) = 106.75$
at $\sim 1 \unit{TeV}$.
In the low energy regime where only photons and gravitons are massless,
as appropriate outside an astronomical black hole,
\begin{equation}
  n_\eff
  =
  4
  \ .
\end{equation}
\cite{Page:1976}
gives effective areas $a_s$ for spins $\tfrac{1}{2}$, $1$, and $2$,
and \cite{Bardeen:2014uaa} for spin $0$,
 \begin{equation}
  a_s
  =
  \{
  14.26
  ,
  18.05
  ,
  6.492
  ,
  0.742
  \}
  \pi
  \mbox{~for~}
  s = \{ 0 , \tfrac{1}{2} , 1 , 2 \}
  \ .
\end{equation}
For only photons and gravitons massless,
the effective dimensionless area $a$ is then, at low energies,
\begin{equation}
  a
  =
  {\sum_s g_s n_s a_s \over \sum_s g_s n_s}
  =
  {14.468 \pi \over 4}
  =
  3.617 \pi
  \ .
\end{equation}
Equating the asymptotic Hawking flux~(\ref{fflux})
to the thermal flux~(\ref{fth})
yields a value for the low-energy anomaly coefficient $q_\eff$,
\begin{equation}
\label{qth}
  q_\eff
  =
  {3 n_\eff a \over 512 \pi c}
  \approx
  1
  \ ,
\end{equation}
with values ranging over $q_\eff = 1.22$ to $0.90$
for $c = \tfrac{5}{72}$ to $\tfrac{3}{32}$,
equation~(\ref{crange}).
The inferred value~(\ref{qth}) of $q_\eff$ seems surprisingly small
given the typically ``large'' values of the spin-dependent coefficients $q_s$
given by equations~(\ref{qsmassless}) and~(\ref{qsmassive}).
As remarked after equation~(\ref{rhoHsing}),
the anomaly coefficient $q_\eff$, equation~(\ref{qeff}),
should be summed over {\em all\/} particle species,
not only massless but also massive.
As emphasized by
\cite{Duff:1993wm},
$q_\eff$ vanishes in maximal ($N \,{=}\, 8$) supergravity
with all particles treated as massless.
But at low energies only photons and gravitons
(and possibly one generation of neutrinos) are massless,
and it is far from clear what $q_\eff$ should be in this regime.
Here I leave the difficulty alone,
and simply proceed on the assumption that the estimate~(\ref{qth}) of $q_\eff$
is correct.

With the value~(\ref{qth}) of $q_\eff$ in hand,
it is possible to estimate the ratio of the predicted Hawking energy-density
$\rho_\iso$
near the singularity to that of a relativistic thermal gas
at temperature $\kappa / (2\pi)$,
with $\kappa$ given by the geometric-optics value~(\ref{kappa0sing}),
\begin{equation}
\label{rhoth}
  \rho_\textrm{th}
  =
  {n_\eff \pi^2 \over 30} \left( {\kappa \over 2\pi} \right)^4
  \rightarrow
  {n_\eff \over 1920 \pi^2}
  {M^2 \over r^6}
  \quad
  \mbox{as $r \rightarrow 0$}
  \ .
\end{equation}
The ratio of Hawking~(\ref{rhoHsing}) to thermal energy~(\ref{rhoth}) is
\begin{equation}
  {\rho_\iso \over \rho_\textrm{th}}
  =
  {12 q_\eff \over n_\eff}
  \quad
  \mbox{as $r \rightarrow 0$}
  \ .
\end{equation}
Near the singularity,
particles besides photons and gravitons will become relativistic,
and $n_\eff$ will increase correspondingly.
But if it is assumed (ad hoc) that $q_\eff \propto n_\eff$,
then $q_\eff / n_\eff$ is independent of energy, and
\begin{equation}
\label{rhoHth}
  {\rho_\iso \over \rho_\textrm{th}}
  \approx
  3
  \quad
  \mbox{as $r \rightarrow 0$}
  \ ,
\end{equation}
with values ranging from
${\rho_\iso / \rho_\textrm{th}} = 3.7$ to $2.7$
for $q_\eff = 1.22$ to $0.90$
corresponding to $c = \tfrac{5}{72}$ to $\tfrac{3}{32}$,
equation~(\ref{crange}).
By comparison, in 1+1 dimensions
the ratio of Hawking to thermal density
is 3 with no adjustable parameters,
equation~(\ref{rhoHth2}).
As emphasized in \S\ref{geometricoptics-sec},
Hawking radiation is expected to be non-thermal near the singularity,
because the acceleration $\kappa$ seen by an infaller
is time-varying rather than constant.
Equation~(\ref{rhoHth}) suggests
that the density of Hawking radiation is of the order of
a few times
the density of a thermal distribution at temperature $\kappa / (2\pi)$,
which is not wholly unreasonable.
The low value of $q_\eff$, equation~(\ref{qth}),
inferred from consistency between the trace anomaly and the
Hawking flux at infinity is crucial here:
if $q_\eff / n_\eff$ were much larger,
then the Hawking energy density near the singularity
would substantially exceed the
energy density of a thermal distribution with the same characteristic energy,
which would be hard to explain.

\section{Back reaction}
\label{backreaction-sec}

In the previous section \S\ref{traceanomaly-sec},
the calculation
of the expectation value $\langle T_{kl} \rangle$
of the energy-momentum ignored the back reaction of
that energy-momentum on the Schwarzschild geometry.

As long as densities or curvatures have not yet hit the Planck scale,
the back reaction can be calculated using the
classical Einstein equations.
In a spherically symmetric spacetime,
2 of the 4 Einstein equations are redundant with energy-momentum conservation,
and may therefore be discarded.
The remaining 2 Einstein equations may be taken to be,
in the free-fall tetrad defined by the line-element~(\ref{ff}),
\begin{subequations}
\label{einsteineqs}
\begin{align}
\label{einsteineq0}
  {\dd \beta_0 \over \dd \tau}
  &=
  - \,
  {M \over r^2}
  -
  4\pi r p
  \ ,
\\
\label{einsteineq1}
  {\dd \beta_1 \over \dd \tau}
  &=
  4\pi r f
  \ ,
\end{align}
\end{subequations}
where $M$ is the interior mass~(\ref{M}).
The coefficient $\beta_0$ equals the rate of change
$\dd r / \dd \tau$
of the radius $r$ with respect to proper time $\tau$
in the infalling tetrad-frame,
so the first Einstein equation~(\ref{einsteineq0})
is an equation for the radial acceleration
${\dd \beta_0 / \dd \tau} = \ddsq r / \dd \tau^2$
of the tetrad frame
(the tetrad frame is by construction freely-falling;
what is accelerating here is the radial coordinate $r$).
Equation~(\ref{einsteineq0}) says that the radial acceleration
is the sum of a term $- M / r^2$
which looks like the ordinary Newtonian gravitational attraction,
and a term $- 4\pi r p$
proportional to the radial pressure $p$.
Together $\beta_0$ and $\beta_1$ form the components of
a tetrad vector $\beta_m = \partial_m r$.
The second Einstein equation~(\ref{einsteineq1})
gives an equation for the second component $\beta_1$
of the tetrad vector.
The equation states that $\dd \beta_1 / \dd \tau$
is proportional to the energy flux $f$.

The Einstein equations~(\ref{einsteineqs})
are valid for any spherically symmetric spacetime,
stationary or otherwise.
For a large black hole, it is an excellent approximation
to treat the spacetime as stationary,
in which case
${\dd \beta_m / \dd \tau} = \beta_0 \, {\dd \beta_m / \dd r}$.
The only non-stationary term in the energy-momentum~(\ref{conservedT})
is the $C^-$ term
(given that $C^+$ must vanish, equation~(\ref{Cp})),
but it is safe to drop this term
since it is proportional to $r^{-2}$ and therefore becomes negligible
compared to the $r^{-6}$ terms near the singularity.
If the $C^-$ term is neglected,
then the flux $f$ is related to the energy density $\rho$ and pressure $p$ by
\begin{equation}
\label{fgen}
  f
  =
  -
  {\beta_0 \beta_1 ( \rho + p )
  \over
  \beta_0^2 + \beta_1^2}
  \ ,
\end{equation}
which is positive since $\beta_0$ is negative.

\rhosingfig

Following the arguments of \S\ref{geometricoptics-sec},
I assume that the energy-momentum near the singularity is isotropic
in the free-fall frame,
\begin{equation}
\label{piso}
  p_\perp
  =
  p
  \ .
\end{equation}
Following
\cite{Christensen:1978},
I assume that the trace anomaly is sourced solely by
the Euler density $E$,
equation~(\ref{TE}),
with for simplicity constant coefficient $\alpha$.
In an arbitrary spherical spacetime,
the Euler density $E$ is
\begin{equation}
\label{Espher}
  E
  \equiv
  48 C^2
  -
  2
  R_{kl} R^{kl}
  +
  \tfrac{2}{3}
  R^2
  \ ,
\end{equation}
where the Weyl scalar $C$,
Ricci tensor squared $R_{kl} R^{kl}$,
and Ricci scalar $R$ are
\begin{subequations}
\begin{align}
  C
  \equiv
  \tfrac{1}{2} C_{vuvu}
  &=
  {4\pi \over 3} ( \rho - p + p_\perp ) - {M \over r^3}
  \ ,
\\
  R_{kl} R^{kl}
  &=
  (8\pi)^2
  ( \rho^2 + p^2 + 2 p_\perp^2 - 2 f^2 )
  \ ,
\\
  R
  &=
  8\pi ( - \rho + p + 2 p_\perp )
  \ .
\end{align}
\end{subequations}

In summary,
the equations to be solved to determine the back reaction
of the quantum energy-momentum are
the Einstein equations~(\ref{einsteineqs}),
the energy conservation equation~(\ref{conservedTpm})
with the isotropic equation of state~(\ref{piso}),
the flux equation~(\ref{fgen})
(which is also a consequence of energy-momentum conservation),
and the trace anomaly equation~(\ref{TE})
with Euler density~(\ref{Espher}).

Unsurprisingly,
the quantum back-reaction becomes important when
the energy density and curvature squared,
both of which are diverging as $M^2 / r^6$ near the singularity,
reach the Planck scale.
At that point the proper time $|\tau|$ left to hit the singularity
is also about a Planck time,
but the radius, which goes as $r \propto M^{1/3} \tau^{2/3}$,
is still much larger than a Planck length
for a black hole of astronomical mass $M$
($10$ solar masses = $10^{39}$~Planck masses).
Of course,
a classical calculation of the back reaction should no longer
be trusted when the density and curvature reach the Planck scale,
but still it is useful to carry out the calculation to see what happens.

Figure~\ref{rhosing}
shows the interior mass $M$,
equation~(\ref{M}),
scaled to the mass $M_\bullet$
of the black hole at infinity,
and the radiation and vacuum densities $\rho_\iso$ and $\rho_\vac$,
when the back reaction of the density on the geometry is taken into account.
One might have thought that the overall positive energy density
near the singularity might reduce the interior mass $M$,
but the opposite is the case.
In the Schwarzschild regime,
the interior mass $M$
equals the mass $M_\bullet$ at infinity,
but once back reaction sets in,
the interior mass starts to increase as
\begin{equation}
  M
  \propto r^{-0.6}
  \ .
\end{equation}
The vierbein coefficients $\beta_m$ satisfy
\begin{equation}
  \beta_0
  =
  -
  \sqrt{2 M \over r}
  \ , \quad
  \beta_1 \approx {M \over M_\bullet}
  \ .
\end{equation}
More precisely, the vierbein coefficient $\beta_1$ varies smoothly from
$\beta_1 = M / M_\bullet$ at $r \gg ( \alpha M )^{1/3}$ to
$\beta_1 = 0.8 M / M_\bullet$ at $r \ll ( \alpha M )^{1/3}$.
Because $\beta_0$ is so much larger than $\beta_1$,
the approximation
$\beta_0 = - \sqrt{2 M / r}$
is very well satisfied.
Thanks to the increasing interior mass,
the infall velocity $\beta_0$ of the free-fall frame diverges
even faster than in Schwarzschild,
as $\beta_0 \propto r^{-0.8}$.
After back reaction sets in,
the densities of radiation and vacuum
diverge as $M / r^3$,
which is more slowly than the $r^{-6}$ of the Schwarzschild regime,
\begin{equation}
  \rho
  \propto
  {M \over r^3}
  \propto
  r^{-3.6}
  \ .
\end{equation}
The ratio $\rho_\vac / \rho_\iso$ of vacuum to radiation density is
$- \frac{2}{3}$
before back reaction sets in,
and
$- \frac{1}{3}$
after back reaction sets in.

The fact that the back reaction hardens rather than softens the singularity,
in the sense that the back reaction causes
the infall velocity $\beta_0$ of the free-fall frame to diverge even faster,
can be traced to the Einstein equations~(\ref{einsteineqs}).
The right hand side of the evolution equation~(\ref{einsteineq0})
for $\beta_0$
shows that $\beta_0$ must increase inward (become more negative) as long as
the interior mass $M$ and radial pressure $p$ are positive.
In the present case
the radial pressure $p$ is sourced by isotropic Hawking radiation,
which has positive pressure,
and by vacuum energy, which also has positive pressure
as long as the trace anomaly is positive and therefore
the vacuum energy is negative.
The right hand side of the evolution equation~(\ref{einsteineq1})
for $\beta_1$
depends on the energy flux $f$, equation~(\ref{fgen}).
The flux $f$ is sourced only by Hawking radiation,
not by vacuum energy, which has $\rho + p = 0$.
The flux is suppressed by a factor of $\sim 1 / \beta_0$
compared to the density $\rho$ or pressure $p$,
so $\beta_1$ evolves more slowly than $\beta_0$.
Thus the evolution of the interior mass $M$
is dominated by the evolution of $\beta_0$.
The result is that the interior mass $M$ starts to increase
once the pressure term
on the right hand side of the evolution equation~(\ref{einsteineq0})
becomes comparable to the mass term.
The increase of interior mass is inevitable as long as the radial pressure $p$
is positive, as here.

The conclusion is that, at the classical level,
the back reaction of quantum energy-momentum
on the singularity in no way softens the singularity
or in any way alleviates the inevitability of singularities
in general relativity.
A resolution of the singularity of a Schwarzschild black hole
requires a full theory of quantum gravity.

\section{1+1 dimensions}
\label{dim2-sec}

As a check on the analysis of this paper,
it is useful to see what happens in 1+1 dimensions.

General relativity is weird in 2 spacetime dimensions.
In 2 spacetime dimensions,
the Einstein tensor vanishes identically,
and the usual Einstein equations relating curvature to energy-momentum
do not hold.
Extensions of general relativity,
such as dilaton gravity \cite{Grumiller:2002nm},
are however well-defined in 2 spacetime dimensions.

Notwithstanding the shortcomings of general relativity
in 2 spacetime dimensions,
models in 2 dimensions have historically been treated extensively
because they admit an exact calculation of the expectation value
$\langle T_{kl} \rangle$ of the quantum contribution to
the energy-momentum associated with any prescribed geometry
\cite{Birrell:1982}.

\subsection{Quantum energy-momentum in 1+1 dimensions}

By a suitable coordinate transformation of the two coordinates,
the line element in 1+1 dimensions can be taken to have the
conformally flat form
\begin{equation}
\label{ds2conformal}
  \dd s^2
  =
  - \ee^{2\xi} \dd v \dd u
  \ ,
\end{equation}
where $v$ and $u$ are null coordinates,
and $\xi$ is some function of the two coordinates.
The trace anomaly in 2 spacetime dimensions is
\cite[eq.~(6.121)]{Birrell:1982}
\begin{equation}
  T
  =
  \alpha R
  \ , \quad
  \alpha
  \equiv
  {q_\eff \over 24 \pi}
  \ ,
\end{equation}
where $q_\eff = 1$ for a single minimally-coupled massless scalar field.
Energy-momen\-tum conservation then determines
the remaining two components of the energy-momentum
up to two arbitrary functions.
The arbitrary functions are associated with residual gauge freedoms
that leave the line element in conformally flat form,
namely transformations
$\xi \rightarrow \xi + \xi^+(u) + \xi^-(v)$
of the conformal function $\xi$.
The Newman-Penrose tetrad-frame components of the energy-momentum are\footnote{
The energy-momenta~(\ref{conservedTxi2}) are tetrad-frame components,
whereas coordinate-frame components are commonly quoted in the literature.
Covariant coordinate-frame energy-momenta $T_{\kappa\lambda}$
(or contravariant $T^{\kappa\lambda}$) are equal
to tetrad-frame energy-momenta multiplied by the conformal factor $\ee^{2\xi}$
(or $\ee^{-2\xi}$).}
\begin{subequations}
\label{conservedTxi2}
\begin{align}
\label{conservedTxivv2}
  T^{vv}
  &=
  4\alpha \,
  \ee^{-2 \xi}
  \left[
  {\partial^2 \xi \over \partial u^2}
  -
  \left(
  {\partial^2 \xi \over \partial u^2}
  \right)^2
  +
  f^+ ( u )
  \right]
  \ ,
\\
\label{conservedTxiuu2}
  T^{uu}
  &=
  4\alpha \,
  \ee^{-2 \xi}
  \left[
  {\partial^2 \xi \over \partial v^2}
  -
  \left(
  {\partial^2 \xi \over \partial v^2}
  \right)^2
  +
  f^- ( v )
  \right]
  \ ,
\\
\label{conservedTxivu2}
  T^{vu}
  &=
  -
  \tfrac{1}{2}
  T
  =
  - \tfrac{1}{2} \alpha R
  =
  -
  4 \alpha \,
  \ee^{-2 \xi}
  {\partial^2 \xi \over \partial v \partial u}
  \ ,
\end{align}
\end{subequations}
where
$f^+(u)$ and $f^-(v)$ are some functions of the null coordinates
$u$ and $v$.

\subsection{Schwarzschild analog in 1+1 dimensions}

Now restrict to the case where the line element~(\ref{ds2conformal})
is quasi-stationary,
meaning that the conformal function $\xi$ is a function only of the
radial coordinate $r$, not of the time coordinate $t$.
If the line element is taken to have the Gullstrand-Painlev\'e form~(\ref{ff}),
with $\beta_0$ and $\beta_1$ functions only of radius $r$,
then the Newman-Penrose tetrad-frame components of
the energy-momentum tensor~(\ref{conservedTxi2})
in the free-fall frame satisfy
($R$ here is the Ricci scalar)
\begin{subequations}
\label{conservedT2}
\begin{align}
\label{conservedTvv2}
  T^{vv}
  &=
  \left( {\beta_1 \over \beta_1 + \beta_0} \right)^2
  \left\{
  \alpha
  \left[
  {\beta_0^3 \over \beta_1^2}
  {\partial \over \partial r}
  \left( \beta_1
  {\partial \over \partial r} {\beta_0 \over \beta_1} \right)
  -
  {R \over 2}
  \right]_\textrm{hor}^r
  +
  C^+
  \right\}
  \ ,
\\
\label{conservedTuu2}
  T^{uu}
  &=
  \left( {\beta_1 \over \beta_1 - \beta_0} \right)^2
  \left\{
  \alpha
  \left[
  {\beta_0^3 \over \beta_1^2}
  {\partial \over \partial r}
  \left( \beta_1
  {\partial \over \partial r} {\beta_0 \over \beta_1} \right)
  -
  {R \over 2}
  \right]_\textrm{hor}^r
  +
  C^-
  \right\}
  \ ,
\\
  T^{vu}
  &=
  -
  \tfrac{1}{2} \alpha R
  =
  -
  \alpha
  \beta_1 {\partial \over \partial r}
  \left(
  \beta_0 {\partial \over \partial r} {\beta_0 \over \beta_1}
  \right)
  \ ,
\end{align}
\end{subequations}
where $C^+$ and $C^-$ are constants.
The condition that $T^{vv}$ be finite at a horizon,
where $\beta_1 + \beta_0 = 0$,
imposes $C^+ = 0$.
If the geometry were exactly stationary,
then the $C^-$ term in $T^{uu}$ would also vanish.
But the geometry is not exactly stationary in the situation being considered,
because the black hole is slowly losing energy in Hawking radiation.
The value of the constant $C^-$ is determined by the requirement
that the flux at infinity be purely outgoing.
Note that equations~(\ref{conservedTvv})--(\ref{conservedTvu}) hold
in 1+1 dimensions
with $C^\pm / r^2 \rightarrow C^\pm$.

In the particular case of the Schwarzschild geometry,
where $\beta_1 = 1$ and $\beta_0 = - \sqrt{2 M / r}$,
equations~(\ref{conservedT2}) reduce to
\begin{subequations}
\label{conservedT2M}
\begin{align}
\label{conservedTvv2M}
  T^{vv}
  &=
  \rho_0
  \left( 1 + \sqrt{2 M \over r} \right)^2
  \left( 1 + {4 M \over r} + {12 M^2 \over r^2} \right)
  \ ,
\\
\label{conservedTuu2M}
  T^{uu}
  &=
  \rho_0
  \left( 1 + \sqrt{2 M \over r} \right)^{-2}
  \left( - \, {32 M^3 \over r^3} + {48 M^4 \over r^4} \right)
  \ ,
\\
  T^{vu}
  &=
  - \rho_0 {32 M^3 \over r^3}
  \ ,
\end{align}
\end{subequations}
where $\rho_0$ is the constant
\begin{equation}
  \rho_0
  =
  {\alpha \over 16 M^2}
  =
  {q_\eff \over 384\pi M^2}
  \ .
\end{equation}
The choice
$C^+ = 0$ ensures that the outgoing energy-momentum $T^{vv}$
is finite at the horizon.
The choice
$C^- = - \rho_0 / 2$
ensures no ingoing energy-momentum at infinity,
$T^{uu} \rightarrow 0$ as $r \rightarrow \infty$.

Analogously to equation~(\ref{rhocomp}),
the energy-momentum may be expressed
as a sum of a non-stationary Hawking flux component $\rho_\flux$,
a stationary Hawking component $\rho_\stat$,
and a vacuum component $\rho_\vac$,
\begin{equation}
  \{ \rho , p \}
  =
  \rho_\flux \{ 1 , 1 \}
  +
  \rho_\stat \{ 1 , 1 \}
  +
  \rho_\vac \{ 1 , -1 \}
  \ .
\end{equation}
The energy densities of the 3 components are
\begin{subequations}
\label{rhos2}
\begin{align}
\label{rhos2flux}
  \rho_\flux
  &=
  -
  f_\flux
  =
  - \rho_0
  {1 \over 2 \bigl( 1 + \sqrt{2 M / r} \bigr)^2}
  \ ,
\\
\label{rhos2stat}
  \rho_\stat
  &=
  \rho_0
  \left( 1 + {2 M \over r} \right)
  \left( 1 + {4 M \over r} + {12 M^2 \over r^2} \right)
  \ ,
\\
  \rho_\vac
  &=
  - \rho_0 {32 M^3 \over r^3}
  \ .
\end{align}
\end{subequations}
As in equation~(\ref{fH}),
the energy flux contributed by the stationary component is
\begin{equation}
\label{fH2}
  f_\stat
  =
  {\sqrt{2 M r} \over r + 2 M}
  ( \rho + p )_\stat
  \ ,
\end{equation}
which tends to zero at infinity.
The energy flux in the stationary component is non-zero
because the frame is that of a free-falling observer,
who sees a finite outgoing flux
where a stationary observer would see zero flux.

\rhotwofig

Figure~\ref{rho2}
shows the radiation and vacuum densities from equations~(\ref{rhos2}).
As in 3+1 dimensions,
the radiation density goes smoothly from being highly anisotropic at
large radii to isotropic at small radii.

As in 3+1 dimensions,
the Hawking energy $\rho_\flux + \rho_\stat$ is everywhere positive,
while the vacuum energy $\rho_\vac$ is everywhere negative.
But whereas the total energy density near the singularity in 3+1 dimensions
was positive, equations~(\ref{rhoHvsing}),
the total energy density $\rho = \rho_\stat + \rho_\vac$
near the singularity in 1+1 dimensions is negative,
\begin{equation}
\label{rhoHvsing2}
  \rho_\stat
  =
  - 3 \rho
  =
  \tfrac{3}{8} T
  \ , \quad
  \rho_\vac
  =
  4 \rho
  =
  - \tfrac{1}{2} T
  \ .
\end{equation}
The total energy $\rho_\flux + \rho_\stat + \rho_\vac$ goes through zero
at $r \approx 0.36 M$.

\subsection{Comparison between 1+1 and 3+1 dimensions}

The asymptotic Hawking flux~(\ref{rhos2flux})
equals a thermal flux of radiation
at the Hawking temperature $1 / ( 8\pi M )$,
\begin{equation}
\label{fth2}
  f_\flux
  \rightarrow
  {n_\eff \pi \over 12} \left( 1 \over 8\pi M \right)^2
  =
  {n_\eff \over 768\pi M^2}
  =
  {\rho_0 \over 2}
  \ .
\end{equation}
It is notable that in 1+1 spacetime dimensions
the calculation of the asymptotic flux of Hawking radiation
from the trace anomaly
yields the correct value with no adjustable parameters,
as first pointed out by \cite{Davies:1976}.

Equations~(\ref{kappa0s})--(\ref{tau})
for the acceleration $\kappa_0$
on the illusory horizon below and on the sky above
hold in 1+1 dimensions.
Analogously to equation~(\ref{rhoth}) in 3+1 spacetime dimensions,
the thermal energy-density of Hawking radiation in 1+1 dimensions
associated with the characteristic acceleration is
\begin{equation}
\label{rhoth2}
  \rho_\textrm{th}
  =
  {n_\eff \pi \over 6}
  \left( {\kappa_0 \over 2\pi} \right)^2
  \rightarrow
  {n_\eff M \over 48\pi r^3}
  =
  8 \rho_0 {M^3 \over r^3}
  \quad
  \mbox{as $r \rightarrow 0$}
  \ .
\end{equation}
By comparison the actual Hawking density near the singularity
from equation~(\ref{rhos2stat}) is
$\rho_\stat \rightarrow 24 \rho_0 {M^3 / r^3}$
as $r \rightarrow 0$.
Thus the Hawking density near the singularity
is 3 times that predicted by the thermal approximation,
\begin{equation}
\label{rhoHth2}
  {\rho_\stat \over \rho_\textrm{th}}
  =
  3
  \quad
  \mbox{as $r \rightarrow 0$}
  \ .
\end{equation}
Interestingly, this is the same ratio estimated in 3+1 dimensions,
equation~(\ref{rhoHth}).
As emphasized in the 3+1 case,
Hawking radiation is expected to be non-thermal near the singularity
because the acceleration seen by an infaller is time-varying rather
than constant.

\cite{Chakraborty:2015nwa}
argued that the calculation of the quantum energy-momentum in 1+1 dimensions
should be extrapolated to 3+1 dimensions by multiplying by $1/(4\pi r^2)$.
If so, the quantum energy-momentum in 3+1 dimensions
would diverge as $M / r^5$ near the singularity.
The results obtained in the present paper indicate that this simple
argument is incorrect.
The correct extrapolation is to interpret the acceleration $\kappa$
as the characteristic frequency of Hawking radiation near the singularity,
in which case the quantum energy density in 1+1 dimensions
diverges near the singularity
as $\kappa^2 \sim M / r^3$,
and in 3+1 dimensions as
$\kappa^4 \sim  M^2 / r^6$,
consistent in both cases with the behavior of the trace anomaly.

In 1+1 dimensions as in 3+1 dimensions,
the non-stationary component
$\rho_\flux$
of the quantum energy-momentum
is an ingoing flux of negative energy,
a point emphasized by \cite{Visser:1997sd}.

\section{Summary and conclusions}

It has been argued that from any observer's perspective,
whether outside or inside the true (future event) horizon,
the boundary of a black hole,
the boundary that separates what an observer can see
from what they cannot see,
is not the true horizon, but rather the illusory horizon
\cite{Hamilton:2010my},
the dimming, redshifting surface of the star or whatever
collapsed into the black hole long ago.
An observer who falls inside the true horizon continues to
see the illusory horizon ahead of them,
still dimming and redshifting away.
Since the illusory horizon is the perceived boundary
of the observer's spacetime,
the illusory horizon must also be the source of Hawking
radiation from the black hole.
Once it is accepted that the illusory horizon is the source
of Hawking radiation,
it becomes possible to calculate Hawking radiation
seen by an observer inside the true horizon.

Brute-force calculations either of Hawking radiation seen by an infaller,
or of the expectation value of the quantum energy-momentum inside
a Schwarzschild black hole, have to date proved prohibitively difficult,
e.g.\ \cite{Candelas:1980,Hodgkinson:2014iua}.
I have therefore resorted to simpler arguments to infer
Hawking radiation inside a Schwarzschild black.

The classic calculation of Hawking radiation from a horizon
shows that the Hawking radiation is thermal,
with a temperature equal to $1/(2\pi)$ times the acceleration (gravity)
$\kappa$ at the horizon.
The acceleration equals the rate at which matter falling through the horizon
appears to redshift.
I therefore start by calculating the acceleration $\kappa$
at the illusory horizon seen by a freely-falling, non-rotating observer.
A freely-falling observer accelerates away from the sky above as they fall,
so an infaller should also see Hawking radiation from the sky above.
In \S\ref{geometricoptics-sec},
the acceleration on the illusory horizon below and in the sky above
is calculated in the geometric-optics limit,
which should be valid for Hawking radiation at high frequencies.
Remarkably, near the singularity
the acceleration becomes isotropic, the same in all directions,
both on the illusory horizon below and in the sky above.
Moreover the acceleration diverges in a power-law fashion near the singularity,
the acceleration going as one third the inverse proper time $|\tau|$
left to hit the singularity,
$\kappa \rightarrow 1 / ( 3 |\tau| ) \propto r^{-3/2}$,
equation~(\ref{kappa0sing}).
The Hawking radiation seen by an observer depends on their motion
(for example, if they accelerate or rotate),
but there is no observer who can move in such a way as to eliminate
Hawking radiation altogether.
Thus regardless of details,
the robust conclusion is that the interior of the black hole
must contain Hawking radiation that contributes an energy-momentum
that near the singularity is isotropic and diverging as
$\rho \propto \kappa^4 \propto r^{-6}$.

In \S\ref{waves-sec}
the wave equation in the Schwarzschild geometry is solved numerically
to confirm that the geometric-optics limit holds as long as
the ``out'' frequency $\omega$ at which a wave of pure ``in'' frequency
is being observed is larger than the inverse time left to hit the singularity,
$| \omega \tau| \gg 1$.

\S\ref{traceanomaly-sec} follows the pioneering work of
\cite{Christensen:1977},
who pointed out that
it is possible to constrain the expectation value
of the quantum energy-momentum tensor in the Schwarzschild black hole
by imposing conservation of energy-momentum,
invoking the known value of the trace of the quantum energy-momentum
(known as the trace anomaly, or conformal anomaly),
and requiring that the energy-momentum go over to the known
Hawking flux far from the black hole.
\cite{Christensen:1977}
applied their arguments to calculate the quantum energy-momentum
near and outside the horizon,
but did not discuss the interior of the black hole,
perhaps because they chose to work in a coordinate rest frame
(as opposed to a free-fall frame),
perhaps because the physical arguments were insufficient to
constrain the inside energy-momentum.
The additional information gained in the present paper,
that the Hawking radiation is isotropic and going as $\rho \propto r^{-6}$
in a free-fall frame near the singularity,
suffices to fix the quantum energy-momentum throughout
the Schwarzschild spacetime.
Encouragingly,
the trace anomaly, which is proportional to the Weyl curvature squared,
diverges as $r^{-6}$
consistent with the predicted behavior of the Hawking radiation.

The energy-momentum near the singularity proves to be a sum
of an isotropic, relativistic fluid of Hawking radiation,
and a vacuum component with negative energy density.
The ratio of vacuum to Hawking energy density is $- \frac{2}{3}$,
so the net energy density is positive near the singularity.
This makes physical sense:
the Hawking radiation diverging as $r^{-6}$ near the singularity
is acquiring its energy from the vacuum, which itself goes into deficit.

\S\ref{backreaction-sec}
uses the classical Einstein equations to
calculate the back reaction of the quantum energy-momentum
on the Schwarzschild geometry.
The back reaction serves merely to exacerbate the singularity,
in the sense that the back reaction causes
the infall velocity of a freely-falling frame to diverge
even faster than it already does in the Schwarzschild geometry.
Thus the back reaction calculated at the classical level does not resolve
the fundamental problem of singularities in general relativity.
To resolve the singularity of a Schwarzschild black hole,
a full theory of quantum gravity is required.

\S\ref{dim2-sec}
shows that all the results derived for 3+1 dimensions in this paper
are consistent with known results in 1+1 dimensions \cite{Birrell:1982}.

The results of this paper raise many questions
which it is beyond the scope of this paper to address:
what are the implications for
generalized thermodynamics,
or for the information paradox,
or holography, or string theory, or quantum gravity in general?


\appendix

\section*{Appendices}


\section{Penrose coordinates}
\label{penrosecoords}

The tortoise, or Regge-Wheeler \cite{Regge:1957},
coordinate $r^\ast$ is defined by
\begin{equation}
\label{rtortoise}
  r^\ast
  \equiv
  \int {\dd r \over 1 - 2 M / r}
  =
  r + 2 M \ln \left| {r \over 2 M} - 1 \right|
  \ .
\end{equation}
Kruskal-Szekeres
\cite{Kruskal:1960,Szekeres:1960}
coordinates
$r_\textrm{K}$ and $t_\textrm{K}$ are defined by
\begin{equation}
  \begin{array}{l}
  \displaystyle
  r_\textrm{K} + t_\textrm{K}
  =
  2 M
  \exp \left( {r^\ast + t \over 4 M} \right)
  \ ,
  \\[2ex]
  \displaystyle
  r_\textrm{K} - t_\textrm{K}
  =
  \pm
  2 M
  \exp \left( {r^\ast - t \over 4 M} \right)
  \ ,
  \end{array}
\end{equation}
where the $\pm$ sign in the last equation
is $+$ outside the horizon, $-$ inside the horizon.
Penrose time and space coordinates $t_\textrm{P}$ and $r_\textrm{P}$
can be defined by any conformal transformation
\begin{equation}
\label{rtPenrose}
  r_\textrm{P} \pm t_\textrm{P}
  =
  f ( r_\textrm{K} \pm t_\textrm{K} )
\end{equation}
for which $f ( z )$ is finite as $z \rightarrow \pm \infty$.
The transformation~(\ref{rtPenrose})
brings spatial and temporal infinity to finite values of the coordinates,
while keeping infalling and outgoing light rays at $45^\circ$
in the spacetime diagram.
It is common to draw a Penrose diagram with the singularity horizontal,
which can be accomplished by choosing the function $f ( z )$ to be odd,
$f ( -z ) = - f ( z )$.
The Penrose dagrams in
Figure~\ref{stpcoll} adopt
\begin{equation}
\label{fPenrose}
  f ( z )
  =
  {2 \over \pi} \tan^{-1}\!z
  \ .
\end{equation}
Different choices of the odd function $f(z)$
leave the Penrose diagrams of Figure~\ref{stpcoll} qualitatively unchanged.

\section{Classical acceleration perceived by an infaller}
\label{tech}

The Schwarzschild metric in polar coordinates
$x^\mu \equiv \{ t , r , \theta , \phi \}$
is
(units $c = G = 1$)
\begin{equation}
  \dd s^2
  =
  - \, \Delta \, \dd t^2
  + \Delta^{-1} \dd r^2
  + r^2 ( \dd \theta^2 + \sin^2\!\theta \, \dd \phi^2 )
  \ ,
\end{equation}
where $\Delta$ is the horizon function
\begin{equation}
  \Delta \equiv 1 - 2 M / r
  \ ,
\end{equation}
whose vanishing defines the horizon,
the Schwarzschild radius at $r = 2 M$.

The radially infalling observer is conveniently
taken to be at the pole $\theta = 0$.
Light rays from an emitter to the observer then follow trajectories of constant
longitude, $\dd \phi = 0$.
In the Schwarzschild metric,
the coordinate 4-velocity
$k^\mu \equiv \dd x^\mu / \dd \lambda$
along a null geodesic at constant longitude satisfies three conservation laws,
associated with
energy,
mass,
and
angular momentum $J$ per unit energy,
\begin{equation}
\label{kschw}
  k^t = {1 / \Delta}
  \ , \quad
  k^r = \pm \sqrt{1 - J^2 \Delta / r^2}
  \ , \quad
  k^\theta = - {J / r^2}
  \ , \quad
  k^\phi = 0
  \ .
\end{equation}
Here
the photon frequency $k^t$ has
been normalized to one as perceived by observers at rest at infinity.
If a radial infaller has specific energy $E$,
then their coordinate 4-velocity $u^\mu \equiv \dd x^\mu / \dd \tau$ is
\begin{equation}
  u^t = {E \over \Delta}
  \ , \quad
  u^r =
  - \sqrt{E^2 - \Delta}
  \ , \quad
  u^\theta =
  u^\phi = 0
  \ .
\end{equation}
This paper considers a radial infaller who falls from zero velocity
at infinity, in which case $E = 1$,	
but the equations given in this Appendix are valid for arbitrary
specific energy $E$.
The photon frequency
$\omega \equiv - u_\mu k^\mu$
emitted or observed by a radial infaller,
relative to unit frequency at rest at infinity,
is
\begin{equation}
\label{omega}
  \omega
  =
  {E
  - u^r k^r
  \over \Delta}
  \ .
\end{equation}
Equation~(\ref{omega}) applies to a radial infaller,
whether emitter or observer.
For an emitter at infinity,
the emitted frequency
relative to unit frequency at rest at infinity
is constant,
\begin{equation}
  \omega_\emit
  =
  E
  \quad
  ( r_\emit \rightarrow \infty )
  \ .
\end{equation}
For an emitter near the illusory horizon,
where $\Delta \rightarrow 0$,
the emitted frequency
relative to unit frequency at rest at infinity
is proportional to $1/\Delta$
regardless of the emitter's orbital parameters,
\begin{equation}
  \omega_\emit
  \propto
  {1 \over \Delta}
  \quad
  ( r_\emit \rightarrow 2 M )
  \ .
\end{equation}

To avoid any ambiguity,
in the remainder of this Appendix
observer and emitter quantities are labelled with explicit subscripts
$\obs$ and $\emit$.
The photon angular momentum $J$ is related to
the infalling observer's radius $r_\obs$ and viewing angle $\chi$ by
\begin{equation}
\label{chi}
  J
  =
  \left|
  {r_\obs \sin\chi \over E_\obs - u_\obs^r \cos\chi}
  \right|
  \ .
\end{equation}
The angle $\theta$ between emitter and observer
is determined by integrating
$\dd \theta / \dd r = k^\theta / k^r$, giving
\begin{equation}
\label{theta}
  \theta
  =
  \int_{r_\emit}^{r_\obs}
  {J \, \dd r \over r^2 \sqrt{1 - J^2 \Delta / r^2}}
  \ ,
\end{equation}
which is an elliptic integral.
The rate $\dd \tau_\emit / \dd \tau_\obs$
at which the observer sees the emitter's proper time elapse
equals the blueshift factor $\omega_\obs / \omega_\emit$,
\begin{equation}
  {\dd \tau_\emit \over \dd \tau_\obs}
  =
  {\omega_\obs \over \omega_\emit}
  \ .
\end{equation}
The acceleration $\kappa$ that the observer perceives
from a (radially moving) emitter at fixed angular location $\theta$ is
\begin{align}
\label{kappatheta}
  \kappa
  &\equiv
  -
  \left.
  {\dd \ln ( \omega_\obs / \omega_\emit ) \over \dd \tau_\obs}
  \right|_\theta
\\
\nonumber
  &=
  - \,
  {\dd r_\obs \over \dd \tau_\obs}
  \left.
  {\partial \ln \omega_\obs \over \partial r_\obs}
  \right|_J
  +
  {\omega_\obs \over \omega_\emit}
  {\dd r_\emit \over \dd \tau_\emit}
  \left.
  {\partial \ln \omega_\emit \over \partial r_\emit}
  \right|_J
  -
  \left.
  {\dd J \over \dd \tau_\obs}
  \right|_\theta
  \left.
  {\partial \ln ( \omega_\obs / \omega_\emit ) \over \partial J}
  \right|_{r_\emit , r_\obs}
  \ ,
\end{align}
the derivative
$\left.  {\dd J / \dd \tau_\obs} \right|_\theta$
of the photon angular momentum
$J$ at fixed $\theta$ being determined by
\begin{equation}
\label{Jtheta}
  {\dd \theta \over \dd \tau_\obs}
  =
  0
  =
  {\dd r_\obs \over \dd \tau_\obs}
  \left.
  {\partial \theta \over \partial r_\obs}
  \right|_{J , r_\emit}
  +
  {\omega_\obs \over \omega_\emit}
  {\dd r_\emit \over \dd \tau_\emit}
  \left.
  {\partial \theta \over \partial r_\emit}
  \right|_{J , r_\obs}
  +
  \left.
  {\dd J \over \dd \tau_\obs}
  \right|_\theta
  \left.
  {\partial \theta \over \partial J}
  \right|_{r_\emit , r_\obs}
  \ ,
\end{equation}
with $\theta$ from equation~(\ref{theta}).
The radial velocities are
$\dd r_\obs / \dd \tau_\obs = u^r_\obs$
and
$\dd r_\emit / \dd \tau_\emit = u^r_\emit$,
which are negative since both emitter and observer are infalling.
The middle term on the right hand side of equation~(\ref{Jtheta})
is negligible for an infalling emitter asymptotically close to the horizon,
or for an emitter at infinity.
In evaluating equation~(\ref{kappatheta}),
a useful intermediate result is
\begin{equation}
\label{udomega}
  {u^r \over \omega}
  \left.
  {\partial \ln \omega \over \partial r}
  \right|_J
  =
  {1 \over 2 k^r}
  \left[
  \left(
  1 - {J^2 \over r^2 \omega^2}
  \right)
  {\dd \Delta \over \dd r}
  +
  {( u^r )^2 \over \omega^2}
  {\dd ( J^2 / r^2 ) \over \dd r}
  \right]
  \ ,
\end{equation}
which holds for both emitter and observer.
For an infalling emitter near the horizon,
the quantity~(\ref{udomega}) tends to a constant $- 1 / (4 M)$.
For an emitter at infinity,
the quantity~(\ref{udomega}) is zero.
In the simple case of zero photon angular momentum,
$J = 0$,
equation~(\ref{udomega}) reduces to
\begin{equation}
\label{udomega0}
  {u^r \over \omega}
  \left.
  {\partial \ln \omega \over \partial r}
  \right|_{J = 0}
  =
  {1 \over 2}
  {\dd \Delta \over \dd r}
  \ ,
\end{equation}
and the acceleration $\kappa$, equation~(\ref{kappatheta}),
simplifies to
\begin{equation}
\label{kappa0omega}
  \kappa_0
  =
  -
  {\omega_\obs \over 2}
  \left(
  {\dd \Delta_\obs \over \dd r_\obs}
  -
  {\dd \Delta_\emit \over \dd r_\emit}
  \right)
  \ .
\end{equation}
Equation~(\ref{kappa0omega}) reproduces equations~(\ref{kappa0s})
when $E_\obs = 1$ and the emitter is either at the illusory horizon,
or at infinity.

\kappathetafig

Figure~\ref{kappathetas}
shows the acceleration $\kappa$ perceived by an infaller
watching a fixed location $\theta$ on the illusory horizon,
calculated from equations~(\ref{kappatheta}) and (\ref{Jtheta}).
The perceived acceleration $\kappa$ is independent of
the orbital parameters of the emitter,
as long as the emitter is asymptotically close to the horizon.
The acceleration is plotted in
Figure~\ref{kappathetas}
not as a function of the location $\theta$,
but rather as a function of the viewing angle $\chi$,
in part to allow comparison to Figure~\ref{kappachi},
and in part to bring out the fact that the
acceleration $\kappa$ at fixed $\theta$
is approximately the same function of $\chi$
at any observer radius $r_\obs$.
Figure~\ref{kappathetas}
shows that the acceleration $\kappa$ is positive (redshifting)
at viewing angles $\chi$ less than about a radian,
and negative (blueshifting) at larger viewing angles.

To watch a fixed angular location $\theta$ on the illusory horizon
or at infinity,
the infalling observer must rotate their view as they fall in.
As argued in \S\ref{accel},
a cleaner measurement of Hawking radiation is made by an inertial observer,
who watches in a fixed direction as they free-fall.
The non-rotating observer's view pans across the illusory horizon
or the sky above
as they fall in.
The acceleration $\kappa$ that the observer perceives
while staring along a fixed viewing direction $\chi$ is
\begin{align}
\label{kappachieq}
  \kappa
  &\equiv
  -
  \left.
  {\dd \ln ( \omega_\obs / \omega_\emit ) \over \dd \tau_\obs}
  \right|_\chi
\nonumber
\\
  &=
  - \,
  \left.
  {\dd \ln ( \omega_\obs / \omega_\emit ) \over \dd \tau_\obs}
  \right|_\theta
  -
  \left.
  {\partial \ln ( \omega_\obs / \omega_\emit ) \over \partial J}
  \right|_{r_\emit , r_\obs}
  \left(
  \left.
  {\dd J \over \dd \tau_\obs}
  \right|_\chi
  -
  \left.
  {\dd J \over \dd \tau_\obs}
  \right|_\theta
  \right)
  \ ,
\end{align}
the first term on the right hand side of which is the acceleration
at fixed $\theta$ given by equation~(\ref{kappatheta}).
The derivative
$\left.  {\dd J / \dd \tau_\obs} \right|_\theta$
of $J$ at fixed $\theta$ is given by equation~(\ref{Jtheta}),
while its derivative
$\left.  {\dd J / \dd \tau_\obs} \right|_\chi$
at fixed $\chi$ is determined by
\begin{equation}
\label{Jchi}
  {\dd \chi \over \dd \tau_\obs}
  =
  0
  =
  {\dd r_\obs \over \dd \tau_\obs}
  \left.
  {\partial \chi \over \partial r_\obs}
  \right|_J
  +
  \left.
  {\dd J \over \dd \tau_\obs}
  \right|_\chi
  \left.
  {\partial \chi \over \partial J}
  \right|_{r_\obs}
  \ ,
\end{equation}
with $\chi$ given in terms of $r_\obs$ and $J$
by equation~(\ref{chi}).
For an emitter at infinity, the emitted frequency $\omega_\emit$ is constant,
and the factor
$\left. {\partial \ln ( \omega_\obs / \omega_\emit ) / \partial J} \right|_{r_\emit , r_\obs}$
on the right hand side of equation~(\ref{kappachieq})
simplifies to
\begin{equation}
  \left.
  {\partial \ln ( \omega_\obs / \omega_\emit ) \over \partial J}
  \right|_{r_\emit , r_\obs}
  =
  \left. {\partial \ln \omega_\obs \over \partial J} \right|_{r_\obs}
  \quad
  ( r_\emit \rightarrow \infty )
  \ ,
\end{equation}
with the observed frequency $\omega_\obs$
being given in terms of $r_\obs$ and $J$
by equation~(\ref{omega}).
For an emitter on the illusory horizon,
the calculation of the factor
$\left. {\partial \ln ( \omega_\obs / \omega_\emit ) / \partial J} \right|_{r_\emit , r_\obs}$
on the right hand side of equation~(\ref{kappachieq})
is more involved.
To ensure that the observer is watching the same emitted in-mode
from the illusory horizon,
the emitted affine distance $\lambda_\emit$,
equation~(\ref{lambdaem}),
must be held fixed as the emitter's angular position is varied
over the illusory horizon
at constant observer position $r_\obs$.
Normalized to a frame at rest at infinity,
the affine distance $\lambda$
between emitter and observer
is obtained by integrating $\dd r / \dd \lambda = k^r$, giving
\begin{equation}
\label{lambdaschw}
  \lambda
  =
  \int_{r_\emit}^{r_\obs}
  {\dd r \over \sqrt{1 - J^2 \Delta / r^2}}
  \ .
\end{equation}
The emitted and observed affine distances
$\lambda_\emit$ and $\lambda_\obs$ are then
\begin{equation}
\label{lambdemobs}
  \lambda_\emit
  =
  \omega_\emit \lambda
  \ , \quad
  \lambda_\obs
  =
  \omega_\obs \lambda
  \ .
\end{equation}
Equations~(\ref{lambdemobs}) imply that fixing the
emitted affine distance $\lambda_\emit$ as the emitter's angular position
is varied imposes the condition
\begin{equation}
\label{lambdacond}
  \left.
  {\partial \ln ( \omega_\obs / \omega_\emit ) \over \partial J}
  \right|_{r_\emit , r_\obs}
  =
  \left.
  {\partial \ln \lambda_\obs \over \partial J}
  \right|_{r_\emit , r_\obs}
  \quad
  ( r_\emit \rightarrow 2 M )
  \ ,
\end{equation}
with $\lambda_\obs$ as a function of
$J$, $r_\obs$, and $r_\emit$ defined by
equations~(\ref{lambdaschw}) and (\ref{lambdemobs}),
together with (\ref{omega}).
Replacing the factor
$\left. {\partial \ln ( \omega_\obs / \omega_\emit ) / \partial J} \right|_{r_\emit , r_\obs}$
in equation~(\ref{kappachieq})
by the right hand side of equation~(\ref{lambdacond})
yields the acceleration $\kappa$ perceived by the observer
staring at a fixed viewing direction on the illusory horizon,
as illustrated in Figure~\ref{kappachi}.

\section{The wave equation and its solution}
\label{waveeqsoln-sec}

\subsection{The wave equation}
\label{waveeq-sec}

A massless wave of spin $s$
(= $0$, $\tfrac{1}{2}$, $1$, $\tfrac{3}{2}$, or $2$)
and definite chirality has $2s+1$ components
\cite{Chandrasekhar:1983}.
With respect to a Newman-Penrose tetrad,
the components have boost weights $\sigma = - s , -s{+}1 , ... , s$,
and spin weights $\varsigma$
equal (right-handed chirality) or opposite (left-handed chirality)
to their boost weights.
A component of boost weight $\sigma$ is multiplied by $\ee^{\sigma \eta}$
under a Lorentz boost by rapidity $\eta$
in the radial direction.
A component of spin weight $\varsigma$ is multiplied by $\ee^{-\im \varsigma \zeta}$
under a right-handed rotation by angle $\zeta$ in the horizontal plane.
Only one of the $2s+1$ components of a spin-$s$ wave (of definite chirality)
is long range and propagating,
namely the component with boost weight $\sigma = -s$ for an outgoing wave,
and boost weight $\sigma = s$ for an ingoing wave.
The remaining components are short-range.
The short-range components are not independent,
but rather oscillate in harmony with the long-range propagating component.

The propagating component $\varphi_\sigma$
(with boost weight $\sigma = -s$ outgoing, $\sigma = + s$ ingoing)
of a spin-$s$ wave in the Schwarzschild geometry satisfies the wave equation
\begin{equation}
\label{phieq}
  \left[
  \left| {\Delta \over r^2} \right|^s \!
  \left(
  {\partial \over \partial r^\ast}
  \mp
  {\partial \over \partial t}
  \right)
  \left| {\Delta \over r^2} \right|^{-s} \!
  \left(
  {\partial \over \partial r^\ast}
  \pm
  {\partial \over \partial t}
  \right)
  -
  V_{\mp s}
  \right]
  r
  | \Delta |^{s/2}
  \varphi_{\mp s}
  =
  0
  \ ,
\end{equation}
where $r^\ast$ is the tortoise coordinate
defined by equation~(\ref{rtortoise}).
Outside the horizon,
the tortoise coordinate increases outward from
$r^\ast \rightarrow - \infty$ at the horizon to
$r^\ast \rightarrow + \infty$ at spatial infinity.
Inside the horizon,
the tortoise coordinate increases inward from
$r^\ast \rightarrow - \infty$ at the horizon to
$r^\ast = 0$ at the singularity at zero radius.
The potential $V_\sigma$
for a wave of angular momentum
$l = s , s{+}1 , ...$
is
\begin{equation}
\label{V}
  V_{\sigma} (r)
  \equiv
  \left[
  l ( l + 1)
  +
  ( s - 1 )^2
  -
  ( s - 1 )
  ( 2 s - 1 )
  \Delta
  \right]
  {\Delta \over r^2}
  \ ,
\end{equation}
independent of the sign $\sigma = \mp s$ of the boost weight.
Angular eigenmodes are the usual spin spherical harmonics
${}_{\varsigma\,} \! Y_{lm} ( \theta , \phi )$
with spin weight $\varsigma = \pm \sigma$ ($+$ for right-handed,
$-$ for left-handed chirality).

The wave equation~(\ref{phieq}) simplifies near the horizon,
where the horizon function goes to zero,
$\Delta \rightarrow 0$,
and hence the potential goes to zero,
$V_{\sigma} \rightarrow 0$.
Near the horizon,
or more generally wherever the potential $V_\sigma$ is small
(which includes spatial infinity $r \rightarrow \infty$
for finite angular momentum $l$,
that is,
$l ( l {+} 1 ) | \Delta / r^2 | \ll 1$),
the propagating eigenmodes of definite frequency $w$ are
(the eigenfrequency is written $w$
to avoid confusion with
the frequency $\omega$ emitted or observed by a freely-falling observer)
\begin{equation}
\label{phihor}
  r
  \varphi_{\sigma}
  \sim
  \ee^{- \im w ( t \mp r^\ast )}
  | \Delta |^{-s/2}
  \ ,
\end{equation}
with sign $-$ for outgoing, $+$ for ingoing eigenmodes.
For non-zero spin $s$,
the propagating wave $\varphi_{\sigma}$
diverges as $| \Delta |^{-s/2}$ at the horizon,
for both outgoing ($\sigma = -s$) and ingoing ($\sigma = +s$) waves.

\subsection{Waves seen in the frames of free-fallers}

Gullstrand-Painlev\'e coordinates are convenient
to recast waves into the frames of free-fallers.
The Gullstrand-Painlev\'e line element is
\begin{equation}
\label{gp}
  \dd s^2
  =
  - \dd \tau^2
  + ( \dd r - \beta \dd \tau )^2
  + r^2 ( \dd \theta^2 + \sin^2\!\theta \, \dd \phi^2 )
  \ ,
\end{equation}
where $\beta = u^r = \dd r / \dd \tau$
is the (negative) infall velocity of a person who
falls radially from zero velocity at infinity
(such a person has specific energy $E = 1$),
\begin{equation}
\label{betagp}
  \beta ( r ) \equiv
  - \sqrt{1 - \Delta}
  =
  - \sqrt{2 M \over r}
  \ ,
\end{equation}
and $\tau$ is the proper time experienced by such a person,
\begin{equation}
  \tau
  =
  t
  -
  \int
  \beta \, \dd r^\ast
  =
  t
  +
  2 M
  \ln \left|
  {1 - \beta \over 1 + \beta}
  \right|
  +
  {4 M \over \beta}
  \ .
\end{equation}
Outgoing and ingoing
Finkelstein null coordinates $t \mp r^\ast$
can be expressed in terms of Gullstrand-Painlev\'e coordinates as
\begin{equation}
  \dd t \mp \dd r^\ast
  =
  {1 \over 1 \pm \beta}
  \left[
  \dd \tau \mp ( \dd r - \beta \dd \tau )
  \right]
  \ .
\end{equation}
Along the path of the infaller, $\dd r - \beta \dd \tau = 0$.
Thus outgoing and ingoing Finkelstein null coordinates $t \mp r^\ast$
along the path of an infaller are
\begin{align}
\label{tr}
  t \mp r^\ast
  &=
  \int {\dd \tau \over 1 \pm \beta}
\\
  \nonumber
  &=
  \mp \, 4 M \ln \left| 1 \mp \sqrt{r \over 2 M} \right|
  - \sqrt{2 M r} \left( 2 \pm \sqrt{r \over 2 M} + {r \over 3 M} \right)
  \ ,
\end{align}
which is shifted so
$t \mp r^\ast = 0$
at the singularity $r = 0$.

A wave $\varphi_\sigma$ of boost weight $\sigma$
is multiplied by $\ee^{\sigma \eta}$ under a radial boost by rapidity $\eta$.
The rapidity $\eta$ is negative for an inward boost,
positive for an outward boost.
The boost factor $\ee^{\eta}$ in the
transformation from the stationary to the infalling frame is
(here $\eta$ is negative)
\begin{equation}
  \ee^{\eta}
  =
  \sqrt{\left| 1 + \beta \over 1 - \beta \right|}
  \ ,
\end{equation}
which tends to zero at the horizon, where $\beta = - 1$.
The boost factor in an outgoing frame is the reciprocal of this.
Whereas the ingoing frame has specific energy $E = 1$
and radial 4-velocity $u^r$ negative
both outside and inside the horizon,
the outgoing frame has $E = 1$ and $u^r$ positive outside the horizon,
but $E = -1$ and $u^r$ negative inside the horizon.
Near the horizon, the infalling and outgoing boost factors are
\begin{equation}
  \ee^{\eta}
  \approx
  \tfrac{1}{2}
  | \Delta |^{\pm 1/2}
  \ ,
\end{equation}
with sign $+$ for infalling, $-$ for outgoing.
Thus in an infalling or outgoing frame,
the propagating eigenmodes~(\ref{phihor}) behave as
\begin{equation}
\label{phihorinout}
  r
  \varphi_{\sigma}
  \sim
  \left\{
  \begin{array}{ll}
  \ee^{- \im w ( t - r^\ast )}
  | \Delta |^{-s}
  &
  \mbox{out wave, in frame,}
  \\
  \ee^{- \im w ( t + r^\ast )}
  &
  \mbox{in wave, in frame,}
  \\
  \ee^{- \im w ( t - r^\ast )}
  &
  \mbox{out wave, out frame,}
  \\
  \ee^{- \im w ( t + r^\ast )}
  | \Delta |^{-s}
  &
  \mbox{in wave, out frame.}
  \end{array}
  \right.
\end{equation}
Equation~(\ref{phihorinout}) says that an ingoing wave
appears to have constant amplitude to an infalling observer,
and likewise an outgoing wave
appears to have constant amplitude to an outgoing observer;
while an outgoing mode ($\sigma = -s$)
appears to an infalling observer to diverge as
$| \Delta |^{-s}$,
and similarly an ingoing mode ($\sigma = +s$)
appears to an outgoing observer to diverge as
$| \Delta |^{-s}$.

The apparent divergence of an outgoing wave seen by an ingoer
should be interpreted with care.
An outgoing wave near the horizon,
whether outside or inside the horizon,
always moves away from the horizon,
so an ingoing observer always sees the amplitude of an outgoing wave
to be smaller than when the outgoing wave was emitted.
Thus an ingoer falling through the horizon
does not actually see any divergence in outgoing waves.

The propagating near-horizon eigenmodes~(\ref{phihorinout}) take the same
functional form regardless of eigenfrequency $w$.
Therefore a general near-horizon wave must take the form
\begin{equation}
  r
  \varphi_{\sigma}
  \sim
  \left\{
  \begin{array}{ll}
  f ( t - r^\ast )
  | \Delta |^{-s}
  &
  \mbox{out wave, in frame,}
  \\
  f ( t + r^\ast )
  &
  \mbox{in wave, in frame,}
  \\
  f ( t - r^\ast )
  &
  \mbox{out wave, out frame,}
  \\
  f ( t + r^\ast )
  | \Delta |^{-s}
  &
  \mbox{in wave, out frame.}
  \end{array}
  \right.
\end{equation}

\subsection{Outgoing modes near the horizon}

A Hawking mode from the illusory horizon
originates as an outgoing mode of pure frequency $\omegain$ in the frame of
a near-horizon infaller.
A propagating outgoing mode
$\varphi_\omegain$
of pure frequency $\omegain$
in the frame of a near-horizon infaller is
(the subscript $\sigma$ is dropped for brevity,
and a subscript $\omegain$ is added to emphasize that
the mode is being assigned a definite ``in'' frequency $\omegain$),
\begin{equation}
\label{phioomegain}
  r
  \varphi_\omegain ( \tauin )
  =
  f \left[ t(\tau) - r^\ast(\tau) \right]
  | \Delta |^{-s/2}
  \ee^{- s \eta}
  =
  \ee^{- \im \omegain \tauin}
  | \Delta |^{-s/2}
  \ee^{- s \eta}
  \ ,
\end{equation}
where $\tauin$
is the infaller's proper time shifted to zero at horizon crossing,
and $t(\tau)$ and $r^\ast(\tau)$
are the time and tortoise coordinate along the path of the infaller.
The boost factor $\ee^{-s \eta}$ in equation~(\ref{phioomegain})
has been factored out to permit flexibility in the choice of frame
(the rapidity $\eta$ is negative for an infalling frame).
Near the horizon, the outgoing null coordinate
$t - r^\ast$
varies with the infaller's proper time $\tauin$ as
\begin{equation}
\label{trh}
  t - r^\ast
  \approx
  -
  \int {\dd \tauin \over \kappa_\hor \tauin}
  =
  -
  {1 \over \kappa_\hor} \ln | \tauin |
  \ ,
\end{equation}
and
$\kappa_\hor \equiv - \dd \beta / \dd \tau = 1/(4 M)$
is the acceleration at the horizon.
Equation~(\ref{trh}) is valid not only outside but also inside the horizon,
with $\tauin < 0$ outside the horizon and $\tauin > 0$ inside the horizon.
Substituting equation~(\ref{trh})
into equation~(\ref{phioomegain})
yields the near-horizon form of the pure-frequency outgoing wave
as a function of $t$ and $r^\ast$,
\begin{equation}
\label{phinearhorizon}
  r
  \varphi_\omegain ( t , r )
  \approx
  \exp \left[ \pm \im {\omegain \over \kappa_\hor} \ee^{- \kappa_\hor ( t - r^\ast )} \right]
  | \Delta |^{-s/2}
  \ee^{- s \eta}
  \ ,
\end{equation}
with the $\pm$ sign being $+$ outside the horizon, $-$ inside the horizon.

The form
$\varphi_\omegain ( t , r )$
of the pure-frequency mode everywhere in the spacetime
follows from solving the wave equation~(\ref{phieq})
subject to the near-horizon initial condition~(\ref{phinearhorizon}).
The wave equation~(\ref{phieq}) is in the stationary frame $\eta = 0$,
so the initial condition~(\ref{phinearhorizon}) should be applied
with $\eta = 0$,
and the resulting wave should then be boosted into the frame of the observer.
The boost affects only the amplitude, not the phase, of the wave.

The situation of interest for calculating Hawking radiation
from the illusory horizon
is what the mode
$\varphi_\omegain$ of pure frequency $\omegain$ in the first infaller's frame
looks like to a second infaller who falls in some time later.
As a function of the proper time $\tauout$
along the path of the second infaller,
the pure-frequency mode evolves as
\begin{equation}
  \varphi_\omegain ( \tauout )
  =
  \varphi_\omegain ( \tout , \rout )
  \ ,
\end{equation}
where $\tout \equiv t(\tauout)$ and $\rout \equiv r(\tauout)$
are the time and radius along the path of the second infaller.

\subsection{Numerical computation of the wave equation for outgoing modes}

Suppose that the second infaller hits the singularity
at some time $t = t_0$.
If the numerical integration of the wave
$\varphi_\omegain$
is started near the horizon at some large negative value $r^\ast_0$
of the tortoise coordinate,
then for high-frequency radial (zero angular-momentum) modes
the past lightcone of the second infaller hitting the singularity at $t = t_0$
encompasses a range of times
$t = t_0 \pm r^\ast_0$.
For non-radial modes,
the past lightcone encompasses a broader range of times $t$,
but still it suffices to integrate the wave equation
over a finite initial range of time $t$.
For a wave of given spin and angular momentum,
it suffices to integrate the wave equation once,
and the result can be used to infer what various infallers see
who hit the singularity at a range of times $t$,
provided that the range of time $t$ over which the wave equation is integrated
is broad enough to encompass the past lightcones of the infallers.

The wave equation was solved using the NDSolve routine in Mathematica
over a domain in time $t$ and radius $r$.
To maintain accuracy near the horizon,
the radial coordinate was taken to be the logarithm $\ln |\Delta(r)|$
of the horizon function rather than the radius $r$ itself.
To avoid ambiguities of $2\pi$ in the phase of the wave,
the wave equation was solved for the wavefunction's logarithm
$\ln \varphi_\omegain$,
whose real and imaginary parts give the log-amplitude and phase of the wave.


\subsection{Confluent Heun eigenfunctions}
\label{heun-sec}

Eigenmodes of the wave equation~(\ref{phieq})
are confluent Heun functions.
Hawking modes from the sky above
originate as modes of definite rest-frame frequency $w$ at infinity,
and are therefore confluent Heun functions.
Hawking modes from the illusory horizon
originate as outgoing waves of pure frequency $\omegain$ from the perspective
of a freely-falling observer near the horizon,
equation~(\ref{phinearhorizon});
these are not eigenmodes of a single eigenfrequency,
but rather are superpositions of eigenmodes
with a continuum of eigenfrequencies $w$.

Confluent Heun functions
$\HeunC ( q ; \alpha, \gamma , \delta , \epsilon ; z )$
are solutions of the differential equation
\begin{equation}
\label{HeunCeq}
  \left[
  {\dd^2 \over \dd z^2}
  +
  \left(
  {\gamma \over z} + {\delta \over z - 1} + \epsilon
  \right)
  {\dd \over \dd z}
  +
  {\alpha z - q \over z ( z - 1 )}
  \right]
  \HeunC ( z )
  =
  0
  \ .
\end{equation}
From the differential equation~(\ref{HeunCeq})
it is straightforward to infer that
the confluent Heun function has a power series expansion about $z = 0$
absolutely convergent inside the circle $| z | < 1$ in the complex plane,
\begin{equation}
\label{heunseries}
  \HeunC ( q ; \alpha , \gamma , \delta , \epsilon ; z )
  =
  \sum_{i=0}^\infty c_i z^i
  \ ,
\end{equation}
with coefficients $c_i$ satisfying
\begin{equation}
  c_{-1} = 0
  \ , \quad
  c_0 = 1
  \ , \quad
  c_{i + 2}
  =
  {\left[ ( i + 1 ) ( i + \gamma + \delta - \epsilon ) - q \right] c_{i+1}
  + ( i \epsilon + \alpha ) c_i
  \over
  ( i + 2 ) ( i + 1 + \gamma )}
  \ .
\end{equation}

Eigenmodes of the wave equation~(\ref{phieq}) of spin $s$
that are respectively outgoing and ingoing at the horizon are
\begin{equation}
\label{phihorH}
  r^s
  \varphi_{\sigma}
  =
  {\ee^{- \im w ( t + r^\ast )}
  \over
  | r^2 \Delta |^{\sigma / 2}}
  \left\{
  \begin{array}{ll}
  \displaystyle
  \HeunC \left[
  q {-} \alpha ; {-} \alpha , \delta , \gamma , {-} \epsilon ; 1 {-} z
  \right]
  &
  \mbox{outgoing,}
  \\
  \displaystyle
  ( 1 - z )^{1 - \delta}
  \HeunC \left[
  q {-} \alpha {-} ( 1 {-} \delta ) ( \gamma {+} \epsilon ) ; {-} \alpha {-} ( 1 {-} \delta ) \epsilon , \delta , \gamma , {-} \epsilon ; 1 {{-}} z
  \right]
  &
  \mbox{ingoing,}
  \end{array}
  \right.
\end{equation}
with
$\sigma = -s$ for outgoing, and $\sigma = s$ for ingoing eigenmodes, and
\begin{equation}
  q
  =
  ( l {-} \sigma ) ( l {+} \sigma {+} 1 )
  \, , \ 
  \alpha
  =
  4 \im w M ( 1 {+} 2 \sigma )
  \, , \ 
  \gamma
  =
  1 {+} \sigma
  \, , \ 
  \delta
  =
  1 {+} \sigma {+} 4 \im w M
  \, , \ 
  \epsilon
  =
  4 \im w M
  \, , \ 
  z
  =
  {r \over 2 M}
  \ .
\end{equation}
The near-singularity solution
that matches the outgoing or ingoing eigensolution~(\ref{phihorH})
is some linear combination of eigenmodes with coefficients $C_1$ and $C_2$,
\begin{equation}
\label{phi0H}
  r^s
  \varphi_{\sigma}
  =
  | r^2 \Delta |^{\sigma / 2}
  \ee^{- \im w ( t - r^\ast )}
  \left[
  C_1
  H_1 ( z )
  +
  C_2
  z^{1-\gamma}
  H_2 ( z )
  \right]
  \ ,
\end{equation}
where the confluent Heun functions $H_1$ and $H_2$ are
\begin{equation}
  H_1 ( z )
  =
  \HeunC \left[
  q ; \alpha , \gamma , \delta , \epsilon ; z
  \right]
  \ , \quad
  H_2 ( z )
  =
  \HeunC \left[
  q {+} ( \gamma {-} 1 ) ( \delta {-} \epsilon ) ; \alpha {+} ( 1 {-} \gamma ) \epsilon , 2 {-} \gamma , \delta , \epsilon ; z
  \right]
  \ .
\end{equation}
The leading order behavior of the eigenfunctions near the singularity is
\begin{equation}
  H_1 ( z ) = 1 + O(z)
  \ , \quad
  z^{1-\gamma}
  H_2 ( z )
  = z^{-\sigma} \bigl( 1 + O(z) \bigr)
  \ .
\end{equation}
For zero spin, $\sigma = 0$, so $\gamma = 1$,
the two eigenmodes coincide, $z^{1-\gamma} H_2(z) = H_1(z)$,
and the independent eigenmodes can be taken to be $H_1(z)$ and
\begin{equation}
\label{H20}
  \lim_{\gamma \rightarrow 1}
  {
  z^{1-\gamma}
  H_2 ( z )
  -
  H_1 ( z )
  \over 1 - \gamma}
  =
  \ln z \bigl( 1 + O(z) \bigr)
  \ .
\end{equation}
If $\sigma$ is a negative integer, so $\gamma$ is an integer $\leq 0$,
then $\lim_{\gamma \rightarrow 1 + \sigma} z^{1 - \gamma} H_2(z) / c_{2 , |\sigma|} = H_1(z)$
where $c_{2 , |\sigma|}$
is the $|\sigma|$'th coefficient of the series expansion of $H_2(z)$,
and the independent eigenmodes can be taken to be $H_1(z)$ and
\begin{equation}
\label{H2m}
  \lim_{\gamma \rightarrow 1 + \sigma}
  \left[
  z^{1-\gamma}
  H_2 ( z )
  -
  c_{2 , |\sigma|} H_1 ( z )
  \right]
  =
  z^{-\sigma} \bigl( 1 + O(z \ln z) \bigr)
  \ .
\end{equation}
Likewise if $\sigma$ is a positive integer, so $\gamma$ is an integer $\geq 2$,
then $\lim_{\gamma \rightarrow 1 + \sigma} H_1(z) / c_{1,\sigma} = z^{1 - \gamma} H_2(z)$
where $c_{1 , \sigma}$
is the $\sigma$'th coefficient of the series expansion of $H_1(z)$,
and the independent eigenmodes can be taken to be $z^{1-\gamma} H_2 ( z )$ and
\begin{equation}
\label{H2p}
  \lim_{\gamma \rightarrow 1 + \sigma}
  \left[
  H_1 ( z )
  -
  c_{1 , \sigma}
  z^{1-\gamma}
  H_2 ( z )
  \right]
  =
  1
  +
  O(z \ln z)
  \ .
\end{equation}
For integral $\sigma$,
the power series expansions
of the eigenmodes~(\ref{H20})--(\ref{H2p})
in $z$ involve terms proportional
to $z^i \ln z$ as well as $z^i$.


\subsection{Asymptotic behavior near the singularity}
\label{asymptotesing-sec}

This sub-Appendix derives the asymptotic behavior
of waves near the singularity
from the series expansion of Heun
eigenfunctions of the wave equation~(\ref{phieq}).
For a wave of eigenfrequency $w$,
the asymptotic behavior holds at radii
\begin{equation}
  {r \over 2 M} \ll ( 2 w M )^{-1/2}
  \ .
\end{equation}

The behavior of eigenmodes near the singularity at $r = 0$
proves to depend on spin $s$ (the absolute value of the boost weight $\sigma$),
but not on either the eigenfrequency $w$ or the harmonic number $l$.
The reason for this generic behavior is that the amplitude
of an eigenmode near the singularity is dominated by
the larger of the two independent eigenfunctions,
which is a diverging factor of $z^{-|\sigma|}$ (or $\ln z$ if $\sigma = 0$) larger
than the smaller eigenfunction,
while for the physically relevant range $|\sigma| \leq 2$ of spins
the phase evolves in a fashion that to leading order
depends only on the smaller eigenfunction
(more precisely,
on the imaginary part of the ratio $C_\textrm{small} / C_\textrm{large}$
of the coefficients of the small to large eigenfunctions).
As long as the complex constants
$C_\textrm{large}$ and $C_\textrm{small}$ do not have the same phase
(in particular, neither $C_\textrm{large}$ nor $C_\textrm{small}$ is zero),
the frequency $\omega$ seen by a radially free-falling infaller
near the singularity is
\begin{equation}
\label{omegaH}
  \omega
  \equiv
  {\dd \psi \over \dd \tau}
  \approx
  | \sigma | \, \Im \left( {C_\textrm{small} / C_\textrm{large}} \right)
  r^{-3/2+|\sigma|}
  \quad
  ( |\sigma| \leq 2 )
  \ ,
\end{equation}
regardless of the eigenfrequency $w$ or harmonic number $l$.
For $\sigma = 0$,
replace the coefficient $| \sigma | \rightarrow 1 / ( \ln r )^2$
in equation~(\ref{omegaH}).
For higher-order spins $|\sigma| > 2$,
not only the amplitude but also the phase of the wave is dominated by
the larger of the two independent eigenfunctions,
and the frequency seen by an infaller is
$\omega \propto r^{1/2}$
regardless of the eigenfrequency $w$ or harmonic number $l$.
As a consequence of relation~(\ref{omegaH}),
the acceleration $\kappa$ of any generic wave
observed by an infaller near the singularity is,
for the physically relevant range $\sigma \leq 2$ of spins,
\begin{equation}
\label{kappa0H}
  \kappa
  \equiv
  -
  {\dd \ln \omega \over \dd \tau}
  =
  {2 |\sigma| - 3 \over 3 | \tau |}
  \quad
  ( |\sigma| \leq 2 )
  \ ,
\end{equation}
regardless of the eigenfrequency $w$ or harmonic number $l$.
For higher-order spins $|\sigma| > 2$, the acceleration is
$\kappa = {1 / ( 3 | \tau | )}$.
The acceleration~(\ref{kappa0H}) reproduces equation~(\ref{kappa0spin}).

\subsection{The wave equation in the geometric-optics limit}
\label{wavegeomoptics}

This sub-Appendix checks that the classical result~(\ref{kappa0s})
for the acceleration $\kappa_0$ on the illusory horizon below or sky above
is recovered from the wave equation~(\ref{phieq}) in
the high-frequency, low angular-momentum limit.

At high enough frequencies and finite angular momentum $l$,
the $\partial^2 / \partial t^2$ term in the wave equation~(\ref{phieq})
dominates the potential,
and the expression~(\ref{phinearhorizon}) then holds not only near the horizon
but also elsewhere.
This is the geometric-optics limit for radially outgoing waves.
Note that the geometric-optics limit inevitably fails near the singularity,
where the potential diverges, $V_\sigma \rightarrow \infty$.
In the geometric-optics limit,
the phase $\psiout ( \tauout )$ of the wave
along the path of a second infaller who hits the horizon
at time $t_0$ after the first infaller hit the horizon is
\begin{align}
\label{psihifreq}
  \psiout
  &\equiv
  -
  \Im \ln \varphi_\omegain
  =
  \mp
  {\omegain \over \kappa_\hor}
  \ee^{- \kappa_\hor ( \tout - \rastout )}
\\
\nonumber
  &=
  {\omegain \over \kappa_\hor}
  \left( 1 - \sqrt{\rout \over 2 M} \right)
  \exp
  \left[
  - \,
  \kappa_\hor t_0
  -
  \frac{11}{6}
  +
  \sqrt{\rout \over 2 M}
  +
  {\rout \over 4 M}
  +
  \frac{1}{3}
  \left( {\rout \over 2 M} \right)^{3/2}
  \right]
  \ .
\end{align}
The phase $\psiout$
increases continuously and smoothly as the infaller crosses the true horizon
at $\rout = 2 M$.
The frequency observed by the infaller is
\begin{align}
  \omegaout
  &\equiv
  {\dd \psiout \over \dd \tauout}
\\
  \nonumber
  &=
  \omegain
  \sqrt{\rout \over 2 M}
  \exp
  \left[
  - \,
  \kappa_\hor t_0
  -
  \frac{11}{6}
  +
  \sqrt{\rout \over 2 M}
  +
  {\rout \over 4 M}
  +
  \frac{1}{3}
  \left( {\rout \over 2 M} \right)^{3/2}
  \right]
  \ .
\end{align}
The frequency observed by the second infaller crossing the horizon
at time $t_0$ after the first infaller
is redshifted
by a factor $\ee^{- \kappa_\hor t_0}$.
The observed acceleration is
\begin{equation}
  \kappa
  \equiv
  - {\dd \ln \omegaout \over \dd \tauout}
  \ ,
\end{equation}
the explicit expression for which agrees with
the classical result~(\ref{kappa0s})
for zero angular momentum photons from the illusory horizon.

Similarly,
for an ingoing wave of definite frequency $w$ emitted from the sky above,
the phase $\psi ( \tau )$ of the wave along the path of an infaller is
\begin{equation}
  \psi
  =
  w ( t + r^\ast )
  =
  {w \over \kappa_\hor}
  \left[
  \ln \left( 1 + \sqrt{r \over 2 M} \right)
  -
  \sqrt{r \over 2 M}
  +
  {r \over 4 M}
  -
  \frac{1}{3}
  \left( {r \over 2 M} \right)^{3/2}
  \right]
  \ .
\end{equation}
The corresponding frequency observed by the infaller is
\begin{equation}
  \omega
  \equiv
  {\dd \psi \over \dd \tau}
  =
  w
  {\sqrt{r / 2 M} \over 1 + \sqrt{r / 2 M}}
  \ .
\end{equation}
The observed acceleration
$\kappa \equiv - \dd \ln \omega / \dd \tau$
agrees with
the classical result~(\ref{kappa0s})
for zero angular momentum photons from the sky above.

\section{Two-term, power-law solutions for conserved relativistic energy-momenta}

In the Schwarzschild geometry,
equations~(\ref{conservedT})
for the conserved energy-momentum tensor
of a relativistic, hence traceless, fluid
have two-term, power-law solutions
proportional to arbitrary powers $r^n$ of radius,
consisting of the sum of
an anisotropic relativistic fluid with
$\{ \rho , p , p_\perp \} = \rho_\aniso \{ 1 , 1 , 0 \}$
and an isotropic relativistic fluid with
$\{ \rho , p , p_\perp \} = \rho_\iso \{ 1 , \tfrac{1}{3} , \tfrac{1}{3} \}$,
with respective energy densities
\begin{subequations}
\label{rhopower}
\begin{align}
  \rho_\aniso
  &=
  C
  r^n
  \left[
  ( 2 + n ) ( 5 + n ) - \tfrac{1}{3} ( 1 + n ) ( 4 + n ) {r \over 2 M}
  \right]
  \ ,
\\
  \rho_\iso
  &=
  C
  r^n
  \left[
  - \, ( 1 + n ) ( 5 + n ) + ( 3 + n ) ( 4 + n ) {r \over 2 M}
  \right]
  \ ,
\end{align}
\end{subequations}
where $C$ is some constant.
The fluxes in the two components satisfy equation~(\ref{fH}).

\begin{acknowledgements}	
This research was supported in part by FQXI mini-grant FQXI-MGB-1626.
I thank Prof.\ M.\ J.\ Duff for confirming
that massive as well as massless fields should be taken
into account in calculating the trace anomaly.
\end{acknowledgements}		


\bibliographystyle{spphys}       
\bibliography{bh}

\end{document}